\documentclass[a4paper,11pt]{article}
\pdfoutput=1
\usepackage{jheppub}
\usepackage[T1]{fontenc} 

\usepackage{graphicx}
\usepackage{xcolor}
\usepackage{multirow}
\usepackage{xspace}
\usepackage{slashed}

\newcommand{\fm}{\,\text{fm}}
\newcommand{\MeV}{\,\text{MeV}}

\newcommand{\bx}{\boldsymbol{x}}
\newcommand{\by}{\boldsymbol{y}}
\newcommand{\bb}{\boldsymbol{b}}
\newcommand{\be}{\boldsymbol{e}}

\newcommand{\calO}{\mathcal{O}}

\newcommand{\calN}{\mathcal{N}}

\newcommand{\calD}{\mathcal{D}}

\newcommand{\nB}{n_{\rm B}}

\newcommand{\Msun}{M_\odot}

\newcommand{\eqdef}{\equiv}
\DeclareMathOperator*{\Regret}{Reg}
\newcommand{\RegInv}[1]{\Regret[#1]^{-1}}
\DeclareMathOperator{\erf}{erf}
\newcommand{\MR}{$M$-$R$\xspace}

\newcommand{\TOVmap}{\Psi_{\text{TOV}}}

\newcommand{\sumint}[1]{{\hbox{$\sum$}\!\!\!\!\!\!\!\int\,}_{\!\!\!\!\raise-0.1ex\hbox{$\scriptstyle{#1}$}}}
\newcommand{\tsumint}[1]{{\hbox{$\Sigma$}\!\!\!\!\raise0.3ex\hbox{$\int$}\,}_{\!\!\raise-0.1ex\hbox{$\scriptstyle{#1}$}}}

\newcommand{\Tinti}[1]{{{\Sigma}\!\!\!\!\raise0.3ex\hbox{$\int$}_\rmii{${#1}$}}}

\title{Extensive Studies of the Neutron Star Equation of State
  from the Deep Learning Inference with the Observational Data
  Augmentation}

\author[a]{Yuki Fujimoto}
\author[a,b]{Kenji Fukushima}
\author[c]{Koichi Murase}

\affiliation[a]{Department of Physics, The University of Tokyo,
  7-3-1 Hongo, Bunkyo-ku, Tokyo 113-0033, Japan}
\affiliation[b]{Institute for Physics of Intelligence (IPI),
  The University of Tokyo, 7-3-1 Hongo, Bunkyo-ku, Tokyo 113-0033, Japan}
\affiliation[c]{Center for High Energy Physics, Peking University, Beijing 100871, China}

\emailAdd{fujimoto@nt.phys.s.u-tokyo.ac.jp}
\emailAdd{fuku@nt.phys.s.u-tokyo.ac.jp}
\emailAdd{murase@pku.edu.cn}

\abstract{
  We discuss deep learning inference for the neutron star equation
  of state (EoS) using the real observational data of the mass and the
  radius.  We make a quantitative comparison between the conventional
  polynomial regression and the neural network approach for the EoS
  parametrization.  For our deep learning method to incorporate
  uncertainties in observation, we augment the training data with
  noise fluctuations corresponding to observational uncertainties.
  Deduced EoSs can accommodate a weak first-order phase transition,
  and we make a histogram for likely first-order regions.  We also
  find that our observational data augmentation has a byproduct to
  tame the overfitting behavior.  To check the performance improved by
  the data augmentation, we set up a toy model as the simplest
  inference problem to recover a double-peaked function and monitor
  the validation loss.  We conclude that the data augmentation could
  be a useful technique to evade the overfitting without tuning the
  neural network architecture such as inserting the dropout.
}

\begin{document}
\maketitle

\section{Introduction}

In the central cores of neutron stars baryonic matter is highly
compressed by the gravitational force.  The inward force by gravity is
balanced by the outward pressure which strongly depends on intrinsic
properties of cold and dense matter subject to the strong interaction.
The study of inner structures of the neutron star requires a relation
between the pressure $p$ and the energy density $\varepsilon$; namely,
the equation of state (EoS), $p=p(\varepsilon)$, at $T=0$ (see, e.g.,
Refs.~\cite{Lattimer:2012nd, Ozel:2016oaf, Baym:2017whm,
  Baiotti:2019sew, Kojo:2020krb} for recent reviews).  At the center
the baryon number density $\nB$ may reach 5--10 times as large as the
saturation density of nuclear matter, i.e., $n_0 \simeq
0.16\,\text{(nucleons)}/\text{fm}^3$.

Towards the model-independent EoS determination, we should solve
quantum chromodynamics (QCD), which is the first-principles theory of
the strong interaction.  At the moment, however, the model-independent
results are only available at limited ranges of densities.  At low
density range of $\nB \sim 1$--$2\,n_0$ we can apply
\textit{ab initio} methods combined with the nuclear force derived
from Chiral Effective Theory ($\chi$EFT) with controlled uncertainty
estimates~\cite{Hebeler:2009iv, Gandolfi:2011xu, Tews:2012fj,
  Holt:2013fwa, Hagen:2013yba, Roggero:2014lga, Wlazlowski:2014jna,
  Tews:2018kmu, Drischler:2020hwi} (see Ref.~\cite{Drischler:2021kxf}
for a recent review).  At asymptotically high densities of
$\nB \gtrsim 50\,n_0$ perturbative QCD calculations reliably
converge~\cite{Freedman:1976xs, Freedman:1976ub, Baluni:1977ms,
  Kurkela:2009gj, Fraga:2013qra, Gorda:2018gpy}
(see Ref.~\cite{Ghiglieri:2020dpq} for a recent review).  The
intermediate density region around $2$--$10\,n_0$, which is relevant
for the neutron star structure, still lacks trustable QCD predictions
(see also Ref.~\cite{Fujimoto:2020tjc} for a recent attempt to improve
the convergence in the resummed perturbation theory).  To tackle the
intermediate density region from QCD, we need to develop
non-perturbative methods such as the Monte-Carlo simulation of QCD on
the lattice (lattice QCD), but the lattice-QCD application to finite
density systems is terribly hindered by the notorious sign problem
(see, e.g., Ref.~\cite{Aarts:2015tyj} for a review).  This is why
neutron-star-physics studies still rely on phenomenological EoS
constructions:  rather \textit{ab initio}
approaches~\cite{Akmal:1998cf, Togashi:2017mjp} near the saturation
density, estimates employing the Skyrme
interactions~\cite{Douchin:2001sv}, the relativistic mean field
theories~\cite{Serot:1997xg}, and the functional renormalization
group~\cite{Drews:2016wpi}, etc.

Meanwhile, we have witnessed significant advances in neutron star
observations over the last decades.  Now these advances have opened an
alternative pathway for the model-independent extraction of the EoS by
statistical methods.  Observations include the Shapiro delay
measurements of massive two-solar-mass pulsars~\cite{Demorest:2010bx,
  Fonseca:2016tux, Antoniadis:2013pzd, Cromartie:2019kug}, the radius
determination of quiescent low-mass X-ray binaries and thermonuclear
bursters~\cite{Ozel:2010fw, Steiner:2010fz, Steiner:2012xt,
  Ozel:2015fia, Bogdanov:2016nle} (see
reviews~\cite{Miller:2013tca, Ozel:2016oaf, Miller:2016pom} for
discussions on the methods and associated uncertainties), detections of
gravitational waves from binary mergers involving neutron stars by the
LIGO-Virgo collaboration~\cite{TheLIGOScientific:2017qsa, Abbott:2020uma}
as well as the X-ray timing measurements of pulsars by the NICER
mission~\cite{Riley:2019yda, Miller:2019cac}.
Typical observable quantities of neutron stars involve mass $M$,
radius $R$, compactness $M/R$, tidal deformability $\Lambda$ (and
their variants, e.g., Love number $k_2$), quadrupole moment
$Q$, moment of inertia $I$, etc.  The gravitational waves from binary
neutron star mergers provide us with the information about the tidal
deformability.  Also, the NICER mission particularly targets at the
compactness $M/R$ of stars by measuring the gravitational lensing of
the thermal emission from the stellar surface.
Some of these neutron star quantities are connected through the
universal relations that are insensitive to the EoS details.  Among
such relations, the most well-known example is the I-Love-Q
relation~\cite{Yagi:2013bca, Yagi:2013awa}, which relates the moment
of inertia $I$, the Love number $k_2$, and the quadrupole moment $Q$
(see Refs.~\cite{Yagi:2016bkt, Baiotti:2019sew} and reference therein
for other universal relations and usages).

These observations have provided some insights on the EoS and invoked
discussions about non-perturbative aspect of QCD:  for example, the
emergence of quark matter in the neutron star~\cite{Annala:2019puf}
and a certain realization of duality between the confined and the
deconfined matter~\cite{Masuda:2012kf, Fujimoto:2019sxg,
  McLerran:2007qj, Fukushima:2015bda, McLerran:2018hbz, Jeong:2019lhv,
  Duarte:2020xsp, Zhao:2020dvu, Fukushima:2020cmk}.
Moreover, these observations, particularly the gravitational wave
observation, may provide a unique opportunity for studying the
hadron-to-quark phase transition occurring in the binary star merger
and even better sensitivity will be achieved with the future
detector~\cite{Most:2018eaw, Bauswein:2018bma, Most:2019onn}.

For the purpose of extracting the most likely EoS from the observational
data, there are diverse statistical technologies.  The commonly used
method is the Bayesian inference~\cite{Ozel:2010fw, Steiner:2010fz,
  Steiner:2012xt, Alvarez-Castillo:2016oln, Raithel:2016bux,
  Raithel:2017ity, Raaijmakers:2019dks, Raaijmakers:2019qny, Blaschke:2020qqj}.
There are still other methods such as the one based on the Gaussian
processes which is a variation of the Bayesian inferences with
nonparametric representation of the
EoS~\cite{Landry:2018prl, Essick:2019ldf, Essick:2020flb}, etc.
Despite numerous efforts to quantitatively constrain the EoS, it is
still inconclusive what the genuine EoS should look like because of
uncertainties from the assumed prior distribution in the Bayesian
analysis.  Therefore, the EoS inference program is in need of
complementary tools.  As an alternative to these conventional
statistical methods, the \textit{deep learning}; namely, the machine
learning devising deep neural networks (NNs), has been successfully
applied to a wide range of physics fields~\cite{Pang:2016vdc,
  Mori:2017pne, Hezaveh:2017sht, Chang:2017kvc, Niu:2018csp,
  Kaspschak:2020ezh, Wang:2020hji}.
Several studies have already put the NN in motion in the context of
gravitational wave data analysis of binary neutron star
mergers~\cite{George:2016hay, Carrillo:2016kvt, George:2017pmj, George:2018awu}.
Along these lines, in our previous
publications~\cite{Fujimoto:2017cdo, Fujimoto:2019hxv} we proposed
a novel approach to the EoS extraction based on the deep machine
learning (see also
Refs.~\cite{Ferreira:2019bny,  Morawski:2020izm, Traversi:2020dho} for
related works).

In the present work we make further comprehensive analyses using the
machine learning method developed
previously~\cite{Fujimoto:2017cdo, Fujimoto:2019hxv}.
Here we give a brief synopsis of our machine learning approach to the
neutron star EoS problem.  Given an EoS, various pairs of stellar mass
and radius, \MR, follow from the general relativistic structural
equation, i.e., Tolman-Oppenheimer-Volkoff (TOV)
equation~\cite{Tolman:1939jz, Oppenheimer:1939ne}.  We will express
the \emph{inverse} operation of solving the TOV equation in terms of
the deep NN, train the NN with sufficiently large data set of mock
observations, and eventually obtain a regressor that properly infers
the most likely EoS corresponding to the observational \MR inputs.
While the other approaches to the neutron star EoS explicitly assume a
prior distribution for the EoS probability distributions, our method
only implicitly depends on the prior assumption.  Therefore, in a
sense that the implementations are based on different principles and
the prior dependences appear differently, we can consider our method
as complementary to other conventional approaches.  We also note that
independent algorithms would help us with gaining some information
about systematics.  For more detailed discussions on the prior, see
Sec.~\ref{sec:basic-strategy}.

This paper is organized as follows.
Section~\ref{sec:nseos} provides a detailed account on our problem
setting, i.e., mapping neutron star observables to the EoS\@.  Then,
we outline our method by setting forth our parametrization for
the EoS and reviewing the machine learning technique in general.
Section~\ref{sec:mock} is devoted for the methodology study using mock
data.  We describe our data generating procedure and training
setups, and then give a brief overview of the evolution of learning
processes, which motivates the later analyses in
Sec.~\ref{sec:curve}.  We showcase the typical examples out of the
whole results, and then we carry out statistical analyses for the data
as a whole.  We confront our NN method with a different method;
namely, the polynomial regression, and compare two methods to show
that our method surpasses the polynomial regression.  In
Sec.~\ref{sec:obs}, we present the EoSs deduced from the several
different neutron star observables in real.  We explain our treatment
of uncertainty quantification along with the possibility of a weak
first-order phase transition in the deduced EoS results.  In
Sec.~\ref{sec:curve} we study the efficiency of data augmentation in
the context to remedy the overfitting problem.  Finally, we summarize
this work and discuss future outlooks in Sec.~\ref{sec:sum}.
Throughout this paper we use the natural unit system with $G = c = 1$
unless otherwise specified.

\section{Supervised Learning for the EoS Inference Problem}
\label{sec:nseos}

In this section we explicitly define our problem and summarize
the basic strategy of our approach with the supervised machine
learning.  We will explain the concrete setup in each subsection where
we adjust the strategy in accordance with the goal of each subsection.
In the present study we want to constrain the EoS from the stellar
observables.  The EoS and the observables are non-trivially linked by
the TOV equation which gives a means to calculate the neutron star
structure from the EoS input.  Thus, constraining the EoS from the
observables is the inverse process of solving the TOV equation, but
this inverse problem encounters difficulties from the nature of the
observations.  In Sec.~\ref{sec:TOV-and-inverse-problem} we formulate
the inverse problem of the TOV equation and discuss its difficulties.
We proceed to our approach to this problem using the supervised
machine learning in Sec.~\ref{sec:basic-strategy}.  We closely
describe the EoS parametrization and the data generation in
Sec.~\ref{sec:datagen}.  Then, we explain the design of the deep NN in
Sec.~\ref{sec:nnmethod} and its training procedures in
Sec.~\ref{sec:general-training}.

\subsection{TOV mapping between the EoS and the \MR relation}
\label{sec:TOV-and-inverse-problem}

\begin{figure}
  \centering
  \includegraphics[width=0.75\textwidth]{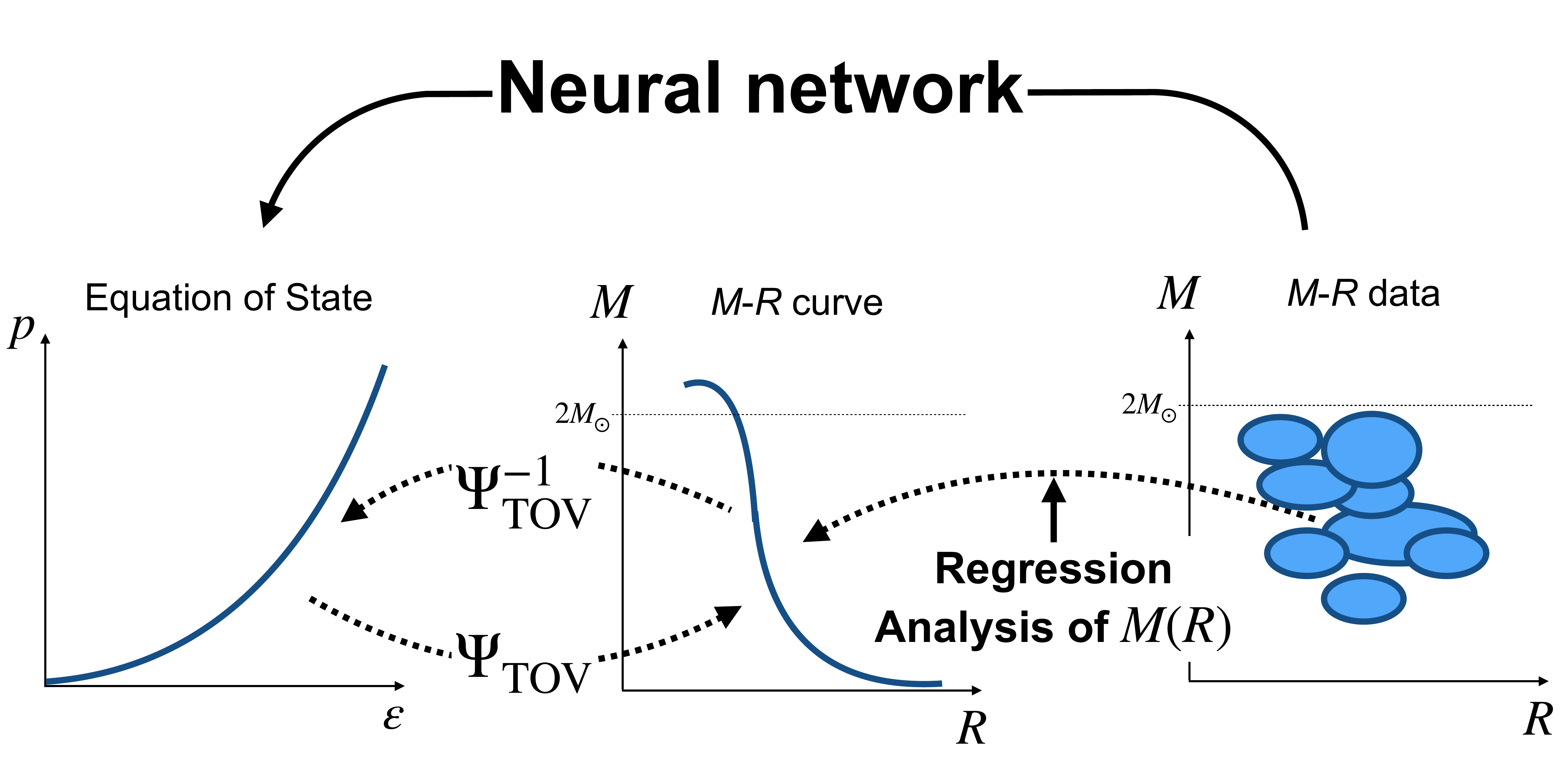}
  \caption{Schematic of the TOV mapping $\Psi_\text{TOV}$ and the
    regression analysis of the inverse TOV mapping from the \MR data.}
  \label{fig:tovreg}
\end{figure}

In the present analysis we focus on the mass $M$ and the radius $R$ as
the neutron star observables.  Given a boundary condition of the core
pressure $p_c$, the observables $M$ and $R$ of such a neutron star can
be determined by solving the TOV
equation~\cite{Tolman:1939jz, Oppenheimer:1939ne}:
\begin{align}
  \frac{dp(r)}{dr} &= -\frac{[\varepsilon(r)+p(r)][m(r)+4\pi r^3 p(r)]}{r[r-2m(r)]}\,, \label{eq:TOV} \\
  m(r) &= 4\pi \int_0^r r'^2 dr'\,  \varepsilon(r')\,, \label{eq:TOV-mass}
\end{align}
where $r$ is the radial coordinate which represents a distance from
the stellar center.  The functions, $p(r)$ and $\varepsilon(r)$, are
the pressure and the energy density (i.e., the mass density),
respectively, at the position $r$.  The function $m(r)$ represents a
mass enclosed within the distance $r$.  We can readily integrate these
integro-differential equations from $r=0$ with the initial condition,
$p(r=0)=p_c$, towards the outward direction.  The radius of the
neutron star, $R$, is defined by the surface condition:  $p(R)=0$.
The mass of the neutron star is the total mass within $R$, i.e.,
$M = m(R)$.

The only missing information in the combined TOV
equations~\eqref{eq:TOV} and \eqref{eq:TOV-mass} is a relationship
between $p$ and $\varepsilon$, which is nothing but the EoS, i.e.,
$p=p(\varepsilon)$.  Once an EoS is fixed, we can draw a $p_c$-parametrized
trajectory of $(M(p_c), R(p_c))$ in the \MR plane.  This
trajectory is called the \MR relation.  Thus, the operation of solving
the TOV equation can be considered as a mapping from the functional space
of the EoSs onto the functional space of the \MR relations.  Here we
call this mapping the ``TOV mapping'' denoted by $\TOVmap$:
\begin{align}
  \Psi_\text{TOV}: &\quad \text{(EoS)}\quad\mapsto\quad\text{(\MR relation)}\,.
\end{align}
It is known that the mapping is invertible, i.e., the inverse mapping
$\TOVmap^{-1}$ exists~\cite{Lindblom:1992}.  Thus, \textit{ideally},
one could uniquely reconstruct the EoS from an observed \MR relation.
However, \textit{practically}, we cannot access the entire continuous
curve on the \MR relation (which we will briefly call the \MR curve)
from neutron star observations.  In reconstructing the EoS from the
observational data, we are facing twofold obstacles:
\begin{enumerate}
\item \textit{Incompleteness} ---
  \MR data from the observation does not form a continuous curve.
  Since one neutron star corresponds to one point on the \MR curve,
  the observation of neutron stars gives a limited number of \MR
  points and only partial information on the original \MR curve is
  available.  Even with a sufficiently large number of neutron stars,
  the information on possible unstable branches of the \MR curve, on
  which the neutron stars do not exist, would be missing.
\item \textit{Uncertainties} ---
  Real data is not a point but a distribution.  Each \MR point is
  accompanied by observational uncertainties represented as a
  credibility distribution.  Moreover, the central peak position of
  the distribution is not necessarily on the genuine \MR relation.
\end{enumerate}
We can introduce the following symbolic representation of the above
complication associated with the observational limitation:
\begin{align}
  \Pi_\text{obs}(\omega): &\quad \text{(\MR relation)}\quad\mapsto\quad\text{(\MR data)}_\omega\,.
\end{align}
This mapping by $\Pi_\text{obs}(\omega)$ is a convolution of a
projection
from the \MR relation to a set of points along the \MR curve and a
probability distribution due to observational uncertainties.  The
distribution generally depends on observational scenarios (how many
and what kinds of neutron stars are detected by which signals) and
such dependence is collectively represented by $\omega$.  With this
mapping, the relation between the EoS and the observed \MR data is
more correctly expressed as
\begin{align}
  \Pi_\text{obs}(\omega)\circ\TOVmap: &\quad \text{(EoS)}\quad\mapsto\quad\text{(\MR data)}_\omega\,.
\end{align}
Because of this additional filter by $\Pi_\text{obs}(\omega)$,
one cannot reconstruct the correct EoS from the observational data
just by calculating $\TOVmap^{-1}$.  Since a part of original
information is lost through $\Pi_\text{obs}(\omega)$, this is an
ill-posed problem, i.e., there is no unique well-defined answer due to
insufficient information and uncertainties inherent in this
problem.

Here, instead of reconstructing the \textit{unique EoS}, we try to
infer the \textit{most likely EoS} from \MR observations with
uncertainty; that is, the regression analysis of the inverse mapping
including $\Pi_\text{obs}(\omega)$ (see Fig.~\ref{fig:tovreg} for a
schematic illustration).  In particular it is crucial to develop an
approach leading to outputs robust against the observational
uncertainties of the \MR data.

\subsection{General strategy for the machine learning implementation}
\label{sec:basic-strategy}

To solve this ill-posed problem we employ a method of machine learning
with deep NNs to infer the neutron star EoS\@.  In particular we
consider the supervised machine learning to train the NN with as many
and different $\omega$'s as possible.  The trained NN can then receive
the \MR data for one particular $\omega$ and return the output of the
most likely EoS correspondingly.  To put it another way, the
``inversion'' of the mapping $\Pi_\text{obs}\circ\TOVmap$ is
approximated by the regression, particularly by the deep NN in our
approach, which would be symbolically written as
$\RegInv{\Pi_\text{obs}\circ\TOVmap}$.

In the supervised learning, we need to prepare the training data and
the validation data composed of pairs of the input \MR data and the
desired output EoS\@.  To generate the training data efficiently, we
make use of the asymmetry between
$\Pi_\text{obs}\circ\TOVmap$ and
$\RegInv{\Pi_\text{obs}\circ\TOVmap}$; that is, we can
straightforwardly calculate the forward mapping of
$\Pi_\text{obs}\circ\TOVmap$ by modeling the observation
$\Pi_\text{obs}(\omega)$, while the latter inverse mapping, which is
what we currently want to know, is more non-trivial.
Thus, first, we randomly generate many possible answer EoSs
represented by several parameters.  We next generate the corresponding
\MR data by applying the forward mapping with various $\omega$'s.
We will explain technical details of handling and simplifying $\omega$
in Sec.~\ref{sec:datagen}.

Now, we turn to the description of the architecture of our NN model
and we will optimize the model parameters so that the model can infer
the answer EoS corresponding to the input training \MR data.  During
the training, it is important to monitor the training quality by
checking the prediction error behavior for the validation data.  After
the optimization process is complete with good convergence of the
validation error, we can test the predictive power using the mock data
for which the true answer EoS is known (see Sec.~\ref{sec:mock} for
actual calculations).  Once the method passes all these qualification
tests, we finally proceed to the application of the real observational
data to obtain the most realistic EoS (see Sec.~\ref{sec:obs} for
details).

Here, we comment on an alternative possibility of the inference
problem formulation.  As already mentioned, the inverse TOV mapping,
$\TOVmap^{-1}$, is well-defined by itself.
Thus, it is also feasible to decompose the inference as
$\RegInv{\Pi_\text{obs}\circ\TOVmap} = \TOVmap^{-1} \circ \RegInv{\Pi_\text{obs}}$
and train the NN aimed to approximate $\RegInv{\Pi_\text{obs}}$; that
is, the NN model would predict the \MR curve from the input \MR data.
We will not adopt this strategy since it is unclear in this approach
how to measure the reconstruction performance of the \MR curve.  Our
target space is the EoS space, but the distance in the \MR space is
not uniformly reflected in the distance in the desired EoS space.  We
are ultimately interested in the EoS, so it is natural to directly
model $\RegInv{\Pi_\text{obs}\circ\TOVmap}$ by maximizing the
performance for the EoS reconstruction.

We also make a remark on the \textit{discriminative modeling} and the
\textit{generative modeling} to clarify a difference between our
machine learning method and other statistical approaches such as the
Bayesian inference.  The approach we are advocating in this paper
belongs to the discriminative modeling: our NN model directly predicts
the response variable, i.e., the EoS, for a fixed set of explanatory
variables, i.e., the \MR data.  In contrast, the statistical
approaches such as the Bayesian inference for the neutron star EoS are
to be categorized as the generative modeling.  The Bayesian inference
first models occurrence of the combination of the explanatory and
response variables by the joint probability,
$\Pr(\text{\MR data}, \text{EoS}) \eqdef \Pr(\text{\MR data}|\text{EoS}) \Pr(\text{EoS})$.
Here, $\Pr(\text{EoS})$ models the prior belief on the EoS,
and $\Pr(\text{\MR data}|\text{EoS})$ models the observation process.
Then, the Bayes theorem gives a posterior distribution of EoS,
$\Pr(\text{EoS}|\text{\MR data})$, from the joint probability, which
corresponds to the likely EoS distribution for given \MR data.

These discriminative and generative approaches can be considered to be
complementary to each other.  They have the pros and cons as follows:
\begin{itemize}
\item
  The advantage in the generative approaches is that the EoS
  prediction for a given \MR data takes a distribution form, so that
  the prediction uncertainties can be estimated naturally.
  In the discriminative model, on the other hand, additional analysis
  is needed to estimate the prediction uncertainties.  We will perform
  the additional uncertainty analysis within our method in
  Sec.~\ref{sec:UQ}.  At the same time, we will utilize our
  uncertainty analysis for a physics implication as discussed in
  Sec.~\ref{sec:uncertainty-quantification}.
\item
  In the discriminative modeling, once the NN training is completed,
  the most likely EoS corresponding to a given \MR data can
  immediately result from the mapping represented by the NN:  the NN
  models the inference procedure by itself.  Thus, the computational
  cost is much smaller than statistical analyses, which enables us to
  perform supplementary estimation on observables and/or quality
  assessments if necessary.  In the generative approaches, on the
  other hand, the posterior calculation is computationally expensive
  because the EoS parameter space is large.  In addition, one needs to
  calculate the posterior for each \MR data separately.  If one
  applies the Bayesian inference to the real observational data only
  once, this computational cost would not matter.  The computational
  cost becomes problematic, however, if one wants to examine the
  statistical nature of the inference using a large number of
  generated EoSs and \MR data.
\end{itemize}

\subsection{Random EoS generation with parametrization by the speed
  of sound}
\label{sec:datagen}

The training data consists of pairs of the input \MR data and the
output EoS, $p(\varepsilon)$, parametrized by a finite number of
variables.  We first generate EoSs by randomly assigning values to the
EoS parameters, solve the TOV equation to find the corresponding \MR
curve, and finally generate the \MR data.  The \MR data generation
involves $\Pi_\text{obs}(\omega)$, namely, we sample observational
points of neutron stars on the \MR curve, and probabilistically
incorporate finite observational uncertainties, which should mimic
observations in the scenario $\omega$.  The real \MR data is composed
of the probability distribution for each neutron star on the \MR plane,
but in practice we should simplify $\omega$ and need to parametrize
the \MR data as well.  We will discuss the parametrization later and
in this section we focus on our EoS parametrization and generation.

In designing the EoS parametrization, it is important to consider its
affinity to the physical constraints such as the thermodynamic
stability and the causality, i.e, $p$ should be a monotonically
non-decreasing function of $\varepsilon$, and the speed of sound,
$c_s$, should not be larger than the speed of the light.  For this reason,
$c_s$, which is derived from
\begin{align}
  c_s^2 = \frac{\partial p}{\partial\varepsilon}\,,
  \label{eq:sound}
\end{align}
is the convenient choice as the EoS parameters to impose the physical
constraints easily.

Also, we need to smoothly connect the parametrized EoS to the
empirical EoS known from the low density region.  Up to the energy
density of the saturated nuclear matter, $\varepsilon_0$, we use a
conventional nuclear EoS, specifically SLy4~\cite{Douchin:2001sv}.
Because the EoS is well constrained by the saturation properties and
the symmetry energy measurements near the saturation density, this
specific choice of SLy4 would not affect our final results.  In the
energy region above $\varepsilon_0$ we employ the standard piecewise
polytropic parametrization and partition a density range into a
certain number of segments.  In our present analysis we choose the
density range, $[\varepsilon_0,\; 8\varepsilon_0]$, and six energy
points, $\varepsilon_{i}$ ($i=0,\dots,5$) with
$\varepsilon_5=8\varepsilon_0$, equally separated in the logarithmic
scale.  Then, $(\varepsilon_{i-1},\; \varepsilon_{i})$ ($i=1,\dots 5$)
form 5 segments.  To be more specific, these values are
$(\varepsilon_0, \varepsilon_1, \varepsilon_2, \varepsilon_3,
\varepsilon_4, \varepsilon_5) = (150, 227, 345, 522, 792, 1200) \MeV/\fm^3$.
The EoS parameter is the average speed of sound,
$c_{s,i}^2 \equiv \langle c_s^2\rangle = \langle\partial p/\partial \varepsilon\rangle$ ($i = 1,\dots,5$),
in $i$-th segment.
From these definitions we see that the pressure values at the $i$-th
segment boundaries, $p_{i}$, are read as%
\footnote{
  We can confirm that $c_{s,i}^2$ is indeed an average in each segment under this construction as \begin{align*}
  \langle c_{s}^{2}\rangle
  &\eqdef \int_{\varepsilon_{i-1}}^{\varepsilon_{i}} \frac{d\varepsilon}{\varepsilon_{i} - \varepsilon_{i-1}} c_{s}^{2}
  = \int_{\varepsilon_{i-1}}^{\varepsilon_{i}}
  \frac{d\varepsilon}{\varepsilon_{i} - \varepsilon_{i-1}}
  \frac{\partial p}{\partial \varepsilon} \\
  &= \frac1{\varepsilon_{i} - \varepsilon_{i-1}} \int_{p_{i-1}}^{p_{i}} dp
  = \frac{p_{i} - p_{i-1}}{\varepsilon_{i} - \varepsilon_{i-1}}
  = c_{s,i}^{2}. \end{align*}}
\begin{align}
p_{i} = p_{i-1} + c_{s,i}^{2} (\varepsilon_{i} - \varepsilon_{i-1})\,,
\end{align}
where $p_{0}$ is determined by $p_0=p(\varepsilon_0)$ from our chosen
nuclear EoS, i.e., SLy4.  We make polytropic interpolation for the EoS
by $p = k_{i} \varepsilon^{\gamma_{i}}$ whose exponent and coefficient
are given by
\begin{align}
\gamma_{i} = \frac{\ln(p_{i} / p_{i-1})}{\ln(\varepsilon_{i} / \varepsilon_{i-1})}\,,
  \qquad
k_{i} = \frac{p_{i}}{\varepsilon_{i}^{\gamma_{i}}}\,.
\label{eq:polytrope}
\end{align}

\begin{figure}
    \centering
    \includegraphics[width=0.48\textwidth]{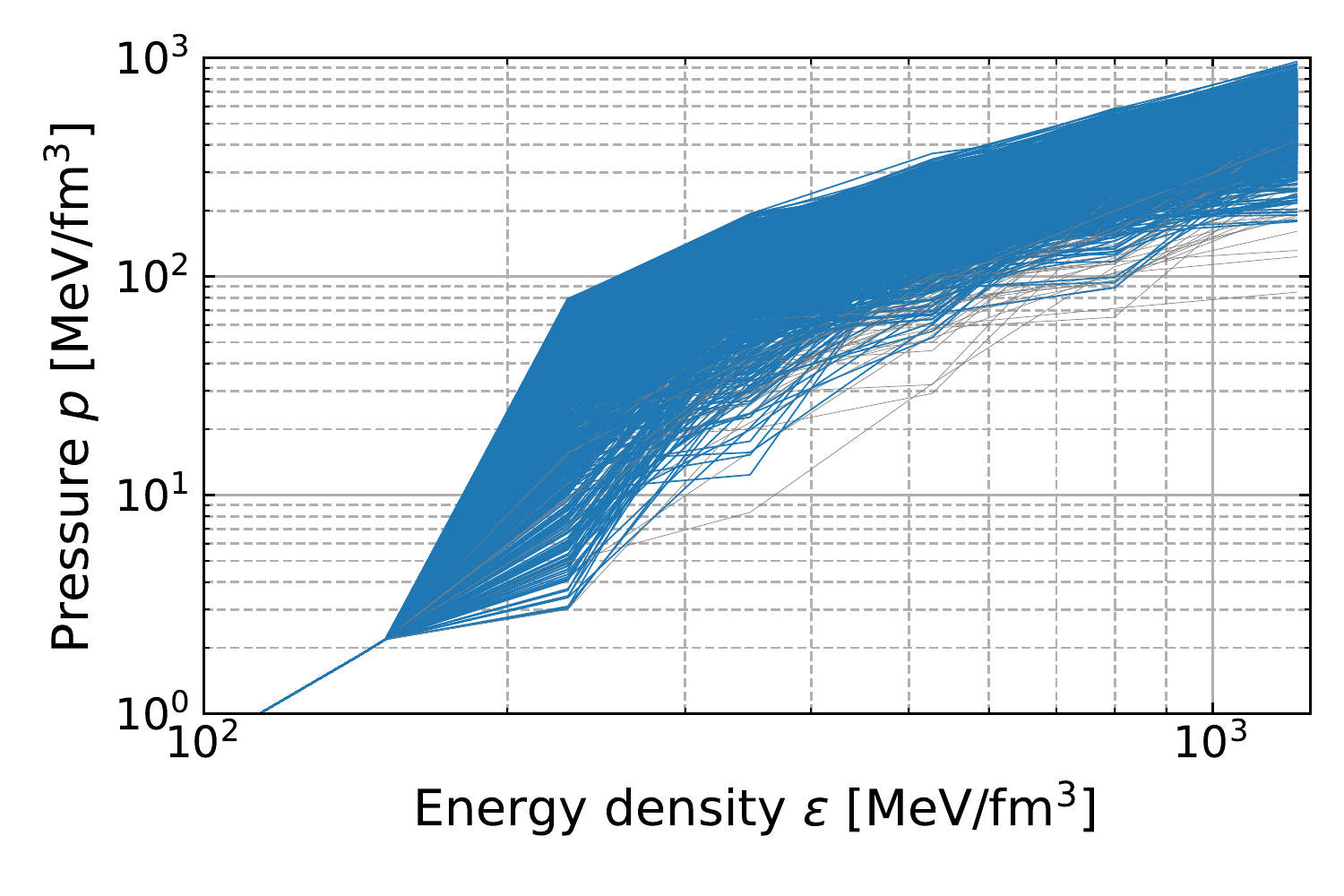} \hspace{0.5em}
    \includegraphics[width=0.48\textwidth]{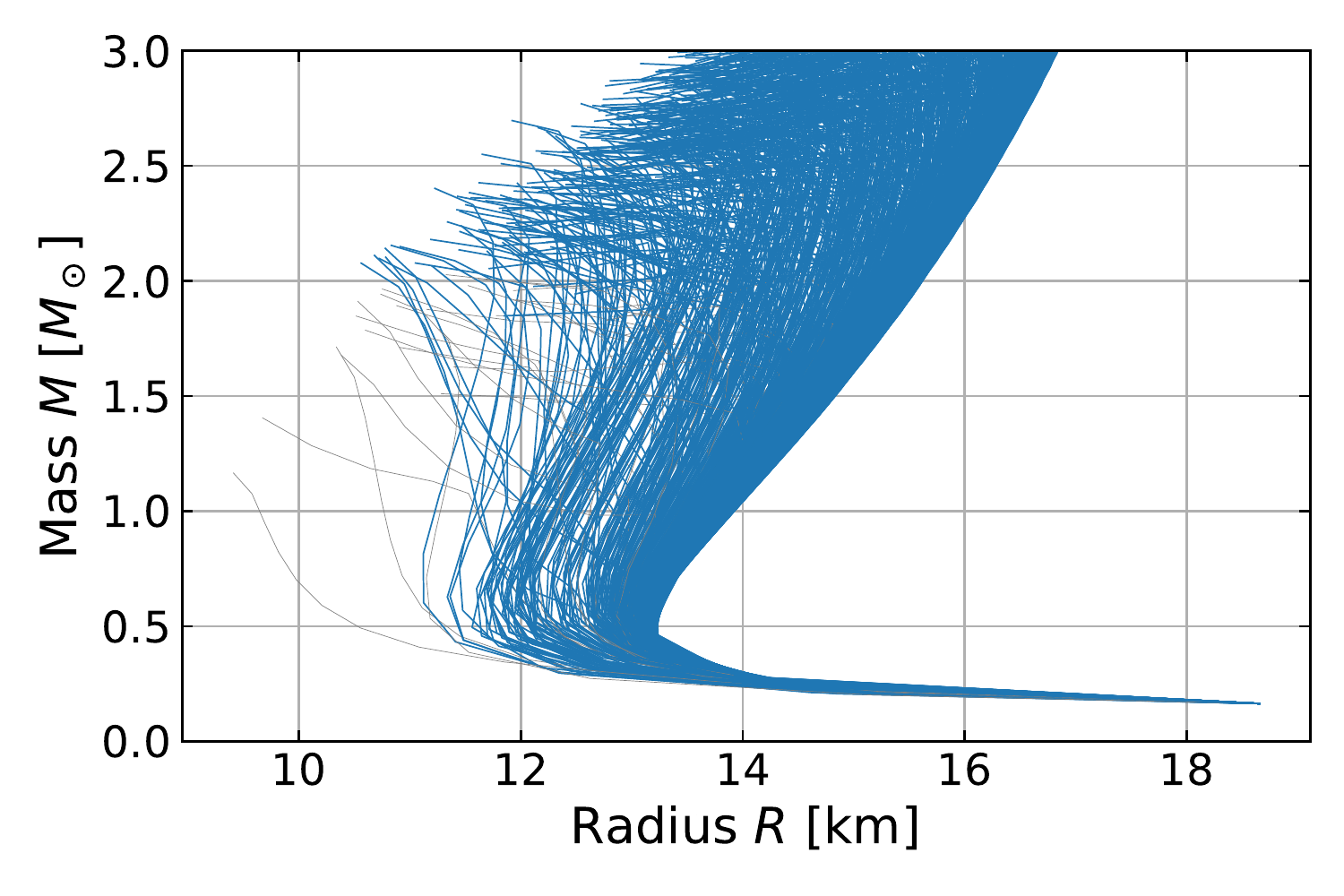}
    \caption{Randomly generated EoSs (left) and the
      corresponding \MR curves (right).}
    \label{fig:train}
\end{figure}

For the EoS generation we randomly assign the average sound
velocities, $c_{s,i}^2$, of the uniform distribution within
$\delta\le c_{s,i}^2 < 1-\delta$.  The upper bound comes from the
causality $c_s^2 < c^2 = 1$, and the lower bound from the
thermodynamic stability $\partial p/\partial \varepsilon \ge 0$.
Here, a small margin by $\delta=0.01$ is inserted as a regulator to
avoid singular behavior of the TOV equation.  We here note that we
allow for small $c_s^2$ corresponding to a (nearly) first-order phase
transition.  Repeating this procedure, we generate a large number of
EoSs.  For each generated EoS, we solve the TOV equation to obtain the
\MR curve as explained in Sec.~\ref{sec:TOV-and-inverse-problem}.  In
Fig.~\ref{fig:train} we show the typical EoS data and the
corresponding \MR curves used in our later analyses.  It should be
noted here that we excluded artificial biases in the EoS generation
and allowed for any possible EoSs including ones disfavored by the
model calculations or the observational data (e.g., an EoS whose
maximum mass exceeds $3\Msun$ in the right panel of
Fig.~\ref{fig:train}).  This is important for the NN to be free from
biases coming from specific model preference.  In this way our
analysis gains an enough generalization ability, covers even exotic
scenario cases, and captures the correct underlying relations between
the input and the output.

Before closing this part, let us mention a possible improvement for
the random EoS generation, though we would not utilize this
improvement to keep our present approach as simple as possible.  The
problem arises from our assumed uniform distribution of $c_{s,i}^2$.
The parametrization and generation algorithm as explained above
definitely allows for a first-order phase transition for sufficiently
small $c_{s,i}^2$'s;  however, due to the uniform distribution, it is
a tiny percentage among the whole data for the generated EoSs to
accommodate a first-order phase transition, and this may be a part of
our ``prior'' dependence.  Since a strong first-order transition
scenario has already been ruled out from phenomenology, this prior
dependence should be harmless.  Nevertheless, if necessary, we can
naturally increase the percentage of the EoSs with a first-order phase
transition in a very simple way as follows:
In Eq.~\eqref{eq:polytrope} we carry out the linear interpolation in
the log-log plane.  Alternatively, we can use the \textit{spline}
interpolation in the log-log plane.  With such a smooth
interpolation, there appear the energy density regions with negative
$\partial p/\partial\varepsilon$, i.e., the EoS can be non-monotonic.
We can then replace this non-monotonic part
by the first-order phase transition using the Maxwell construction.
This is one effective and natural way to enhance a finite fraction of
EoSs with a first-order phase transition while keeping the same
$c_{s,i}^2$'s.  We also note that one more merit in this procedure is
that the end points of the first-order energy regions are not
restricted to be on the segment boundaries.  Thus, the strength of
the first-order phase transition is not necessarily discretized  by
the segment width but an arbitrarily weak first-order phase transition
could be realized with the same parametrization.  We say again that we
would not adopt this procedure:  the present observational precision
gives a tighter limitation and for the moment we do not benefit from
this EoS improvement.  In the future when the quality and the quantity
of the observational data ameliorate, the above mentioned procedure
could be useful.

\subsection{Neural network design: a general introduction}
\label{sec:nnmethod}

The neural network, NN, is one representation of a function with fitting
parameters.  The deep learning, i.e., the machine learning method
using a deep NN, is nothing but the optimization procedure of the parameters
contained in the function represented by the NN\@.  In particular, the
supervised learning can be considered as the fitting problem (namely,
\textit{regression}) for given pairs of the input and the output, i.e.,
the training data.  We often refer to a NN ``model'' to mean the
optimized system of the function given in terms of the NN\@.

The advantage of the deep learning, as compared to naive fitting
methods, lies in the generalization properties of the NN\@.  We need not
rely on any preknowledge about reasonable forms of fitting functions:
the NN with a sufficient number of neurons is capable of capturing any
continuous functions~\cite{citeulike:3561150, HORNIK1991251}.
With a large number of neurons (and layers), there are many fitting
parameters, and recent studies of machine learning have developed
powerful optimization methods for such complicated NNs.

\begin{figure}
  \centering
  \includegraphics[width=0.6\textwidth]{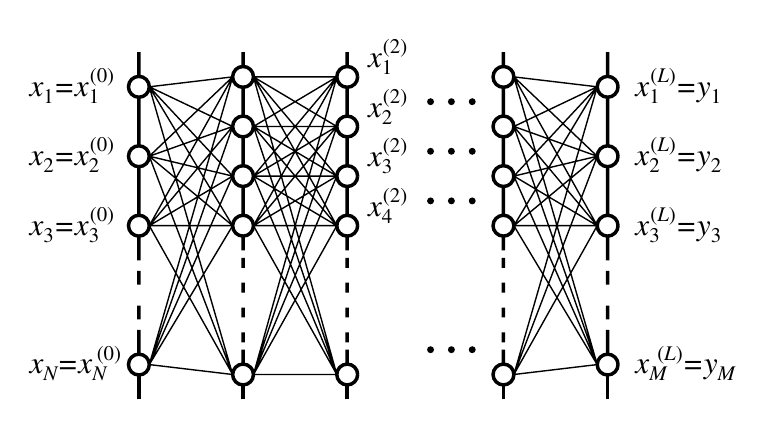}
  \caption{Schematic illustration of the feedforward neural network.}
  \label{fig:ffnn}
\end{figure}

The model function of feedforward NN can be expressed as follows:
\begin{equation}
  \by = f(\bx|\{W^{(1)},\bb^{(1)},\dots,W^{(L)},\bb^{(L)}\})\,,
\end{equation}
where $\bx$ and $\by$ are input and output, respectively.
The fitting parameters, $\{W^{(\ell)},\bb^{(\ell)}\}_{\ell = 1}^L$,
are given in the form of matrices and vectors, respectively.  For
calculation steps, we first prepare $(L+1)$ \textit{layers}
(including the input and the output layers).  
Each layer $\bx^{(\ell)}$ ($\ell=0,\dots L$) consists of nodes
called \textit{neurons} corresponding to the vector components,
$x_1^{(\ell)},\dots,x_{n_\ell}^{(\ell)}$.  We note that the number of
neurons, $n_\ell$, can be different for different $\ell$.  The input
$\bx$ is set to the first layer $\bx^{(0)}$ (which is labeled as
$\ell=0$ in this paper), and the subsequent layers are calculated
iteratively as
\begin{align}
  \bx^{(0)} &= \bx\,, \\
  \bx^{(\ell)} &= \sigma^{(\ell)} \bigl( W^{(\ell)} \bx^{(\ell-1)}
                 + \bb^{(\ell)} \bigr) \quad (\ell = 1,\dots,L)\,.
\end{align}
The $L$-th layer becomes the final output, $\by = f(\bx) = \bx^{(L)}$.
Here, $\sigma^{(\ell)}(x)$'s are called \textit{activation functions},
whose typical choices include the sigmoid function:
$\sigma(x) = 1/(e^x+1)$, the ReLU: $\sigma(x) = \max\{0, x\}$,
hyperbolic tangent: $\sigma(x)=\tanh(x)$, etc.  In the above notation
$\sigma^{(\ell)}$ returns a vector by applying the activation to each
vector component of the function argument, i.e.,
$\sigma(\bx) = \bigl(\sigma(x_1), \dots,\sigma(x_n)\bigr)^\mathrm{T}$.
The fitting parameters, $W^{(\ell)}$ and $\bb^{(\ell)}$, on the $\ell$-th
layer, denote the {\it weights} between nodes in the two adjacent layers
and the activation offset at each neuron called {\it bias},
respectively.  These fitting parameters have no physical
interpretation at the setup stage.  The general architecture is
schematically depicted in Fig.~\ref{fig:ffnn}, in which the
calculation proceeds from the left with input $\bx$ to the right with
output $\by$.

The complexity choice of the NN structure, such as the number of
layers and neurons, has a tradeoff problem between the fitting power
and the overfitting.  To capture the essence of the problem, the
complexity of layers and neurons should be sufficiently large.  At the
same time, to avoid the overfitting problem, and to train the NN
within reasonable time, the number of layers and neurons should not be
too large.  There is no universal prescription and the performance
depends on the problem to be solved. For our setup we found overall
good performance when we chose the neuron number on the first layer
greater than the input neuron number.  A more sophisticated NN may be
possible, but for our present purpose a simple network structure as in
Fig.~\ref{fig:ffnn} is sufficient.

\subsection{Loss function and training the neural network}
\label{sec:general-training}

For the actual procedure of machine learning we need to choose a
\textit{loss function} and minimize it, which is called
\textit{training}.  We define a loss function as
\begin{align}
  \mathcal{L}(\{W^{(\ell)}, \bb^{(\ell)}\}_{\ell})
    \equiv \int d\bx \Pr(\bx) \ell\Bigl(\by, f(\bx|\{W^{(\ell)},
  \bb^{(\ell)}\}_{\ell})\Bigr)\,,
  \label{eq:loss}
\end{align}
where $\ell(\by, \by')$ quantifies the distance or error between the
NN prediction $\by'$ and the answer $\by$ provided in the training
data.  The exact formula for the distribution $\Pr(\bx)$ is not
necessarily known in general.  Moreover, even if it is known,
multidimensional integration in Eq.~\eqref{eq:loss} is practically
intractable, so let us approximate it with the sum over the samples in
the training data, $\calD = \{(\bx_n, \by_n)\}_{n=1}^{|\calD|}$, with
$|\calD|$ denoting the sample size:
\begin{align}
  \mathcal{L}(\{W^{(\ell)}, \bb^{(\ell)}\}_{\ell}) \approx \frac1{|\calD|}
  \sum_{n=1}^{|\calD|} \ell\Bigl(\by_n, f(\bx_n|\{W^{(\ell)},
  \bb^{(\ell)}\}_{\ell})\Bigr)\,.
  \label{eq:lossreal}
\end{align}
The optimization problem we deal with here is to minimize the above
loss function by tuning the network parameters $\{W^{(\ell)},
\bb^{(\ell)}\}_{\ell}$.  The minimization is usually achieved with the
stochastic gradient descent method or its improved versions.  The
deterministic methods would be easily trapped in local minima or saddle
points of the loss function.  In practice, therefore, the stochastic
algorithms are indispensable.  In gradient descent method the
parameters $\{W^{(\ell)}(t), \bb^{(\ell)}(t)\}_{\ell}$ at the iteration step $t$ are
updated along the direction of the derivative of the loss
function with respect to the parameters:
\begin{align}
  W^{(\ell)}(t+1) =W^{(\ell)}(t) - \eta \frac{\partial
    \mathcal{L}(W^{(\ell)}(t))}{\partial W^{(\ell)}(t)}\;,
\end{align}
where $\eta$ is called learning rate.  We can update $\bb^{(\ell)}$ in
the same way.  We usually evaluate the derivatives by using the
\textit{mini-batch method}.  In the mini-batch method, the training
data set is first randomly divided into multiple subsets, which is
called \textit{mini-batches}. Then, the derivative of the loss
function is estimated within a mini-batch $\mathcal{B}$.
Using the explicit expression~\eqref{eq:lossreal}, the approximated
derivatives read:
\begin{align}
  \frac{\partial \mathcal{L}(W^{(\ell)})}{\partial W^{(\ell)}} \approx
  \frac1{|\mathcal{B}|} \sum_{n=1}^{|\mathcal{B}|}
  \frac{\partial\ell\Bigl(\by_n, f(\bx_n|\{W^{(\ell)},
  \bb^{(\ell)}\}_{\ell})\Bigr)}{\partial W^{(\ell)}}\,,
  \label{eq:minibatch}
\end{align}
where the \textit{batch size} $|\mathcal{B}|$
denotes the number of sample points in $\mathcal{B}$, whose optimal choice
differs case-by-case.  The \textit{epoch} counts the number of scans
over the entire training data set $\calD$.
The parameters are updated for each mini-batch, so one epoch amounts to
$|\calD|/|\mathcal{B}|$ updates.  Usually, the derivative,
$\partial\ell /\partial W$ appearing in
Eq.~\eqref{eq:minibatch}, is evaluated by a method called
backpropagation.

For numerics we basically use a Python library,
Keras~\cite{software:Keras} with TensorFlow~\cite{arXiv:1605.08695} as
a backend.  In this work we specifically employ \texttt{msle}, i.e.,
the mean square logarithmic errors, for the choice of the loss
$\ell(\by, \by')$ appearing in Eq.~\eqref{eq:loss}; that is,
\begin{align}
  \ell_{\texttt{msle}} (\by, \by') &\equiv \left|\log \by - \log
  \by'\right|^2\,.
  \label{eq:msle}
\end{align}
We use a specific fitting algorithm,
Adam~\cite{DBLP:journals/corr/KingmaB14}, with the mini-batch size
being 100 for the analysis in Sec.~\ref{sec:mock} and 1000 for the
analysis in Sec.~\ref{sec:obs}.  We initialize our NN parameters
with the Glorot uniform distribution.

\section{Extensive Methodology Study and the Performance Tests}
\label{sec:mock}

This section is an extensive continuation from our previous work on
the methodology study~\cite{Fujimoto:2017cdo}.  We here present more
detailed and complete investigations of various performance tests.

In this section, to put our considerations under theoretical control,
we assume $\omega$ to be a hypothetical observational scenario that 15
neutron stars are observed and $(M_i,\, R_i)$ ($i=1,\dots,15$) are
known with a fixed size of observational uncertainty following the
previous setup in
Refs.~\cite{Ozel:2015fia,Bogdanov:2016nle,Ozel:2016oaf,Fujimoto:2017cdo}.
We design a NN, generate the validation data as well as the training data,
and train the NN to predict the EoS parameters, $c_{s,i}^2$,
from the artificially generated \MR data,
$(M_i,\, R_i)$ ($i=1,\dots,15$) (i.e., mock data).  We discuss the training
behavior of the NN, the reconstruction performance of the EoS and the
\MR curve, and the distribution of reconstruction errors.

\subsection{Mock data generation and training}
\label{sec:uni5p}

For the current study in this section, we deal with the input \MR
data, $(M_i,\, R_i)$ ($i=1,\dots 15$), and the output EoS parameters,
$c_{s,i}^2$ ($i=1,\dots 5$), as described in the previous section.
A combination of these input and output data forms a sample point in
the training and the validation data, and we need to generate many
independent combinations to prepare the training and the validation
data.

\begin{figure}
  \centering
  \includegraphics[width=0.7\textwidth]{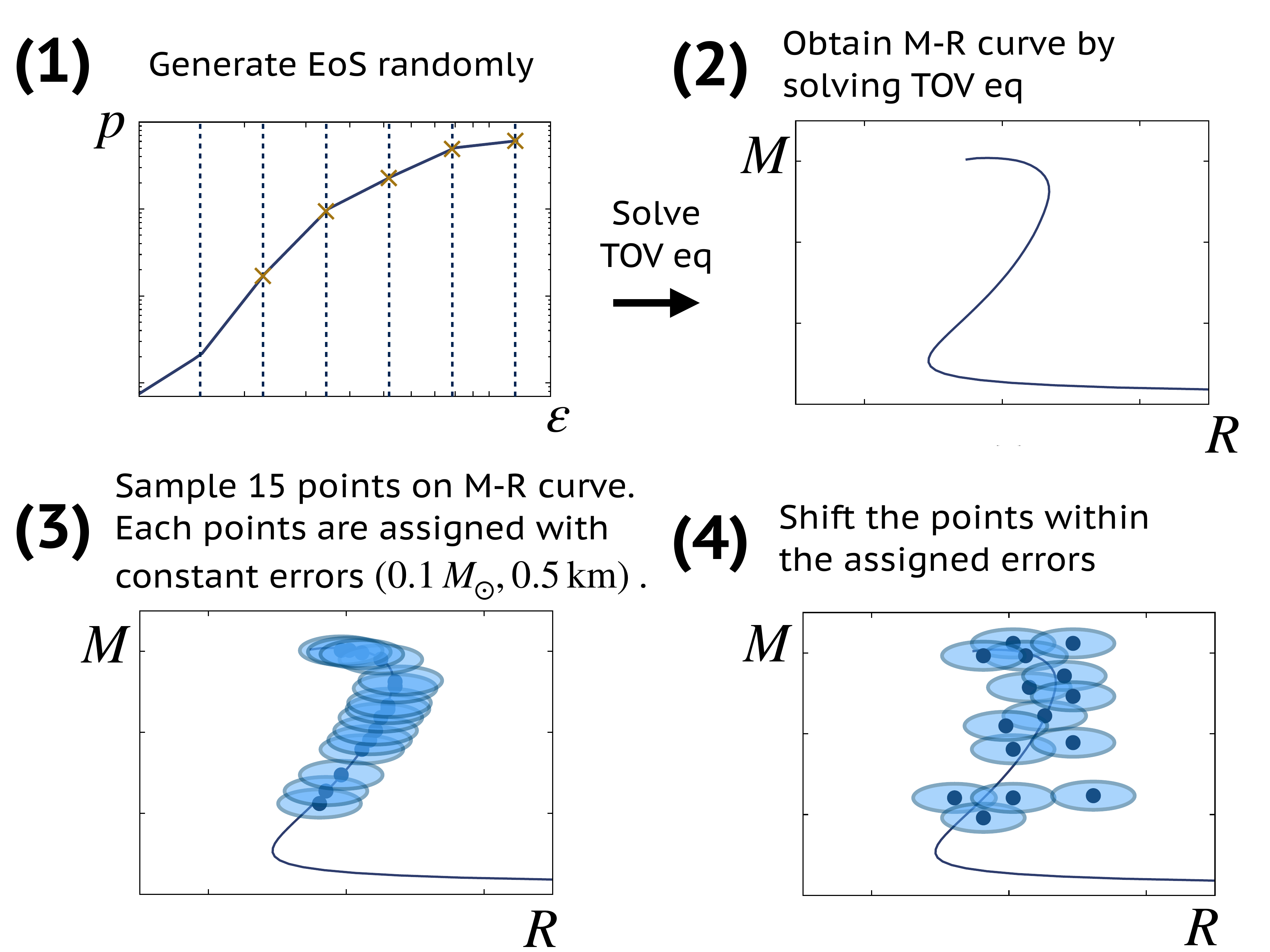}
  \caption{Schematic flow of data generation procedures for the
    analysis in Sec.~\ref{sec:mock}.}
  \label{fig:procedure_a}
\end{figure}

The procedure for sampling the \MR points is sketched in
Fig.~\ref{fig:procedure_a}.  We follow the basic strategy described in
Sec.~\ref{sec:datagen} and introduce minor modifications for the
current setup.  For each randomly generated EoS, $p=p(\varepsilon)$,
we solve the TOV equation to get the \MR curve and identify the
maximum mass $M_{\rm max}$ of the neutron star, corresponding to (1)
and (2) in Fig.~\ref{fig:procedure_a}.  If $M_{\rm max}$ does not
reach the observed maximum, such an EoS is rejected
from the ensemble.
Then, for each EoS, we randomly sample 15 pairs of $(M_i,\, R_i)$ on
the corresponding \MR curve, as shown in (3) in
Fig.~\ref{fig:procedure_a}, assuming a uniform distribution of $M_i$
over the interval $[\Msun,\, M_{\rm max}]$.  If there are multiple values of $R$
corresponding to one $M$, we take the largest $R$ discarding others
belonging to unstable branches.  In this way we select 15 points of
$(M^*_i,\, R^*_i)$ on the \MR curve.

We also expect the NN to learn that observational data should include
uncertainties, $\Delta M_i$ and $\Delta R_i$.  We randomly generate
$\Delta M_i$ and $\Delta R_i$ according to the normal distribution
with assumed variances, $\sigma_M=0.1\Msun$ for the mass and
$\sigma_R=0.5\,\text{km}$ for the radius.  These variances should be
adjusted according to the real observational uncertainty as we will
treat in Sec.~\ref{sec:obs}, but for the test purpose in this
section, we simplify the treatment by fixing $\sigma_M$ and $\sigma_R$
by hand.  Furthermore, the distributions
are not necessarily centered on the genuine \MR curve as in (3) in
Fig.~\ref{fig:procedure_a}.  So, we shift data points from
$(M^*_i,\, R^*_i)$ to
$(M_i=M^*_i+\Delta M_i,\, R_i=R^*_i+\Delta R_i)$.  Importantly, we
repeat this procedure to generate an ensemble of randomly shifted data
for each EoS, and the repeating number (and thus the number of obtained
shifted data) is denoted by $n_s$ in this work.  Our typical choice is
$n_s=100$, whose effects on learning will be carefully examined in
later discussions.  For studies in this section, we have
prepared 1948 EoSs (2000 generated and 52 rejected), so the total size
of the training data set is $1948\times n_s$.  We find that the learning
proceeds faster if the data is normalized appropriately to be
consistent with the Glorot initialization; we use
$M_i/M_{\rm norm}$ and $R_i/R_{\rm norm}$ with
$M_{\rm norm}=3\Msun$ and $R_{\rm norm}=20\,\text{km}$.

The NN is optimized to fit the training data so as to have
a predictive power.  To test it, we use the validation loss
calculated for the validation data, i.e., independent mock data of the
neutron star observation.  We separately generate 200 EoSs, among
which 196 EoSs pass the two-solar-mass condition, and sample an
$n_s=1$ observation for each EoS to generate the validation data set.

It is worth mentioning here that calculating \MR curves requires
repeated numerical integration of the TOV equation with various
boundary conditions, which is computationally expensive.  Generally
speaking, in the machine learning approach, the preparation of the
training data is often the most time-consuming part.
In contrast, increasing the sample size by repeating the
observation procedure does not need further computational cost.
This procedure can actually be regarded as a kind of the data augmentation
by the noise injection, where the observational errors play the role of the injected noise.
Let us call this procedure \textit{observational data augmentation} hereafter.
This observational data augmentation has actually another virtue,
which will be discussed in later sections.

\begin{table}
  \centering
  \begin{tabular}{ccc} \hline
    Layer index & Neurons & Activation Function \\ \hline
    0 & 30 & N/A \\
    1 & 60 & ReLU \\
    2 & 40 & ReLU \\
    3 & 40 & ReLU \\
    4 & 5  & $\tanh$ \\ \hline
  \end{tabular}
  \caption{Neural network architecture used in Sec.~\ref{sec:mock}.
    In the input layer, 30 neurons correspond to input 15 points of
    the mass and the radius.  In the last layer 5 neurons correspond
    to 5 output parameters of the EoS\@.}
  \label{tab:nn_a}
\end{table}

We are constructing a NN that can give us one EoS in the output side
in response to one ``observation'' of 15 neutron stars,
($M_i$,\, $R_i$) ($i=1,\dots,15$) in the input side.  Thus, the number
of neurons in the input and the output layers should be matched with
the size of the observational data and the number of EoS parameters,
respectively.  The input layer with 30 neurons matches 15 \MR points
(30 input data).  We sorted 15 \MR pairs by their masses in ascending
order, but we found that the sorting does not affect the overall
performance.  The output neurons in the last layer correspond to the
prediction target, i.e., 5 parameters of the speed of sound.

The concrete design of our NN is summarized in Tab.~\ref{tab:nn_a}.
We choose the activation function at the output last layer as
$\sigma^{(4)}(x) = \tanh(x)$ so that the speed of sound is
automatically bounded in a causal range $c_s^2 < 1$.  In principle the
output could be unphysical: $c_s^2 < 0$ may occur, and then we
readjust it as $c_s^2 = \delta$.  For the other layers we choose the
ReLU, i.e., $\sigma^{(\ell)}(x) = \max\{0, x\}$
($\ell = 1,\dots, L-1$), which is known to evade the vanishing
gradient problem.

\subsection{Learning curves with the observational data augmentation}
\label{sec:learningcurve}

\begin{figure}
  \centering
  \includegraphics[width=0.5\textwidth]{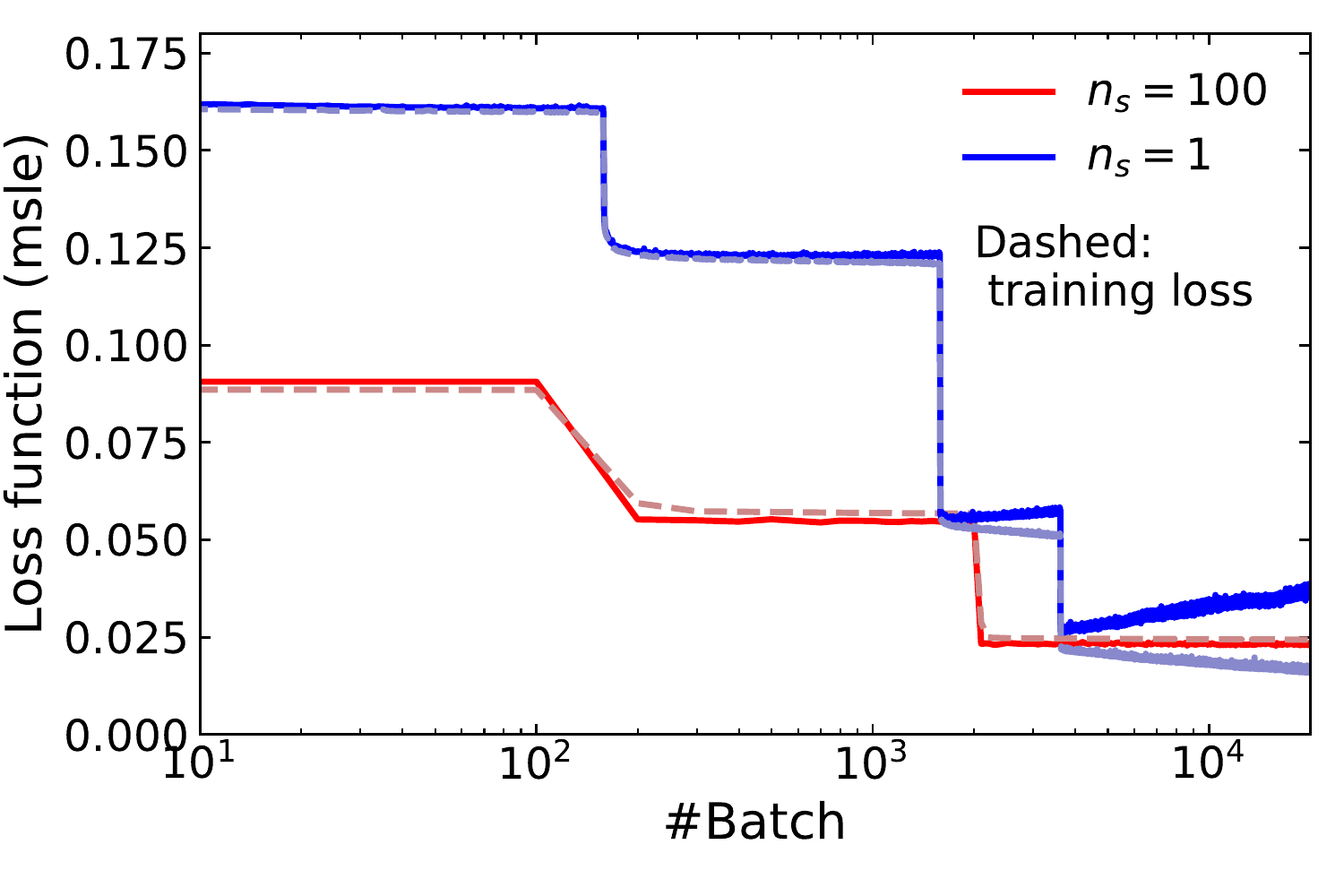}
  \caption{Typical behavior of learning curves in our setup.
  Solid lines show the validation loss and the dashed lines show the training loss.}
  \label{fig:learningcurve}
\end{figure}

The stochastic gradient descent method is an iterative procedure to
adjust the NN parameters step by step and to gradually make the loss
function smaller.  To judge the point where we stop the iteration, we
need to monitor the convergence of the loss function in training.  For
this purpose we plot the loss function as a function of
\textit{training time} (i.e., the number of iteration units), which we
call the \textit{learning curve} in this work.

The actual shape of the learning curve depends on the initial
parameters of the NN and also on the choice of
mini-batches in the stochastic gradient descent method.  Although the
learning curve differs every time we start the new training, we can still
find some common tendency.  In Fig.~\ref{fig:learningcurve} we show
the typical behavior of learning curves for our neutron star problem.
It should be noted that the horizontal axis in
Fig.~\ref{fig:learningcurve} is the number of mini-batches.  In this
study we will compare the training speeds with different $n_s$ but
with the fixed mini-batch size.  The number of epochs -- the number of
scans over the whole training data set -- is often used as the unit of
training time, but in the present study for the training speeds, we
need to adopt the number of mini-batches which is proportional to the
actual number of parameter updating and the computational time of the
training.

In Fig.~\ref{fig:learningcurve} we plot two different learning curves
for a single training: one for the \textit{training loss} (dashed
line) and the other for the \textit{validation loss} (solid line).
The training and the validation losses are the loss functions
calculated for the training and the validation data sets,
respectively, by using Eq.~\eqref{eq:lossreal}.  The former
diagnoses how well the model \textit{remembers} the training data, and
the latter diagnoses how well the model properly \textit{predicts}
unseen data by generalizing the learned information.

Here, we compare the learning curves for two different training data
sets with $n_s = 100$ and $n_s = 1$.  By observing the plot like
Fig.~\ref{fig:learningcurve} for many trials on the training, we can
read out two general tendencies:
\begin{enumerate}
\item
  The value of validation loss of $n_s = 1$ is larger than that of
  $n_s = 100$.  A possible interpretation is that the loss function
  can be more trapped in local minima/saddle points for smaller
  $n_s$.
\item
  For the number of batches $\gtrsim 10^3$, for $n_s=1$, the training
  loss keeps decreasing with increasing number of batches, whilst the
  validation loss turns to increasing behavior.  The loss function of
  $n_s = 1$ apparently shows overfitting behavior but the overfitting
  is not seen for $n_s = 100$.
\end{enumerate}
From these results, we may argue that the data augmentation with
sufficiently large $n_s$ could be useful to improve the problems of
local minimum trapping and overfitting.  To justify or falsify this
speculation, we will perform further quantitative analyses on the
observational data augmentation under a simple setup in Sec.~\ref{sec:curve}.
We note in passing that recent developments show a possibility of over-parametrized
networks, which has a very large number of parameters,
to reconcile a good generalization ability and an
elusion of overfitting~\cite{2016arXiv161103530Z,NEURIPS2019_62dad6e2}.  Here, our point is that the data augmentation
is needed to incorporate the physical uncertainty rather than to
improve the learning quality, but it may have such a byproduct.

\subsection{Typical examples -- EoS reconstruction from the neural network}
\label{sec:typical}

\begin{figure}
  \centering
  \includegraphics[width=0.48\textwidth]{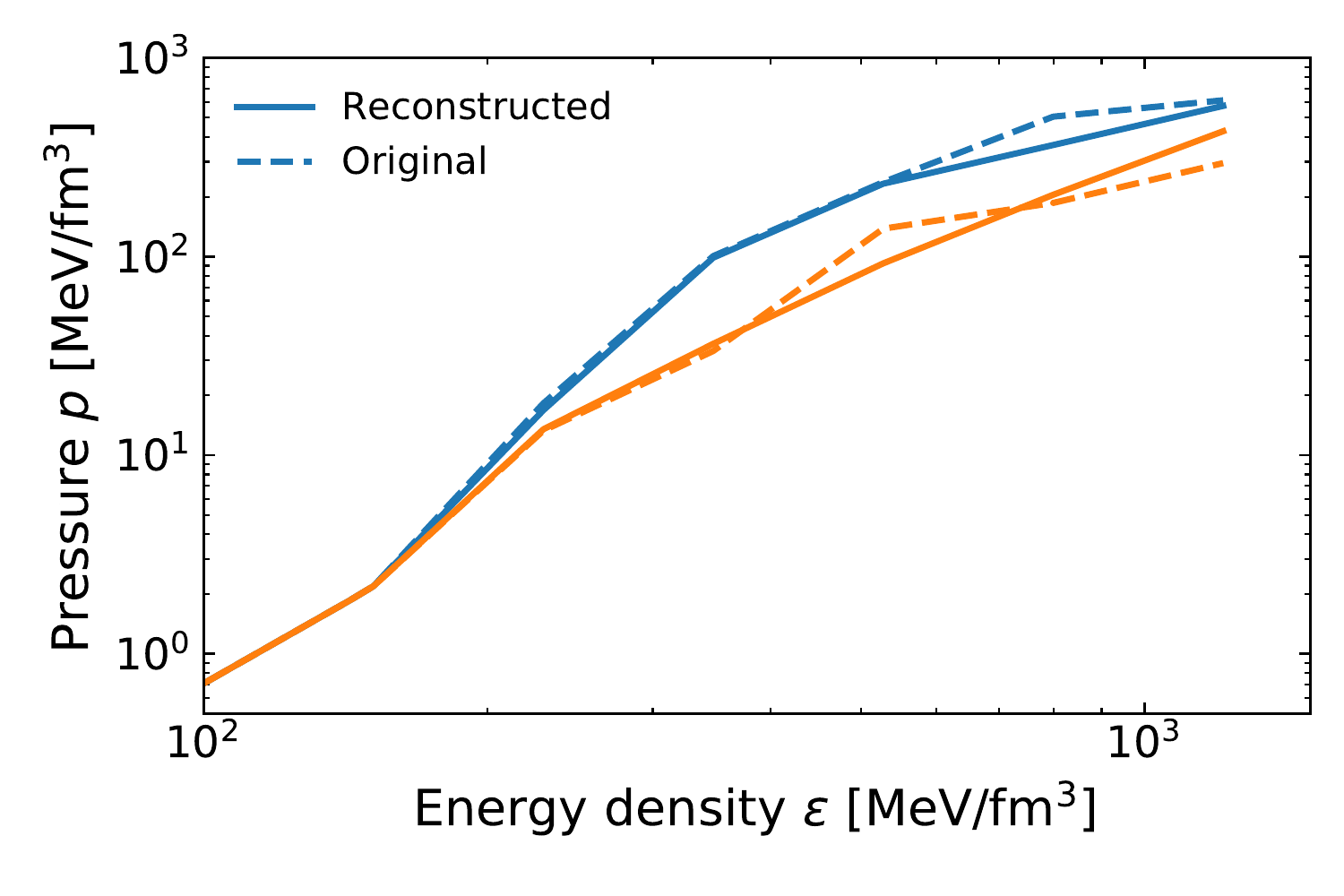} \hspace{0.5em}
  \includegraphics[width=0.48\textwidth]{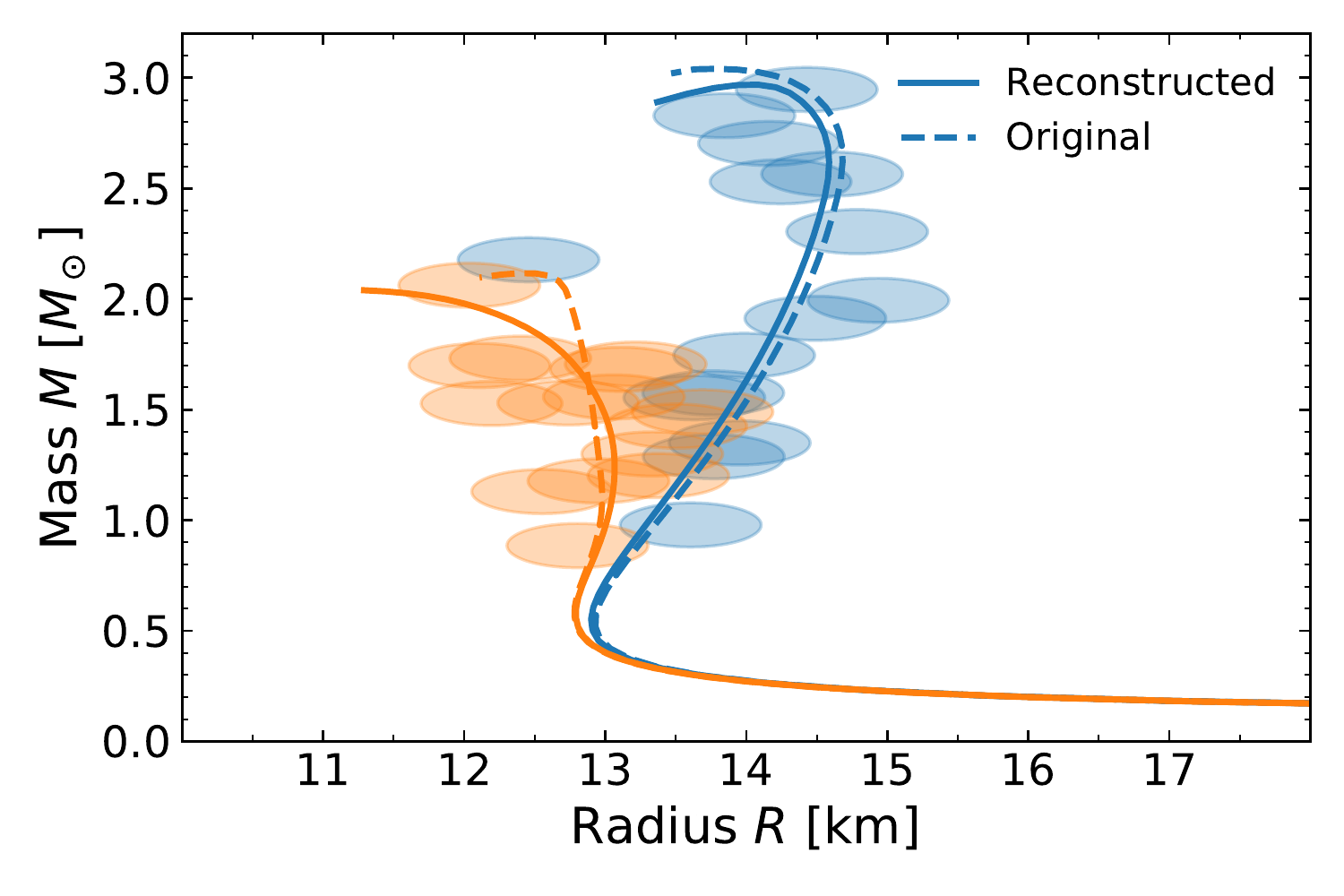}
  \caption{(Left) Two typical examples of the randomly generated EoSs
    (dashed line) and the NN outputs (solid lines) reconstructed from
    one observation of 15 $M$-$R$ points [see the right panel for
    actual ($M_{i}$,\, $R_{i}$)].\quad (Right) Randomly sampled 15
    data points (centers of the ellipses) and the \MR curve from the
    reconstructed EoS (solid lines) and the \MR curve from the
    original EoS (dashed lines).  The orange and blue colors
    correspond to two EoSs shown in the same color in the left
    panel.}
  \label{fig:inferred}
\end{figure}

Once the loss function converges, we can use the trained NN model to
infer an EoS from an observation of 15 \MR points.  We picked two
typical examples represented by orange and blue in
Fig.~\ref{fig:inferred}.  In the left panel of Fig.~\ref{fig:inferred}
the dashed lines represent randomly generated EoSs.  We note that two
EoSs are identical in the low density region because the SLy4 is
employed at $\varepsilon\leq\varepsilon_0$.  The corresponding genuine
\MR relations are shown by the dashed lines in the right panel of
Fig.~\ref{fig:inferred}.  Randomly sampled mock observations
consisting of 15 \MR points are depicted by ellipses where the centers
are the ``observed'' $(M,\, R)$ and the major and minor axes show
observational uncertainties in the $R$ and $M$ directions,
respectively.  The reconstructed EoSs are depicted by solid lines in
the left panel of Fig.~\ref{fig:inferred}.  We can see that the
reconstructed EoSs agree quite well with the original EoSs for these
examples.  It would also be interesting to make a comparison of the
\MR relations corresponding to the original and the reconstructed
EoSs.  The solid lines in the right panel of Fig.~\ref{fig:inferred}
represent the \MR relations calculated with the reconstructed EoSs.
As is expected from agreement in the EoSs in the left panel of
Fig.~\ref{fig:inferred}, the original and the reconstructed \MR
relations are close to each other.  More details are already reported
in our previous publication~\cite{Fujimoto:2017cdo}.

\subsection{Error correlations}
\label{sec:error-correlation}

We make the detailed error analysis of the reconstructed EoSs in terms
of the speed of sound.   To increase the statistics, for the present
analysis, we set $n_s=1000$ (remember that increasing $n_s$ is almost
costless).

\begin{figure}
  \centering
  \includegraphics[width=.95\textwidth]{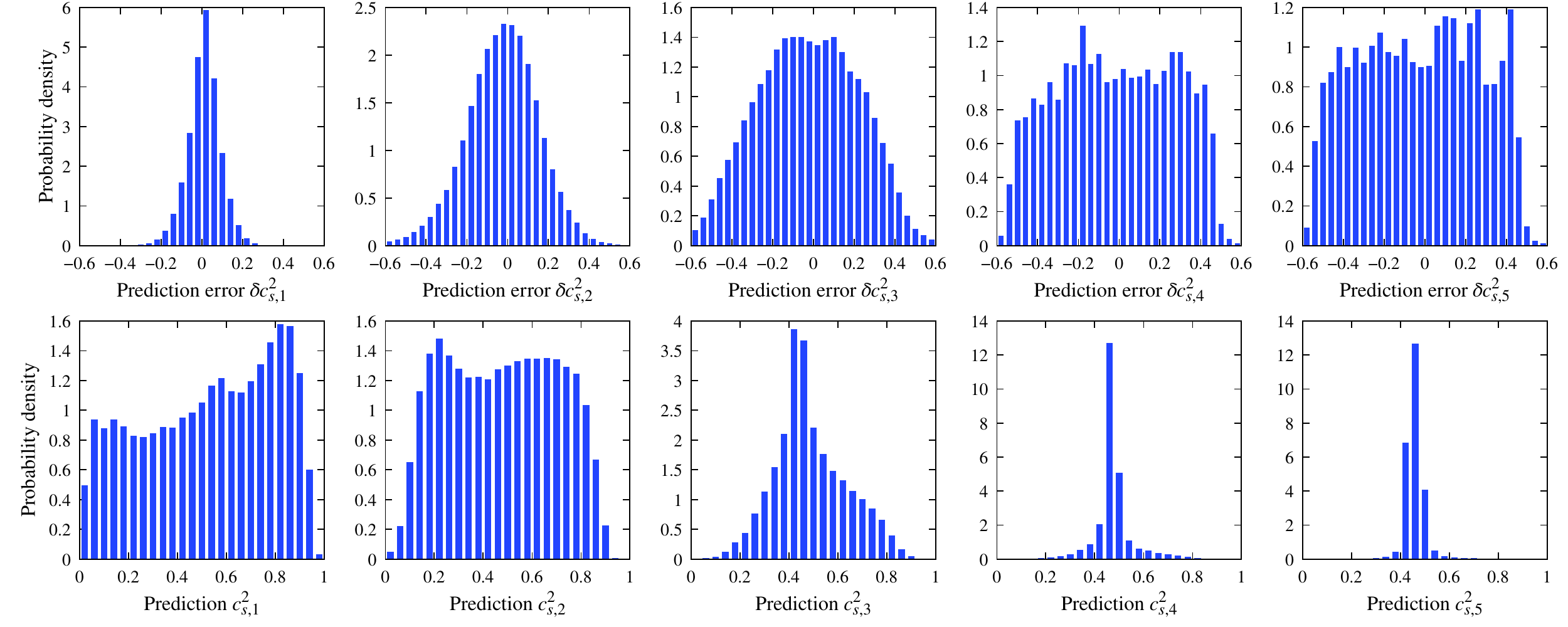}
  \caption{
    The upper and the lower panels show the histograms of prediction
    errors and values, respectively, of $c_{s,i}^2$.  The vertical
    axis is the probability density, i.e., the count in each bin is
    divided by the bin width and the total count.}
  \label{fig:error-1d-histogram}
\end{figure}

Now, we introduce the prediction error in the speed of sound:
\begin{equation}
  \delta c_{s,i}^2 \equiv (c_{s,i}^2)^\text{(rec)} - (c_{s,i}^2)^* \quad
  (i=1,\dots,5)\,,
  \label{eq:cs2err}
\end{equation}
where $(c_{s,i}^2)^\text{(rec)}$ and $(c_{s,i}^2)^*$ are the
reconstructed and original values, respectively, for the speed of
sound.  The first row of Fig.~\ref{fig:error-1d-histogram} shows the
histograms of $\delta c_{s,i}^2$ ($i=1,\dots 5$).  We see that the
prediction error at the lower density (i.e., smaller $i$) is centered
at the origin while that at the higher density (i.e., larger $i$) is
distributed widely, which means that the observational data
efficiently constrains lower density regions.  At the highest density
with $i=5$, the prediction error roughly follows the uniform
distribution within the physical range.  This behavior can be
understood as follows.  The original values of the speed of sound in
the validation data are generated by the uniform distribution in the
range $[0,1]$ while the distribution of the predicted $c_{s,i}^2$
($i=1,\dots 5$) are shown in the second row of
Fig.~\ref{fig:error-1d-histogram}.  At lower density, the distribution
of the prediction roughly follows the original uniform distribution.
However, at higher density, the distribution of the prediction peaks
around $c_{s,i}^2 \sim 0.5$.   At the highest density, therefore, the
NN always predicts $c_{s,i}^2 \sim 0.5$ for any inputs.  This is
because $c_{s,i}^2 \sim 0.5$ is an optimal prediction to minimize the
loss function when the constraining information is inadequate.  The
exact peak position is slightly smaller than 0.5, which is because
our choice of the loss function, \texttt{msle}, gives larger weights
for smaller values.  Thus, the approximate uniform distribution
of the prediction errors at the highest density (as seen in the upper
rightmost figure in Fig.~\ref{fig:error-1d-histogram}) just reflects
the distribution of the original values relative to the fixed prediction.

From this analysis we conclude that the observation data should
contain more information on the lower density EoS but not much on the
high density EoS around $\varepsilon \sim 8\varepsilon_0$.  This is
quite natural, but such a quantitative demonstration in
Fig.~\ref{fig:error-1d-histogram} is instructive.  Also, the results
in Fig.~\ref{fig:error-1d-histogram}  tell us two important aspects of
our NN approach.
First, the NN tends to give a moderate prediction, not too far from
all possible values, when the uncertainty is large.  This is
nontrivial:  the posterior distribution of EoS is random, but such
information is lost after the NN filtering.  This is reasonable
behavior in a sense that we want to design the NN to predict the most
likely EoS instead of the posterior distribution.
Secondly, the moderate prediction for unconstrained cases may
explicitly depend on the choice of the loss function and also on the
prior distribution, i.e., the distribution of the EoSs in the training
data set in our problem.  Thus, we should be careful of possible
biases when we discuss unconstrained results in the high density
regions.

\begin{figure}
  \centering
  \includegraphics[width=.85\textwidth]{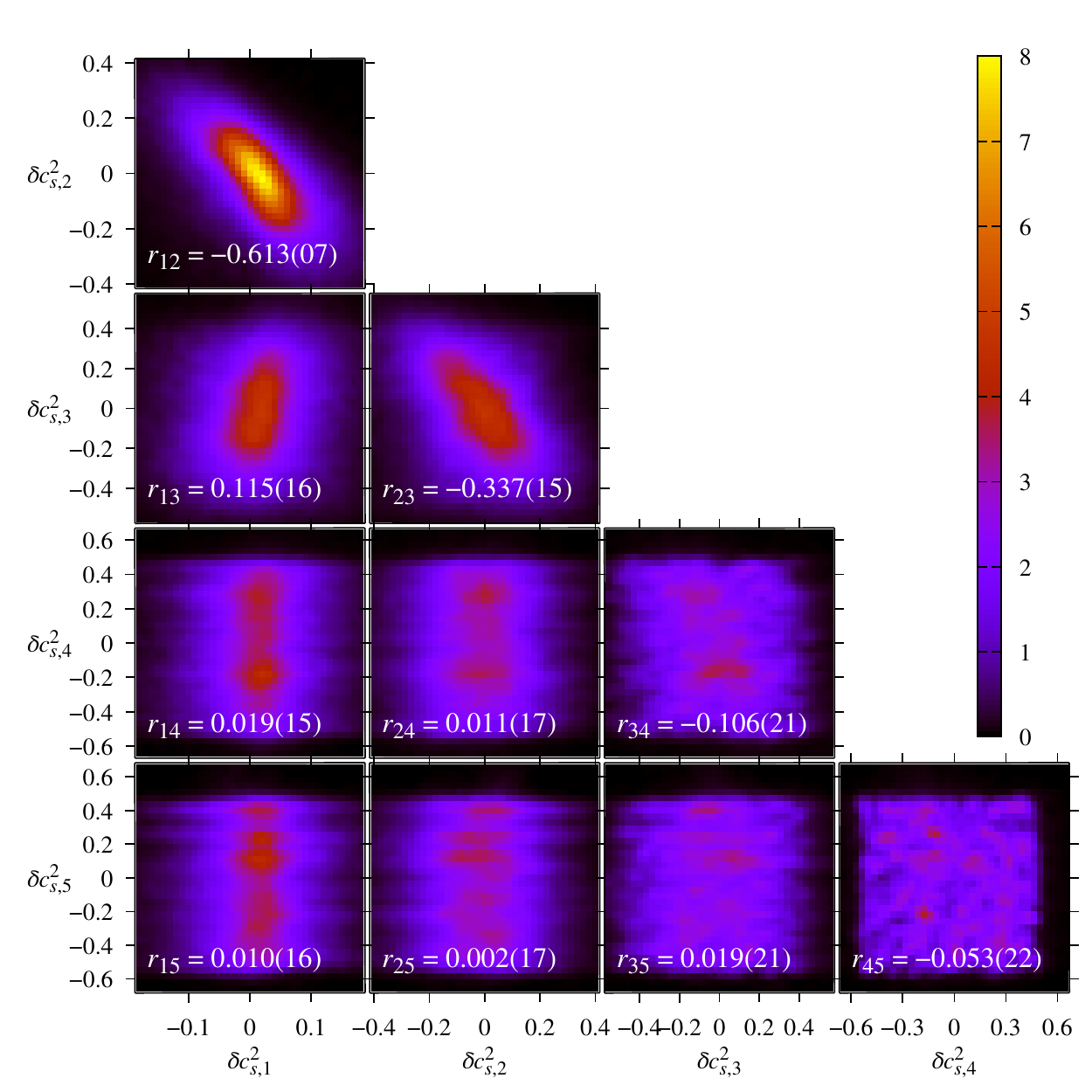}
  \caption{
    Predicted errors marginalized for different pairs of EoS parameters.
    The first to fourth columns correspond to
    $\delta c_{s,i}^2$ ($i=1,\dots 4$).  Similarly, the first to
    fourth rows correspond  to
    $\delta c_{s,i}^2$ ($i=2,\dots 5$).
    Colors show the normalized counts in each bin, i.e.,
    the counts in the bin divided by the total count and multiplied by
    the bin number.  Pearson's correlation coefficients $r_{ij}$
    for respective parameter pairs are shown in panels
    with the statistical uncertainty enclosed in parentheses.}
  \label{fig:error-2d-histogram}
\end{figure}

The error analysis so far is separately made for each EoS parameter.
In Fig.~\ref{fig:error-2d-histogram}, to extract more detailed
information, we plot the marginalized distributions for different
pairs of the EoS parameters with corresponding Pearson's correlation
coefficients $r_{ij}$.  We see that the speeds of sound of neighboring
density segments have negative correlations, which is similar to the
behavior seen in the Bayesian analysis~\cite{Raithel:2017ity}.
The underlying mechanism for this behavior can be explained in the
same way as in Ref.~\cite{Raithel:2017ity}:  an overprediction in a
segment can be compensated by an underprediction in the adjacent
segment, and vice versa.  The correlations are clearer in the lower
density regions in which predicted $c_{s,i}^2$'s have better
accuracy.  We also observe a small and positive correlation, e.g.,
$r_{13}>0$, between the parameters separated by two segments.  This is
also expected from two negative correlations in between.  The other
correlations involving parameters in the high density regions are
consistent with zero within the statistical error.

\subsection{Conventional polynomial regression}
\label{sec:poly}

It is an interesting question how the machine learning method could
be superior to other conventional approaches as a regression tool.  In
this section we make a detour, introducing a \text{polynomial} model,
and checking performance of a traditional fitting method.  We can
model the regression of EoS from the \MR data with the following
polynomial fitting of degree $n$:
\begin{align}
  y_i &= g_i(\{x_j\}|\{ a_{ij_1 j_2 \cdots j_{N}} \}) \notag\\
  &\equiv \sum_{\substack{j_k \ge 0\; (k=1,\dots,N)\\ j_1 + j_2 + \cdots + j_{N} \le n}}
  \!\!\! a_{ij_1 j_2 \cdots j_{N}} x_1^{j_1} x_2^{j_2} \cdots x_{N}^{j_{N}}\,,
\end{align}
where $\{x_i\}$ and $\{y_i\}$ are the input and the output of the
training data set, respectively, with $N=30$ being the size of the
input vector $\{x_i\}$, and $\{a_{ij_1 j_2 \cdots j_{N_1}}\}$ are the
fitting parameters.  For the optimization we employ the traditional
least-square method with the loss function in Eq.~\eqref{eq:loss}
specified as \texttt{mse}, i.e., the mean square errors:
\begin{align}
  \ell_{\texttt{mse}}(\by, \by') \equiv |\by - \by'|^2\,.
  \label{eq:mse}
\end{align}
In this case the mapping is linear with respect to the fitting
parameters, which means the problem is essentially the (multivariate) linear regression.  Thus, we can
find an analytical solution of the exact global minimum of the loss
function~\eqref{eq:mse} in this traditional method.

Here we comment on the number of parameters in the polynomial model,
which is given by the multiset coefficient:
$n_\text{param}
  = \bigl(\kern-0.2em\bigl(\genfrac{}{}{0pt}{}{N+1}{n}\bigr)\kern-0.2em\bigr)
  = \genfrac{(}{)}{0pt}{}{N+n}{n}$.
For $N=30$ in the present problem,  as long as $n\ll N$, we can
approximate $n_\text{param} = \calO(N^n)$ which explodes as the degree
$n$ gets larger.  Explicit calculations read:
$n_\text{param} = 31$ ($n=1$), $496$ ($n=2$), $5456$ ($n=3$), $46376$ ($n=4$), etc.
The drawback of the linear regression is a huge computational cost of
$\calO(n_\text{param}^2|\calD|)$.

\subsection{Comparison between the neural network and the
  polynomial regression}
\label{sec:EoSrec}

In Sec.~\ref{sec:typical} we already observed two successful examples
of the EoS reconstruction.  For other EoSs in the validation data, the
reconstructed \MR curves agree well with the genuine ones.  In this
section we shall quantify the overall agreement, and for this purpose
we enlarge the size of the validation data; namely, we randomly
generate 1000 EoSs, among which 967 EoSs pass the two-solar-mass
condition, and we carry out statistical analyses with 967 EoSs.

We define the overall reconstruction accuracy by calculating the
radius deviation, i.e., the distance between the solid and dashed
lines as already shown in the right panel of Fig.~\ref{fig:inferred}:
\begin{align}
  \delta R(M) \equiv R^\text{(rec)}(M)-R^*(M)\,,
  \label{eq:dr}
\end{align}
where $R^\text{(rec)}$ and $R^*(M)$ are the reconstructed radius
(solid) and the genuine original radius (dashed), respectively, at the
mass $M$.  Then, we estimate the root mean square (RMS) of radius
deviations~\eqref{eq:dr} using 967 data at masses ranging
from 0.6~$\Msun$ to 1.8~$\Msun$ with an interval of 0.2~$\Msun$.
We calculate $\delta R(M)$ for both the NN and the polynomial
regression to make a quantitative comparison.

\begin{figure}
  \centering
  \begin{tabular}{ccc}
    \includegraphics[width=.315\textwidth]{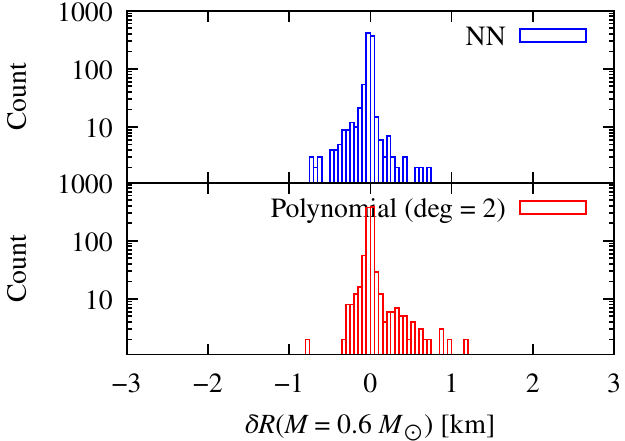} &
    \includegraphics[width=.315\textwidth]{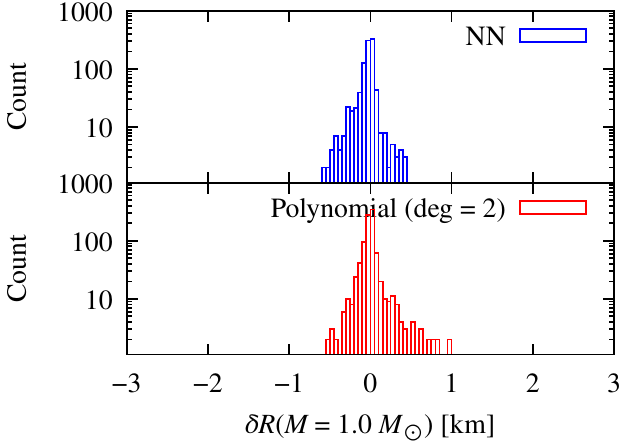} &
    \includegraphics[width=.315\textwidth]{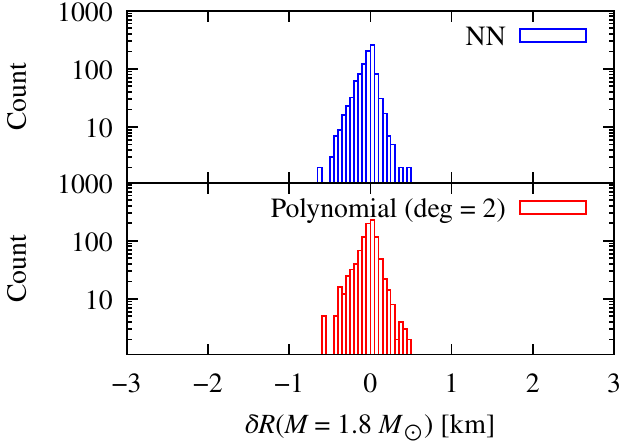} \\
    (a) & (b) & (c)
  \end{tabular}
  \caption{Histograms of $\delta R$ from the NN (upper) and the
    polynomial (lower) regression for (a) $0.6 \Msun$, (b) $1.0
    \Msun$, and (c) $1.8 \Msun$.}
  \label{fig:dr}
\end{figure}

The logarithmic histogram of $\delta R(M)$ is plotted in
Fig.~\ref{fig:dr} for $M= 0.6 \Msun$ [as shown in (a)] $1.0 \Msun$ [as
shown in (b)], and $1.8 \Msun$ [as shown in (c)].  In the plot of the
logarithmic probability density the Gaussian distribution or the
normal distribution takes a quadratic shape, so the $\delta R$
histograms from both the NN and the polynomial regression have a
strong peak at $\delta R = 0$ for all $M$.  The tails are wider than
the normal distribution, and the polynomial results (lower panels)
exhibit even slightly wider tails than the NN results (upper panels).

\begin{figure}
  \centering
  \begin{tabular}{ccc}
    \includegraphics[width=.32\textwidth]{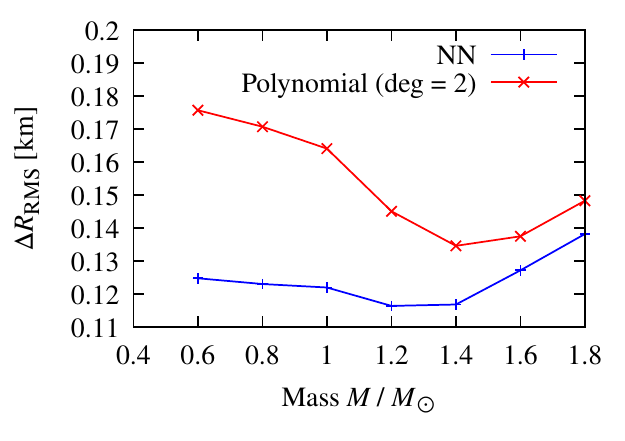} &
    \includegraphics[width=.32\textwidth]{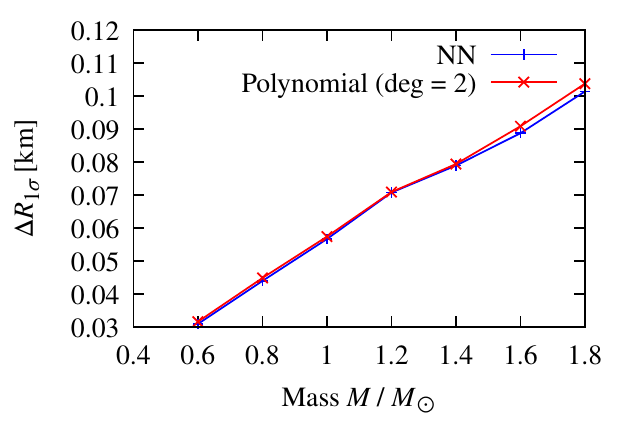} &
    \includegraphics[width=.32\textwidth]{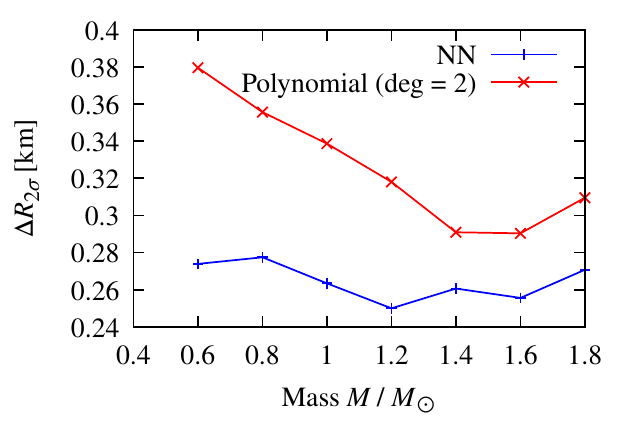} \\
    (a) & (b) & (c)
  \end{tabular}
  \caption{Overall performance measured with (a) $\Delta R_{\rm RMS}$,
    (b) $\Delta R_{1\sigma}$, and (c) $\Delta R_{2\sigma}$.}
  \label{fig:rms}
\end{figure}

We can look into this tendency in more quantitative manner in
Fig.~\ref{fig:rms}.  In the leftmost (a) of Fig.~\ref{fig:rms} we plot
the RMS of $\delta R(M)$, i.e., $\Delta R_{\rm RMS}$.  We see that
$\Delta R_{\rm RMS}$ from the NN is around $\sim 0.12\,\text{km}$ for
a wide range of masses.  This indicates that our NN method works
surprisingly good; remember that data points have random fluctuations
by $\Delta R\sim 0.5\,\text{km}$.  It should be noticed that, even
without neutron stars around $M=0.6$--$0.8\Msun$ in mock \MR data of
our setup, the RMS of the corresponding radii are still reconstructed
within the accuracy $\sim 0.1\,\text{km}$.  One might think that this
should be so simply because we use the SLy4 in the low energy region,
but in view of Fig.~\ref{fig:inferred}, different EoSs lead to
significantly different radii even in this region of
$M=0.6$--$0.8\Msun$.  Figure~\ref{fig:rms}~(a) clearly shows
that the NN results surpass the polynomial outputs with $n=2$.

When the distribution has long tails, as seen in Fig.~\ref{fig:dr},
the RMS may not fully characterize the data behavior.  To quantify the
deviation of the reconstruction more differentially, we can define
quantile-based widths, $\Delta R_{1\sigma}$ and $\Delta R_{2\sigma}$,
corresponding to $\simeq 68\%$ and $\simeq 95\%$ of $\delta R$
contained within the half widths, respectively.  They can be
explicitly expressed as
\begin{align}
  \Delta R_{\alpha\sigma} &\eqdef\frac12\left[
    F_{\delta R}^{-1}\biggl(\frac{1 + \erf\frac{\alpha}{\sqrt2}}2\biggr) -
    F_{\delta R}^{-1}\biggl(\frac{1 - \erf\frac{\alpha}{\sqrt2}}2\biggr)\right],
  \label{eq:rsigma}
\end{align}
where $F_{\delta R}^{-1}(p)$ is the quantile function of the
$\delta R$ distribution, i.e., the value of $\delta R$ at the fraction
$p$ integrated from small $\delta R$.  The error function,
$\erf(\alpha/\sqrt{2}) \equiv (1/\sqrt{2\pi})\int_{-\alpha}^{\alpha}dx e^{-x^2/2}$,
is used to translate the statistical significance by $\sigma$ to the
probability, e.g., $\erf(1/\sqrt2) \simeq 68\%$ and
$\erf(2/\sqrt2) \simeq 95\%$.  If $\delta R$ obeys the normal
distribution $\calN(0,\sigma^2)$, in general,
$\Delta R_{\alpha\sigma} = \alpha\sigma$ follows.  Therefore, we can
use $\Delta R_{\alpha\sigma} / \alpha$ as alternatives to
$\Delta R_\text{RMS} = \sigma$.  For more general distributions it is
important to note that these quantile-based widths have a clear
meaning as the 68\% and 95\% confidence intervals unlike
$\Delta R_\text{RMS}$.

In Figs.~\ref{fig:rms}~(b) and (c) we plot $\Delta R_{1\sigma}$ and
$\Delta R_{2\sigma}$.  It is interesting that the $\Delta R_{2\sigma}$
results appear very different while the $\Delta R_{1\sigma}$ results
are almost indistinguishable between the NN and the polynomial
results.  The polynomial results are consistently larger than the NN
results.  This behavior of $\Delta R_{2\sigma}$ implies that the
polynomial results may be contaminated by outliers of wrong values far
from the true answer.  We can understand this by relating it to
Runge's phenomenon of the polynomial approximation;  higher order
terms can cause large oscillation in the interpolation and the
extrapolation.  In particular, the extrapolation may easily break down
by higher order terms that blow up for large inputs.  In contrast, the
NNs with ReLU activation only contain up to the linear terms so that
the outputs are stable.

\section{EoS Estimation from the Real Observational Data}
\label{sec:obs}

In the previous section we have studied our methodology to infer the
most likely EoS from observational data and quantitatively confirmed
satisfactory performance to reproduce the correct results.  Now, we
shall apply the method to analyze the real observational data and
constrain the real neutron star EoS\@.

For practical purposes it is important to develop a way to estimate
uncertainties around the most likely EoS within the framework of the
machine learning method.  As we have already discussed, the EoS
inference from the \MR data is an ill-posed problem and the solution
cannot be uniquely determined.  Thus, the prediction from the NN
method cannot avoid uncertain deviations from the true EoS\@.
To employ a predicted EoS as a physical output, we need to quantify
the uncertainty bands on top of the results.
In this section we consider two different approaches for the
uncertainty quantification.

\subsection{Compilation of the neutron star data}
\label{sec:NSdata}

For the real data we use the \MR relation of the neutron stars
determined by various X-ray observations.  There are three sorts of
sources we adopt here: 8 neutron stars in quiescent low-mass X-ray
binaries (qLMXB)~\cite{Ozel:2015fia, Bogdanov:2016nle, Ozel:2016oaf},
6 thermonuclear bursters with better constraints than the
qLMXBs~\cite{Ozel:2015fia, Ozel:2016oaf} as well as a rotation-powered
millisecond pulsar, PSR J0030+0451, with thermal emission from hot
spots on the stellar surface~\cite{Riley:2019yda, Miller:2019cac}.
The data from the first two sources is tabulated in
Refs.~\cite{Ozel:2015fia,Bogdanov:2016nle,Ozel:2016oaf}, especially in
Fig.~4 of Ref.~\cite{Ozel:2016oaf}\footnote{The data is distributed
  at the website: \texttt{http://xtreme.as.arizona.edu/NeutronStars/}}.
The data from the last source is recently investigated in the NICER
mission, and there are two independent analyses based on the same
observation, among which here we use the data given in
Ref.~\cite{Riley:2019yda}.
All of these \MR relations are provided in the form of the Bayesian
posterior distribution, which takes the two-dimensional probability
distribution on the $R$-$M$ plane (see also Fig.~1 in
Ref.~\cite{Fujimoto:2019hxv} for a graphical representation of the
probability distribution).

Ideally, with sufficient computational resources, the machine learning
would be capable of directly dealing with such two-dimensional
probability distribution using, for example, the convolutional neural
network (CNN).  In the present work, however, we simplify our analysis
by approximately characterizing a two-dimensional probability
distribution with four parameters following the prescription proposed
in our previous publication~\cite{Fujimoto:2019hxv}:  the means and
the variances of the marginal distributions with respect to $M$ and
$R$.  More specifically, we make projection of the two-dimensional
distribution onto the one-dimensional $M$-axis (and $R$-axis) by
integrating over $R$ (and $M$, respectively).  In this way we
marginalize the two-dimensional distribution with respect to $M$ and
$R$ into two one-dimensional distributions, which are eventually
fitted by the Gaussians with the mean and the variance.

\subsection{Training data generation with observational uncertainty}

\begin{figure}
  \centering
  \includegraphics[width=0.7\textwidth]{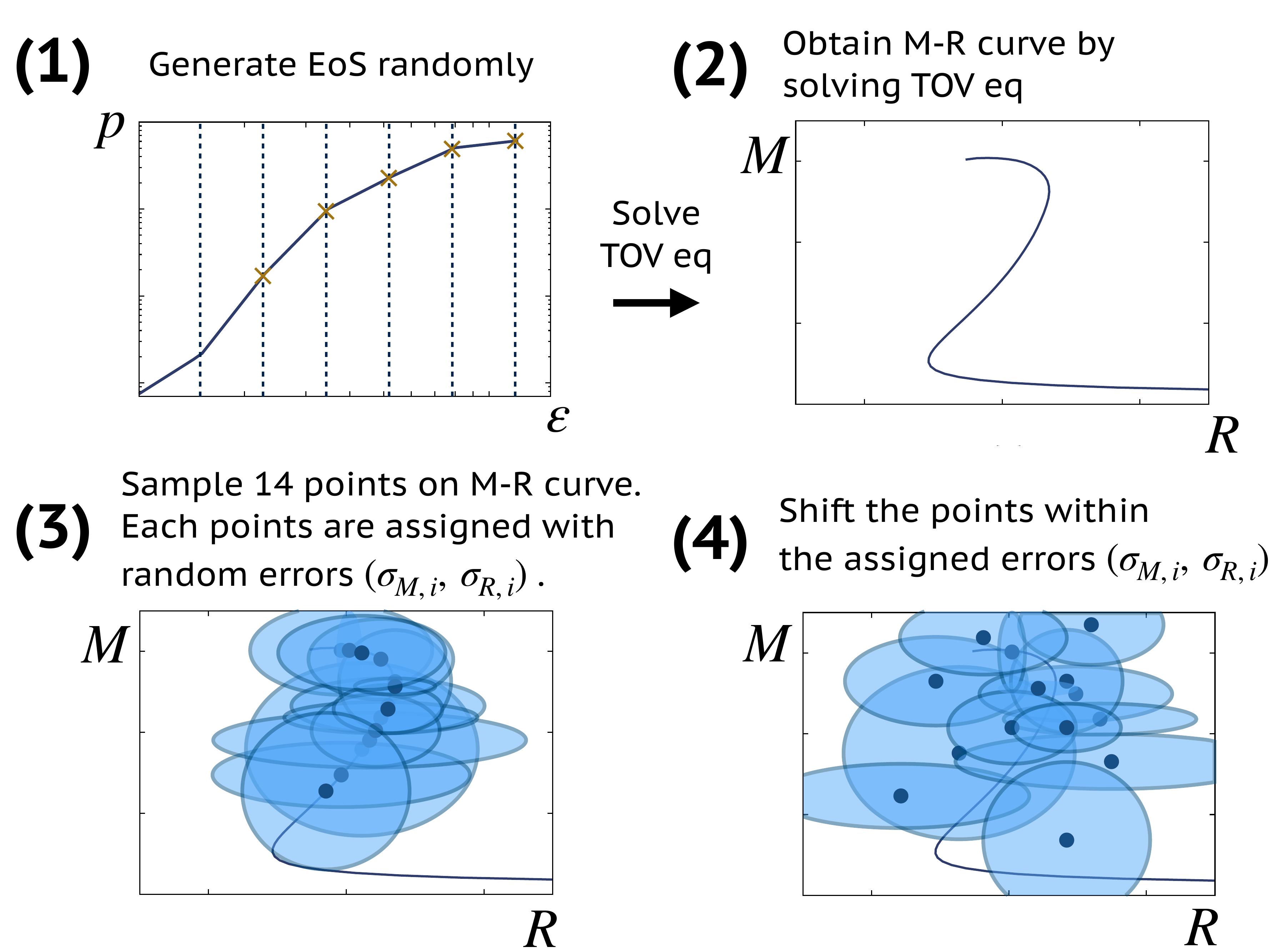}
  \caption{Schematic flow of data generation procedure for the
    analysis in Sec.~\ref{sec:obs}}
  \label{fig:procedure_b}
\end{figure}

We should revise the data generation procedure previously illustrated
in Fig.~\ref{fig:procedure_a} in Sec.~\ref{sec:uni5p}.  Previously,
for the mock data analysis, we assumed the universal variances,
$\sigma_M=0.1\Msun$ and $\sigma_R=0.5\,\text{km}$, for all
$\Delta M_i$ and $\Delta R_i$.  In reality, however, they should vary
for different $i$, i.e., $\sigma_{M,i}$ and $\sigma_{R,i}$ should
correspond to the observational uncertainties of ($M_i, R_i$).  To
deal with the real observational data, the revised procedure for
sampling the \MR points is sketched in Fig.~\ref{fig:procedure_b}.  We
need to design the NN with an input of information including
$\sigma_{M,i}$ and $\sigma_{R,i}$:  the input variables are extended
to ($M_{i}$, $R_{i}$; $\sigma_{M,i}$, $\sigma_{R,i}$).

We shall recapitulate the data generation scheme as follows.  In the
same way as in Sec.~\ref{sec:mock} we prepare 5 EoS parameters
$c_{s,i}^{2}$ ($i=1,\dots 5$) in the output side.  In this section the
training data comprises $14\times4$ input parameters, i.e.,
($M_{i}$, $R_{i}$; $\sigma_{M,i}$, $\sigma_{R,i}$) ($i=1,\dots,14$).
We note that $i$ runs to not 15 but 14 corresponding to the number of
observed neutron stars as explained in Sec.~\ref{sec:NSdata}.  We
calculate the \MR curve for each EoS, and then select 14 points of
$(M^*_i, R^*_i)$ on the \MR curve and add statistical fluctuations of
$\Delta M_{i}$ and $\Delta R_{i}$ [see
Fig.~\ref{fig:procedure_b}~(3)].  Let us go into more detailed
procedures now.  Unlike $\sigma_M$ and $\sigma_R$ in
Sec.~\ref{sec:mock} here we randomly generate $\sigma_{M,i}$ and
$\sigma_{R,i}$ differently for $i=1,\dots 14$.  These variances,
$\sigma_{M,i}$ and $\sigma_{R,i}$, are sampled from the uniform
distributions, $[0,\Msun)$ and  $[0,5~\text{km})$, respectively.  In
view of the observational data, these ranges of the distributions
should be sufficient to cover the realistic situations.  Then,
$\Delta M_i$ and $\Delta R_i$ are sampled according to the Gaussian
distributions with these variances, $\sigma_{M,i}$ and $\sigma_{R,i}$.
Finally we obtain the training data,
$(M_i=M^*_i+\Delta M_i, R_i=R^*_i+\Delta R_i; \sigma_{M,i}, \sigma_{R,i})$
($i=1,\dots 14$) [see Fig.~\ref{fig:procedure_b}~(4)].  Hereafter we
call these 14 tetrads of $(M_i, R_i; \sigma_{M,i}, \sigma_{R,i})$ an
\textit{observation}.

Now we prepare the training data set by taking multiple observations.
For each EoS we randomly generate 100 different pairs of
($\sigma_{M,i}, \sigma_{R,i})$, and then we make another 100
observations for each $(\sigma_{M,i}, \sigma_{R,i})$.  From the former
100 pairs the NN is expected to learn that the observational
uncertainties may vary, and the latter tells the NN that the genuine
\MR relation may deviate from the observational data.  In total we
make $n_s=10000$ ($=100\times100$) ``observations'' per one EoS\@.
The size of the whole training data set is thus 100 times larger than
before.

\begin{table}
  \centering
  \begin{tabular}{ccc} \hline
    Layer index & Neurons & Activation Function \\ \hline
    0 & 56 & N/A \\
    1 & 60 & ReLU \\
    2 & 40 & ReLU \\
    3 & 40 & ReLU \\
    4 & 5  & $\tanh$ \\ \hline
  \end{tabular}
  \caption{Neural network architecture used in Sec.~\ref{sec:obs}.  In
    the input layer, 56 neurons correspond to the input 14 points of
    the mass, the radius, and their variances.  The design of the
    other layers is kept the same as in Tab.~\ref{tab:nn_a}.}
  \label{tab:nn_b}
\end{table}

We modify the architecture of the NN used in this section
accordingly.  The number of neurons in the input layer becomes
$56\, (= 4 \times 14)$ (the performance test was done with 15 points,
but we have numerically confirmed that the same level of performance
is achieved with 14 points as well).  Since we already know from the
mock data analysis that the mass sorting does not affect the
performance, we keep the mass ordering as generated randomly, unlike
in Sec.~\ref{sec:uni5p}.  We normalize the input data as
$M_{i}/M_{\mathrm{norm}}$, $R_{i}/R_{\mathrm{norm}}$,
$\sigma_{M,i} / \sigma_{M\,\mathrm{norm}}$, and
$\sigma_{R,i} / \sigma_{R\,\mathrm{norm}}$ with
$M_{\mathrm{norm}}=3\Msun$,
$R_{\mathrm{norm}}=20\,\text{km}$,
$\sigma_{M\,\mathrm{norm}}=1\Msun$, and
$\sigma_{R\,\mathrm{norm}} = 5~\text{km}$.
Aside from the input layer, the NN design of the other layers is
chosen to be the same as before, as summarized in Tab.~\ref{tab:nn_b}.
For the NN optimization we adopt Adam, the same as in
Sec.~\ref{sec:mock}, but with different mini-batch size, 1000.
Incorporating the observational uncertainties the training data set
becomes larger as compared to before, and we expect that a larger
mini-batch size would fasten the training.

\subsection{Two ways for uncertainty quantification}
\label{sec:UQ}

Here we prescribe two independent methods to quantify uncertainties in
the output EoS based on different principles.

The first one utilizes the validation data, and the procedure is
similar to that in Sec.~\ref{sec:EoSrec}.  The basic idea is that,
once we have trained the NN, we can evaluate the prediction accuracy
from the validation data.  We generate 100 samples of the validation
data set whose input variances are chosen to be in accord with the
real observational uncertainties as explained in
Sec.~\ref{sec:NSdata}.  For the validation data, we know the true \MR
relation so that we can evaluate the deviation $\delta R(M)$ as
defined in Eq.~\eqref{eq:dr}.  Using the whole validation data set, we
calculate the root-mean-square deviation, $\Delta R_{\rm RMS}$, and
regard it as the systematic error inherent in the NN approach.  Later
we show this uncertainty by the label ``validation'' as seen in
Fig.~\ref{fig:resulteos}.

\begin{figure}
  \centering
  \includegraphics[width=0.7\textwidth]{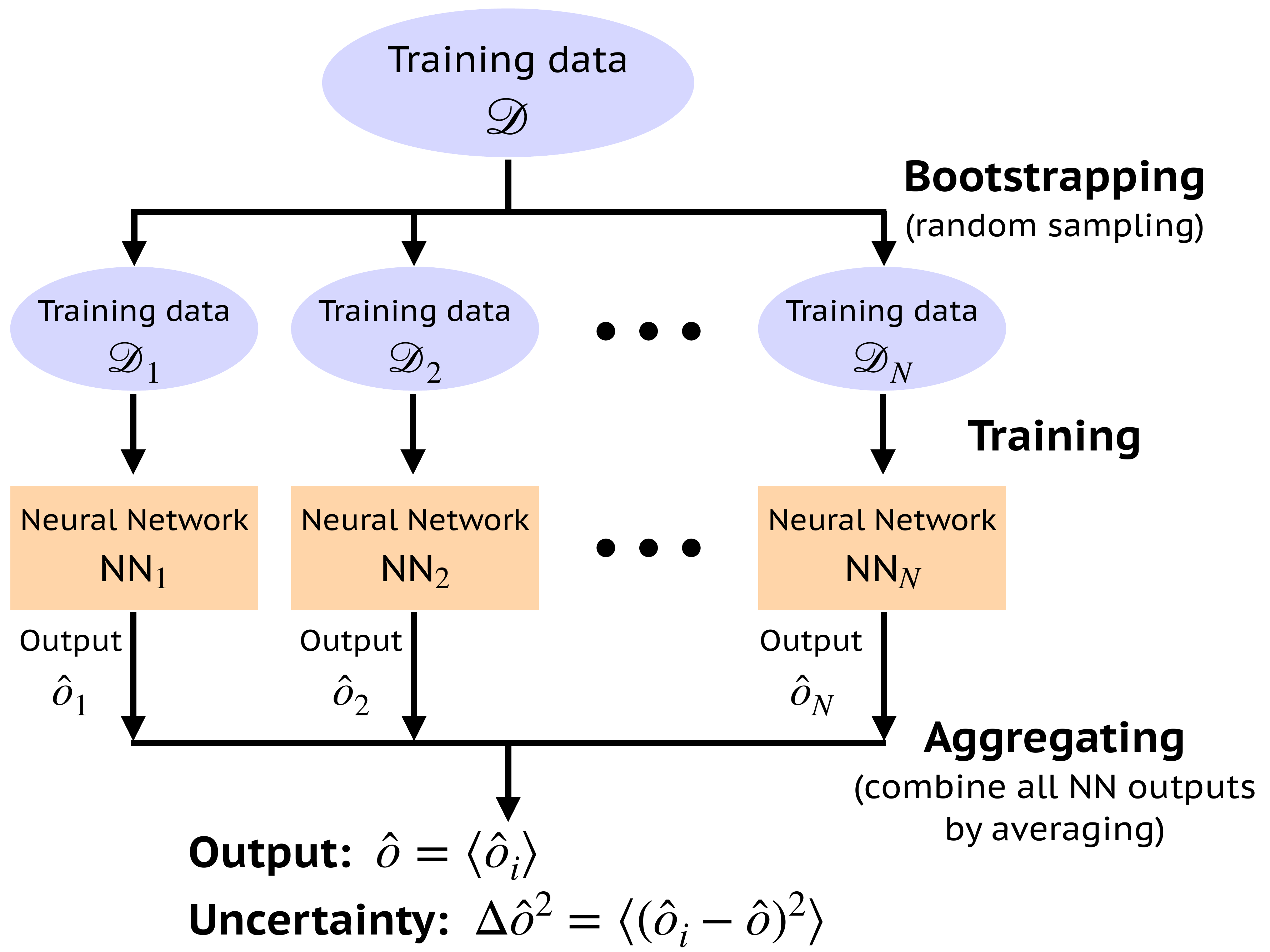}
  \caption{Illustration of the ensemble method procedure for
    uncertainty quantification.}
  \label{fig:bagging}
\end{figure}

The second one is the ensemble method in machine learning.  This
method is usually used to enhance the stability and performance of the
predicted output from NNs.  Here we repurpose it to quantify
uncertainty of the predicted output.  The idea is concisely
summarized in Fig.~\ref{fig:bagging}.  In this method we set up
multiple NNs independently: NN$_1$, NN$_2,\dots$,NN$_N$ with $N$ being
the number of prepared NNs.  We perform random sampling from $\calD$
to generate different subsets, $\calD_1, \calD_2, \cdots, \calD_N$ and
train each NN using $\calD_1, \calD_2, \dots, \calD_N$.  This random
sampling is commonly referred to as \textit{bootstrapping}.  After the
training, by feeding the input data, each NN predicts output values,
which we symbolically denote by
$\hat{o}_1, \hat{o}_2, \dots, \hat{o}_N$ as in Fig.~\ref{fig:bagging}.
Finally, aggregating the outputs from these multiple NNs, i.e.,
averaging all the outputs, we get the overall output
$\hat{o} = \langle\hat{o}_i\rangle \equiv N^{-1}\sum_i \hat{o}_i$.
Here we can also calculate the variance by
$\Delta\hat{o}^2 = \langle (\hat{o}_i - \hat{o})^2 \rangle$, and
regard it as ``uncertainty''.  This whole procedure, comprising
\underline{b}ootstrapping and \underline{agg}regat\underline{ing}, is
named \textit{bagging}\footnote{In the strict sense, the term
  ``bagging'' is used for the random sampling procedure \textit{with}
  replacement of data.  Here we use ``bagging'' even without
  replacement in a loose sense.}~\cite{10.1023/A:1018054314350}.
In this work, we choose $N=10$.  If some regions of the EoS are
insensitive to the \MR observation, independently trained 10 NN models
would lead to different EoSs in such unconstrained regions.  From
$\Delta\hat{o}$ that quantifies how much 10 output EoSs vary, we
can estimate the uncertainty around the output $\hat{o}$.  We use this
bagging for most of the analyses as shown by the band labeled by
``Bagging'' in Figs.~\ref{fig:resulteos}.

We have introduced the two natural ways to quantify uncertainties, but our
working prescriptions are still to be developed.  In fact there is no
established method yet.  A more systematic way for the uncertainty
quantification might be possible.  This is an interesting problem
left for future research in the general context of machine learning.

\subsection{The most likely neutron star EoS}
\label{sec:real}

\begin{figure}
  \centering
  \begin{tabular}{cc}
    \includegraphics[width=.48\textwidth]{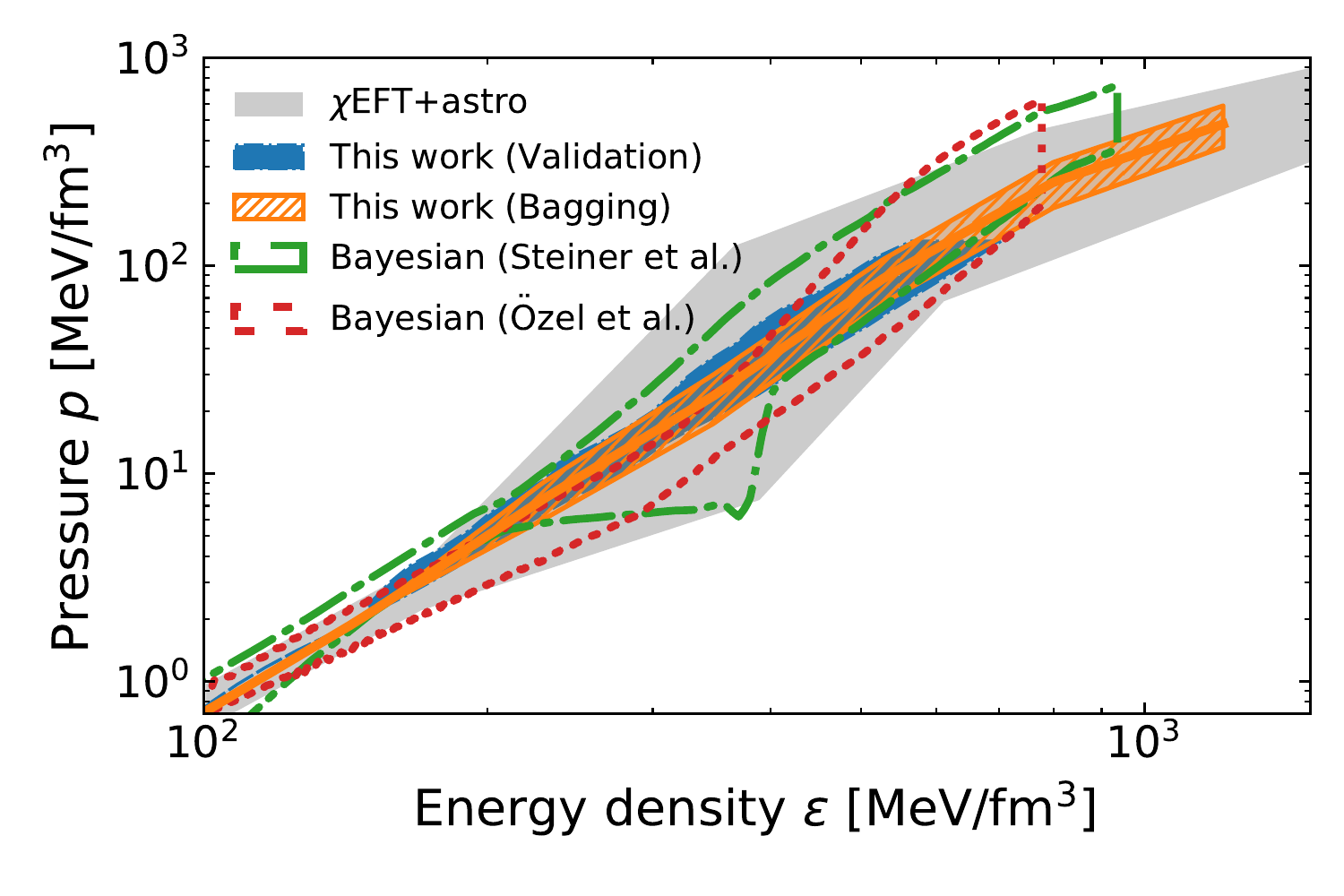} &
    \includegraphics[width=.47\textwidth]{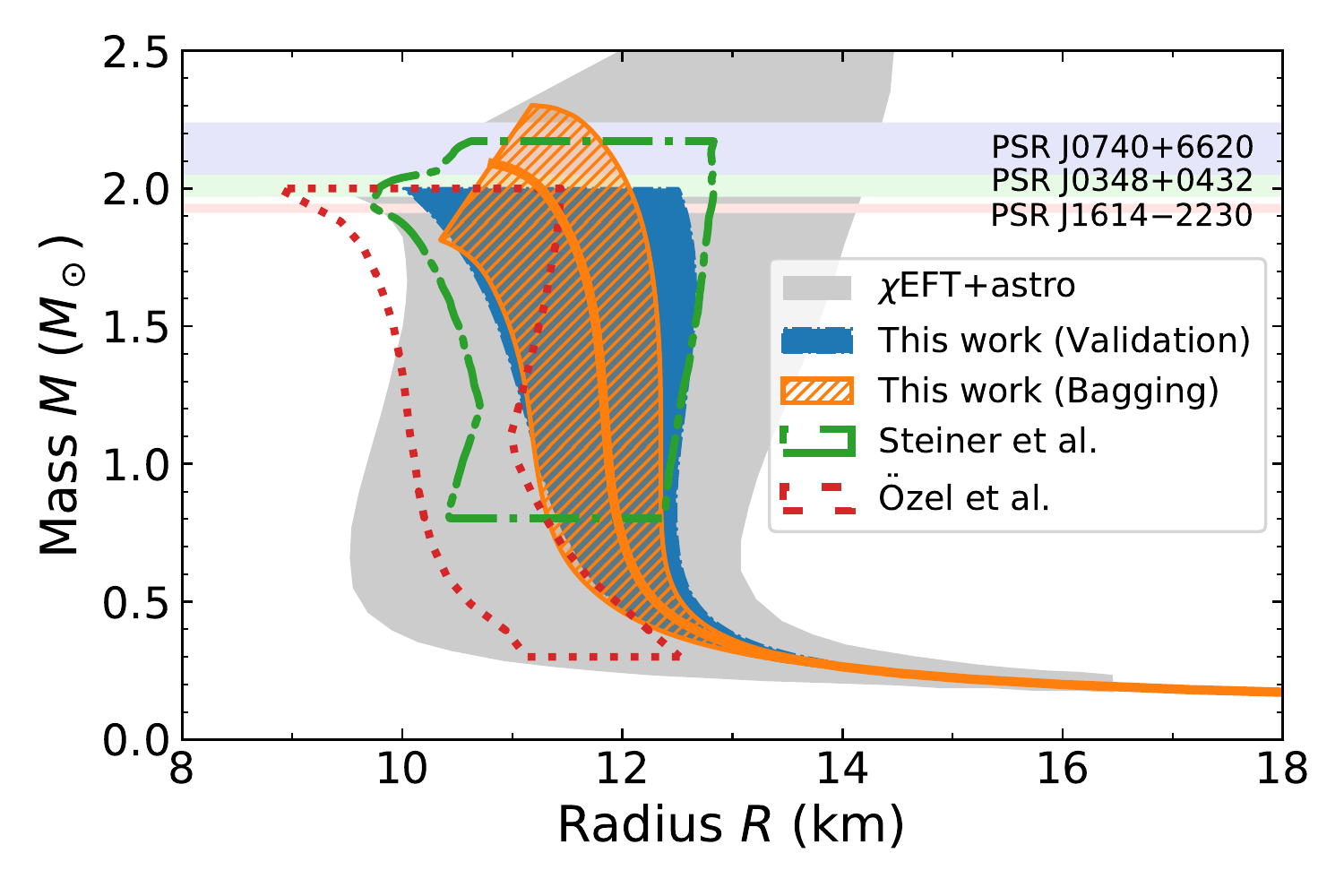} \\
    (a) & (b)
  \end{tabular}
  \caption{(a) EoSs deduced from the observational \MR data of qLMXBs
    and thermonuclear bursters.  The
    shaded blue and hatched orange bands represent our 68\%
    credibility bands from the validation and the bagging estimations.
    The $\chi$EFT prediction and the Bayesian results (Steiner
    {\it et al}.~\cite{Steiner:2010fz, Steiner:2012xt} and
    {\"Ozel} {\it et al.}~\cite{Ozel:2015fia,Bogdanov:2016nle,Ozel:2016oaf})
    are overlaid for reference.  The former band represents 68\% CL,
    while the latter shows the contour of $e^{-1}$ of the maximum
    likelihood.
    (b) \MR relations corresponding to the deduced EoSs from this work
    with references to other approaches.
  }
  \label{fig:resulteos}
\end{figure}

In Fig.~\ref{fig:resulteos} the orange line is the most likely neutron
star EoS deduced from our NN approach.  We estimated uncertainty from
the bagging (shown by the band with a hatch pattern labeled by
``bagging'') and the validation (shown by the blue band labeled by
``validation'').  We plot bands from other works in the figure for
comparison.  The gray band represents an estimate from the $\chi$EFT
calculation combined with polytropic extrapolation and the
two-solar-mass pulsar constraint~\cite{Hebeler:2013nza} (labeled by
``$\chi$EFT+astro'').  Because $\chi$EFT is an \textit{ab initio}
approach, any reasonable predictions should lie within the gray band,
and indeed our results are found to be near the middle of the gray band.
The preceding Bayesian analyses~\cite{Steiner:2010fz, Steiner:2012xt,
  Ozel:2015fia,Bogdanov:2016nle,Ozel:2016oaf} are also overlaid on
Fig.~\ref{fig:resulteos}.  While
{\"Ozel} {\it et al}.~\cite{Ozel:2015fia, Bogdanov:2016nle, Ozel:2016oaf}
and our present analysis use the same astrophysical data,
Steiner {\it et al}.~\cite{Steiner:2010fz, Steiner:2012xt} employs a
subset of the data, i.e., 8 of X-ray sources.  One may think that
our prediction gives a tighter constraint than the others, but the
narrowness of the band may be related with the implicit assumption in
our EoS parametrization; we will come back to this point in
Sec.~\ref{sec:PT} (see Fig.~\ref{fig:result_uni7r}).
Figure~\ref{fig:resulteos}~(b) shows the \MR curves corresponding to
the EoSs in (a).  We see that our EoS (blue curve) certainly supports
neutron stars with
$M>2\Msun$~\cite{Demorest:2010bx, Fonseca:2016tux, Antoniadis:2013pzd,
  Cromartie:2019kug}.

\begin{figure}
  \centering
  \begin{tabular}{cc}
    \includegraphics[width=.48\textwidth]{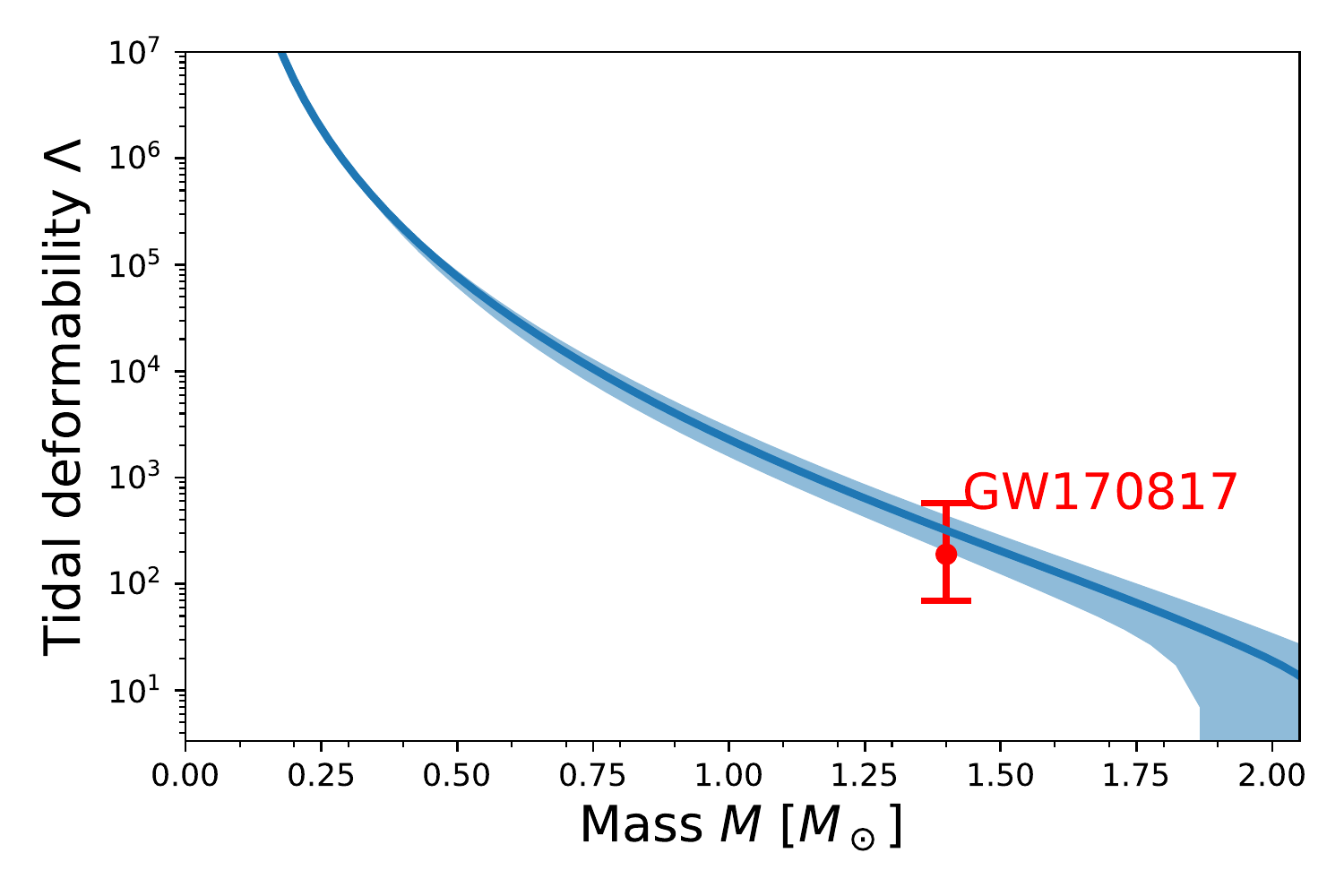} &
    \includegraphics[width=.47\textwidth]{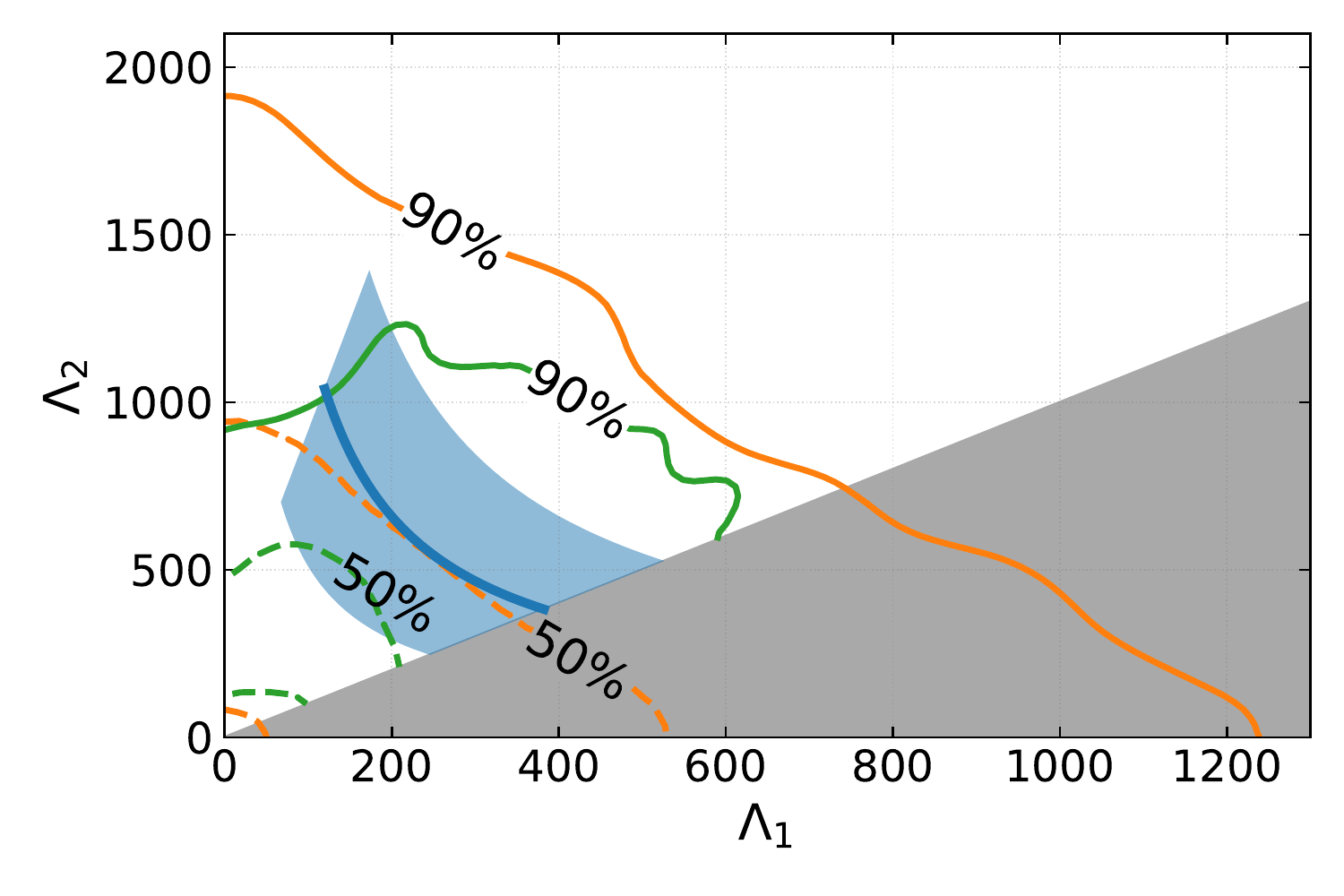} \\
    (a) & (b)
  \end{tabular}
  \caption{(a) Tidal deformability $\Lambda$ calculated
    from our EoS\quad (b) Correlation of tidal deformabilities,
    $\Lambda_1$ and $\Lambda_2$; see the text for details.}
  \label{fig:tidal}
\end{figure}

Figures~\ref{fig:tidal}~(a) and (b) show the tidal deformability and
their correlation,  respectively, in the binary neutron star merger
GW170817.  Once an EoS is given, the dimensionless tidal
deformability, $\Lambda$, results from a quantity called the Love
number $k_2$, which is derived from the Einstein equation under static
linearized perturbations to the Schwarzschild metric due to external
tidal fields.  Practically, we solve a second-order ordinary
differential equation in combination with the TOV equation;  see
Refs.~\cite{Hinderer:2007mb, Hinderer:2009ca} for the explicit form of
the equations.  The blue band in Fig.~\ref{fig:tidal}~(a) represents
$\Lambda$ from the EoS we inferred in the present work, which is
consistent with the merger event GW170817 indicated by the red bar.
In Fig.~\ref{fig:tidal}~(b) we show the correlation of the tidal
deformabilities, $\Lambda_1$ of the star 1 and $\Lambda_2$ of the star
2, using the relation between $\Lambda$ and $M$ as given in
Fig.~\ref{fig:tidal}~(a).  The orange lines in Fig.~\ref{fig:tidal}~(b)
refer to the constraints (solid:90\% and dashed:50\%) for which
$\Lambda_1$ and $\Lambda_2$ are sampled
independently~\cite{TheLIGOScientific:2017qsa}, while the green lines
refer to the constraints for which $\Lambda_1$ and $\Lambda_2$ are
related through $\Lambda_a(\Lambda_s, q)$ with
$\Lambda_a = (\Lambda_2 - \Lambda_1) / 2$,
$\Lambda_s = (\Lambda_2 + \Lambda_1) / 2$, and $q = M_2 / M_1$.
In Fig.~\ref{fig:tidal}~(b) we clearly see that our predicted band is
located within the 90\% contours of the LIGO-Virgo
data~\cite{TheLIGOScientific:2017qsa, Abbott:2018exr}.

\begin{figure}
  \centering
  \begin{tabular}{cc}
    \includegraphics[width=.47\textwidth]{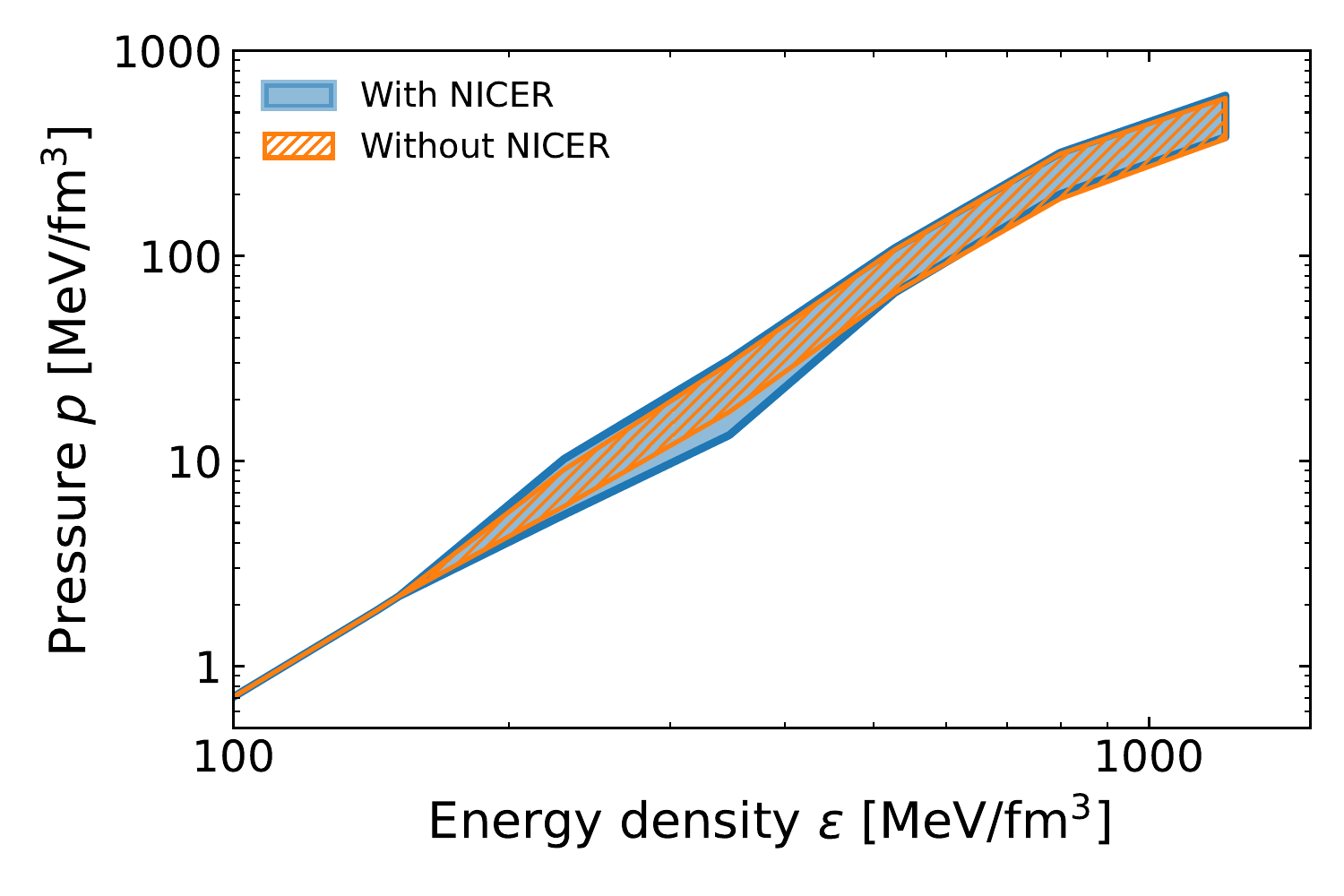} &
    \includegraphics[width=.49\textwidth]{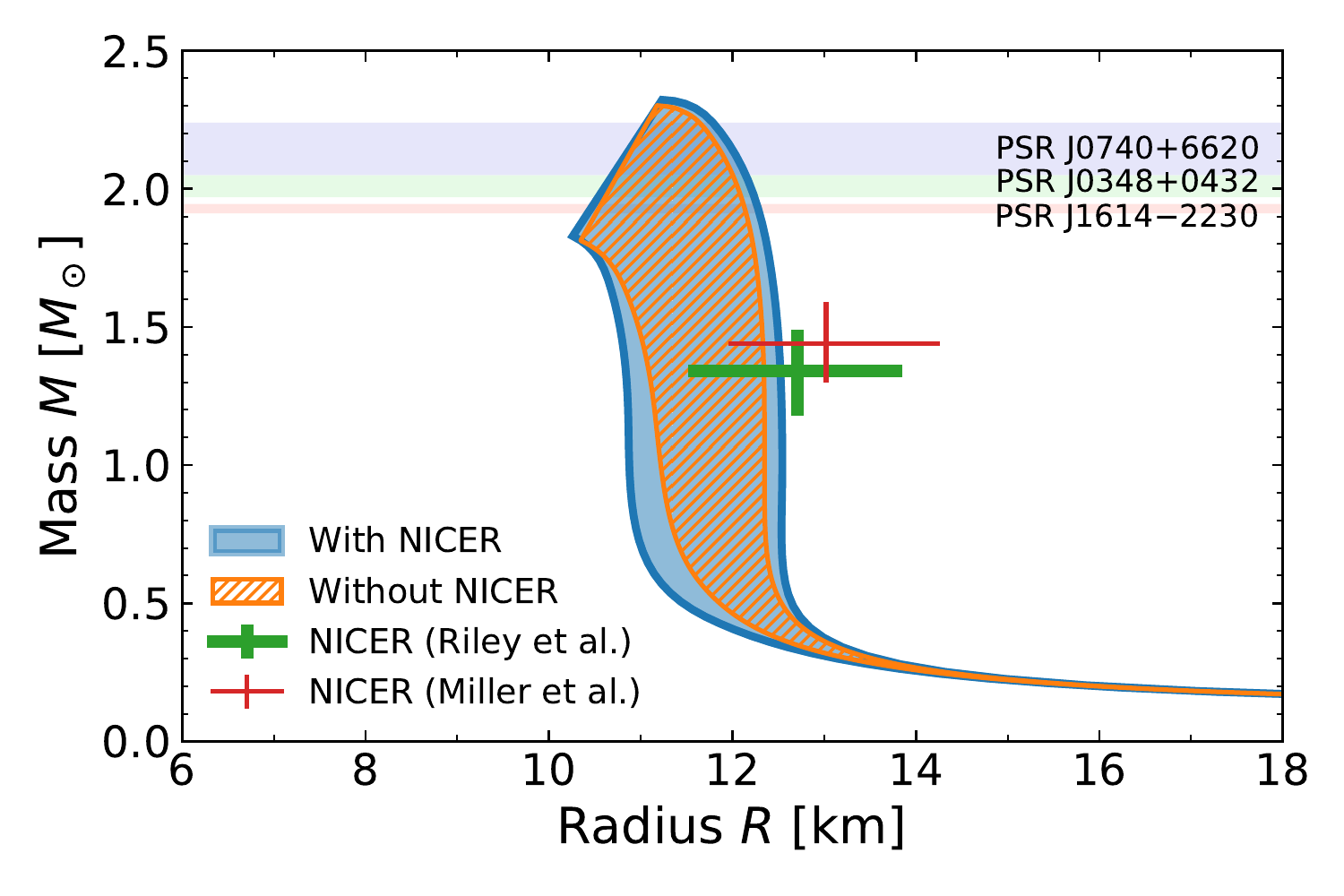} \\
    (a) & (b)
  \end{tabular}
  \caption{(a) EoSs deduced from the observational \MR data of qLMXBs,
    thermonuclear
    bursters~\cite{Ozel:2015fia, Bogdanov:2016nle, Ozel:2016oaf},
    and NICER data of PSR J0030+0451~\cite{Riley:2019yda}.  The shaded
    blue and the hatched orange regions represent the 68\% uncertainty
    band (evaluated in the bagging method) for the analyses with and
    without the NICER data, respectively.
    (b) \MR relations corresponding to the deduced EoSs shown in (a).}
  \label{fig:result_nicer}
\end{figure}

So far we have only used the observational data from the thermonuclear
bursters and qLMXBs, which may contain large systematic errors related
with the uncertain atmospheric model of neutron stars.  The results in
Figs.~\ref{fig:result_nicer}~(a) and (b) include the \MR constraint
from the NICER mission as well.  There are two independent analyses,
as spotted on Fig.~\ref{fig:result_nicer}~(b), on the same observation
of PSR J0030+0451~\cite{Riley:2019yda, Miller:2019cac}, and we adopt
the one [green bar in (b)] in Ref.~\cite{Riley:2019yda}.  We see that
the uncertainty becomes slightly larger by the inclusion of the NICER
data, which is attributed to the relatively large deviation of the
NICER data from others.

\subsection{Possible EoSs with a weak first-order phase transition}
\label{sec:PT}
\label{sec:uncertainty-quantification}

\begin{figure}
  \centering
  \includegraphics[width=0.47\textwidth]{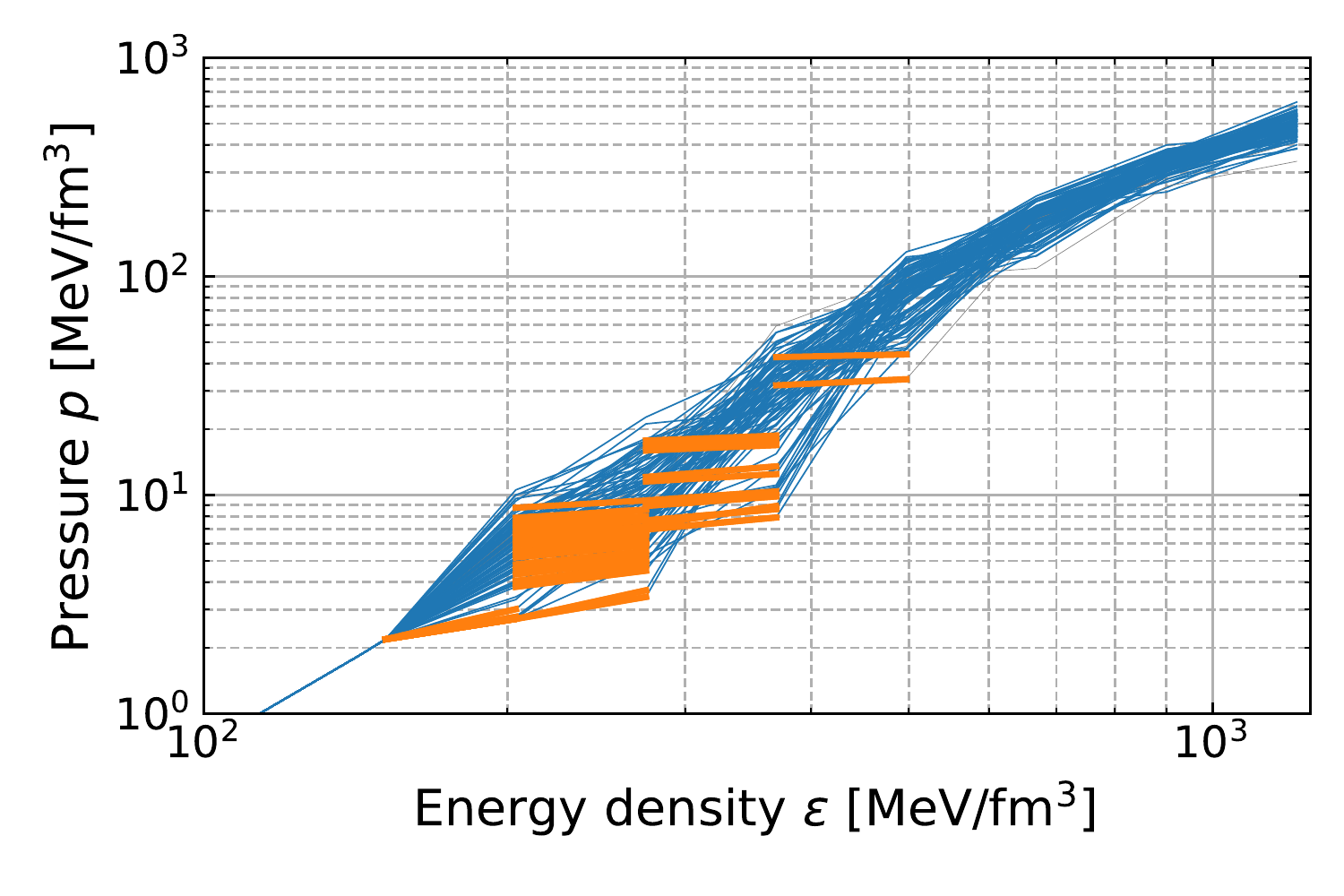} \hspace{1.5em}
  \includegraphics[width=0.47\textwidth]{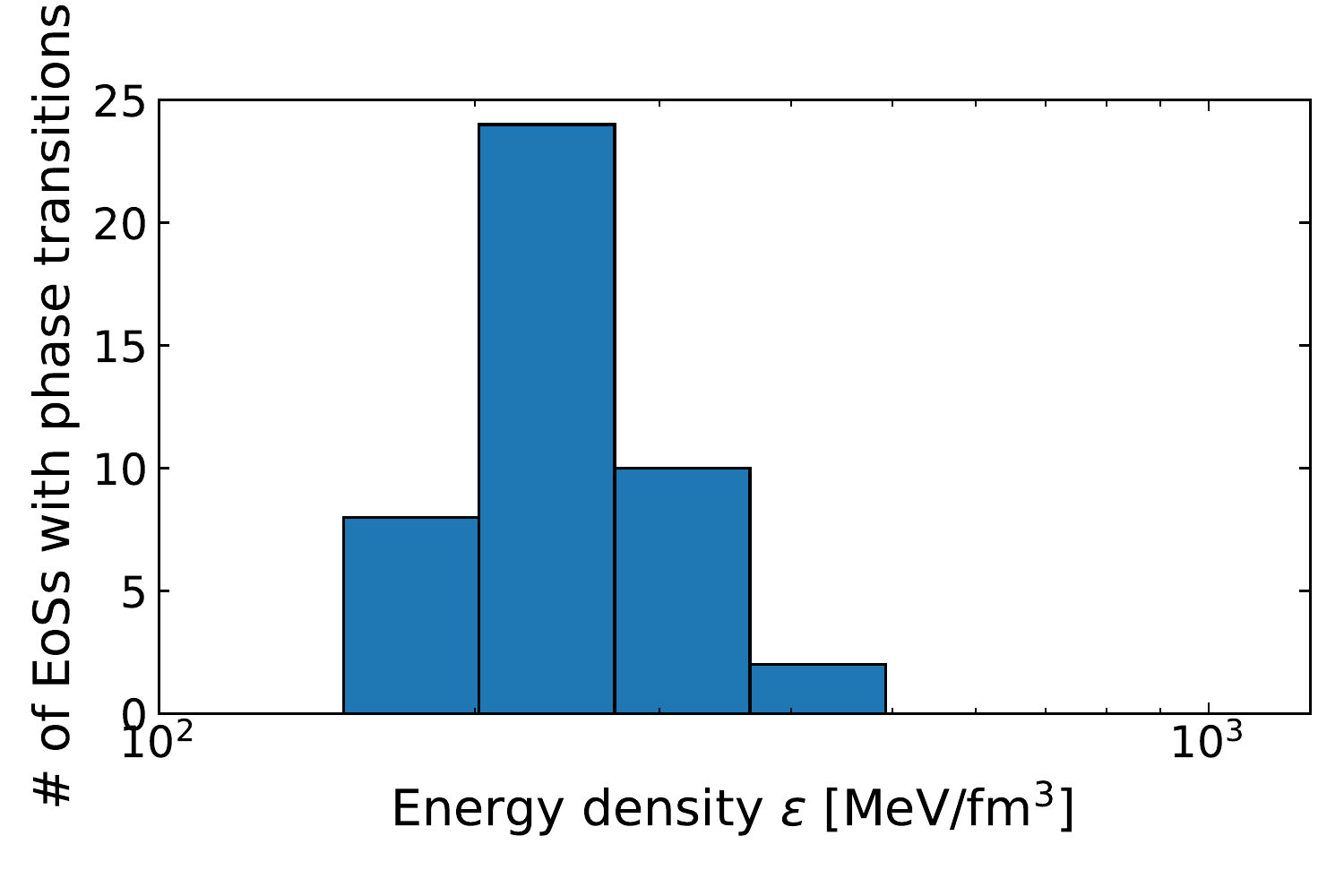}
  \caption{(Left) 100 output EoSs predicted from each bagging
    predictor with a first-order transition highlighted by the orange
    thick lines.
    (Right) Histogram of the first-order phase transitions in each
    energy density region of piecewise polytropes.}
  \label{fig:1stPT}
\end{figure}

In the analyses we have presented so far, we used the piecewise
polytrope with 5 segments of density.  Here, we change the number
of segments from 5 to 7 and repeat the inference with the finer bins.
There are mainly two issues argued in this subsection:  a possibility
of a weak first-order phase transition with the finer bins and its
implication on the uncertainty quantification.

To this end we prepared $N=100$ NNs in the bagging outlined in
Sec.~\ref{sec:UQ}.  We note that each NN is trained so as to predict
an EoS in response to the real observational data.  Here we use the
\MR data of qLMXBs and thermonuclear bursters without the
NICER data.  In the left panel of Fig.~\ref{fig:1stPT} we show 100
EoSs predicted from 100 independent NNs.  We remind that the activation function in the
output layer is chosen to be $\tanh$ which takes a value over $[-1,1]$.
When the predicted values is smaller than the threshold: $c_s^2<\delta=0.01$,
we adjust it to $c_s^2=\delta$ and identify a first-order phase transition.
We found 44 predicted EoSs out of 100 that have a first-order phase transition.
We highlight the region of
first-order phase transition with orange thick lines in
Fig.~\ref{fig:1stPT}.
From this plot we can
understand why we increased the number of segments.  If we use the EoS
parametrization with 5 segments, weak first-order phase transitions
are too strongly prohibited by coarse discretization.

We also make a histogram in the right panel of Fig.~\ref{fig:1stPT} to
show a breakdown of the EoS regions with a first-order phase
transition.  This histogram counts the number of first-order
transition EoSs in each energy density region.  It is interesting to
see that the most of the first-order phase transition is centered
around the energy region $202$--$272\MeV$.  On the one hand, in the
lower energy region $150$--$202\MeV$ the first-order phase transition
is less likely, and this tendency is consistent with the fact that a
stronger first-order phase transition in a lower energy region is more
disfavored by the two-solar-mass pulsar
constraint~\cite{Alford:2015dpa}.  In the higher energy region, on the
other hand, there are also less EoSs with a first-order phase
transition.  One may think that a first-order phase transition would be
more allowed in the higher energy region, but it is not the case in
the NN analysis.  In Sec.~\ref{sec:error-correlation} we already
discussed that the NN model tends to predict the most conservative
value around $c_s^2\sim 0.5$ in the high energy density regions where
the constraints are inadequate.  Therefore, the correct interpretation
of the absence of the first-order transition in the high density
regions as shown in Fig.~\ref{fig:1stPT} should be, not that our
results exclude a first-order transition there, but merely that the
observational data analyzed in our NN method does not favor a
first-order transition there.  Another artificial factor in the high
energy density region is that our piecewise polytrope is equally
spaced in the log scale, so the higher density segments have larger
energy widths in the linear scale.  Then, the parametrization does
not have a resolution to represent a weak first-order phase transition
in the high energy region.  Here, we also make a comment that our NN
approach does not take account of the third family
scenario~\cite{Gerlach:1968zz, Schertler:2000xq} in which separate
branches may appear in the \MR relation (see also
Refs.~\cite{Alford:2017qgh, Blaschke:2020qqj} for recent discussions).

\begin{figure}
  \centering
  \begin{tabular}{cc}
    \includegraphics[width=.46\textwidth]{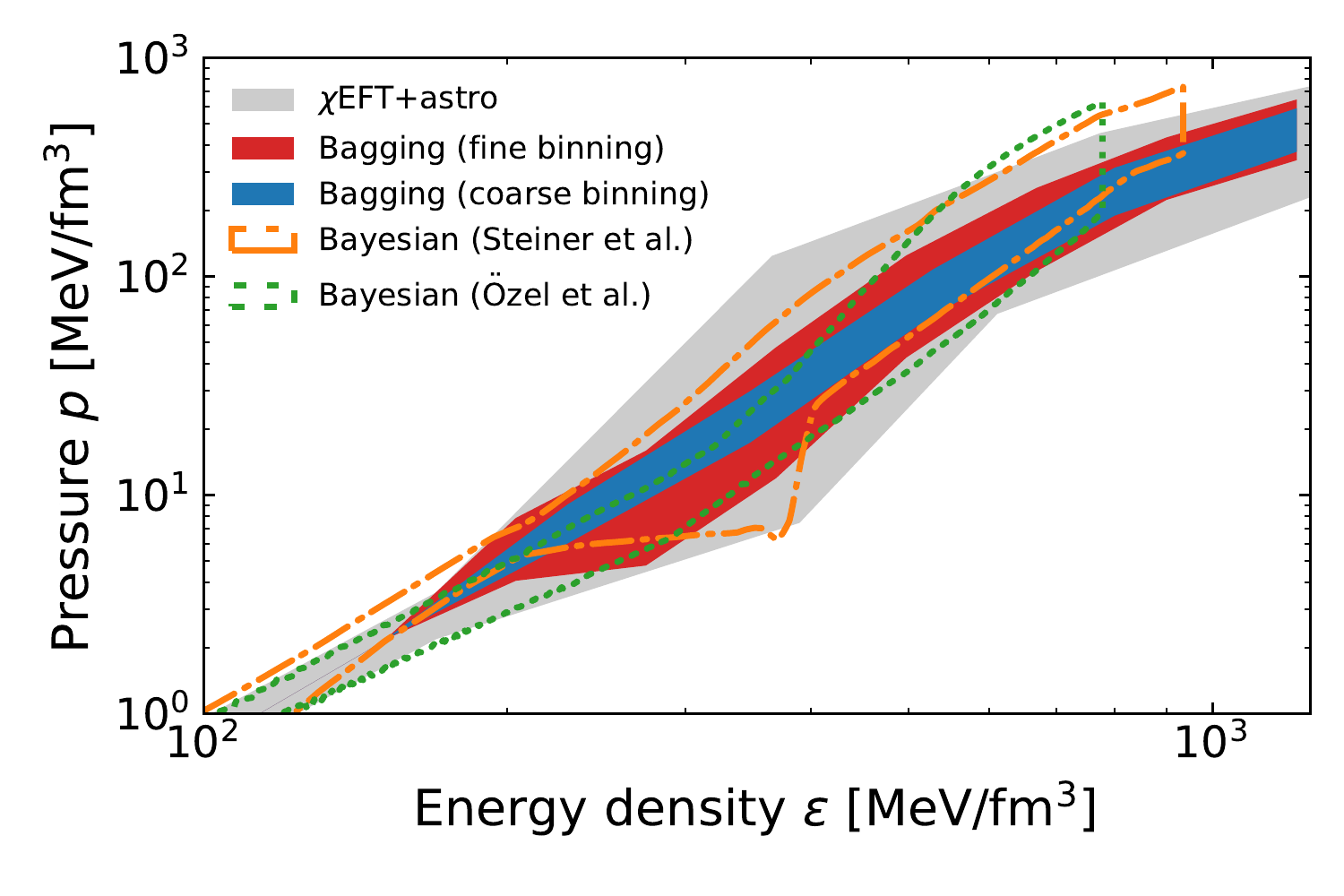} &
    \includegraphics[width=.47\textwidth]{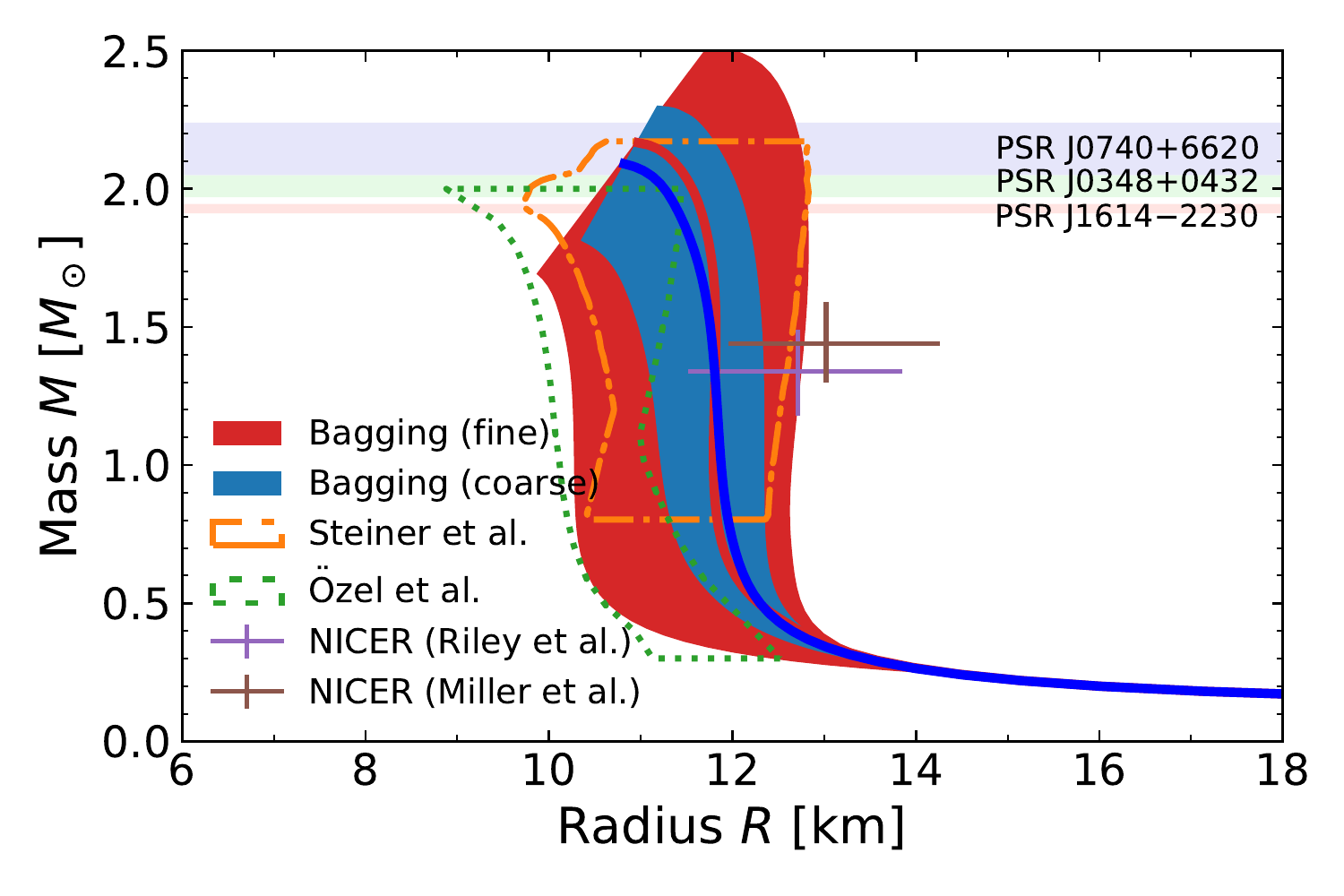} \\
    (a) & (b)
  \end{tabular}
  \caption{(a) EoSs deduced from the NN inference.  The red and the
    blue shades represent the 68\% credibility band evaluated in the
    bagging method with piecewise polytrope with 7 (fine bin) and 5
    (coarse bin) segments, respectively.  Other bands are the same as
    in Fig.~\ref{fig:resulteos}.
    (b) \MR relations corresponding to the deduced EoSs and
    uncertainty bands in (a).}
  \label{fig:result_uni7r}
\end{figure}

In Fig.~\ref{fig:result_uni7r} we show the EoS results with 7 segments
(fine binning) after bagging.  We see that uncertainty is widened
compared with the results with 5 segments (coarse binning).  This is
partially because the number of output parameters, $c_{s,i}^2$
($i=1,\dots,7$), is increased, while the amount of the input data is
unchanged, so the corresponding uncertainty should increase.
Furthermore we see that the uncertainty is enhanced by the effect of
first-order phase transition.  If there is a first-order phase
transition, $c_s^2$ becomes (nearly) zero and these zero values
enlarge the variance of the output in the bagging and increase the
uncertainty accordingly.  Now the uncertainty appears comparable with
other Bayesian methods as seen in Fig.~\ref{fig:result_uni7r}.  Our
previous results with 5 segments as shown in
Ref.~\cite{Fujimoto:2019hxv} and in Sec.~\ref{sec:real} of this paper
had smaller uncertainty, which implies that the choice of 5 segments
turns out to be optimal.

\begin{figure}
  \centering
  \includegraphics[width=.46\textwidth]{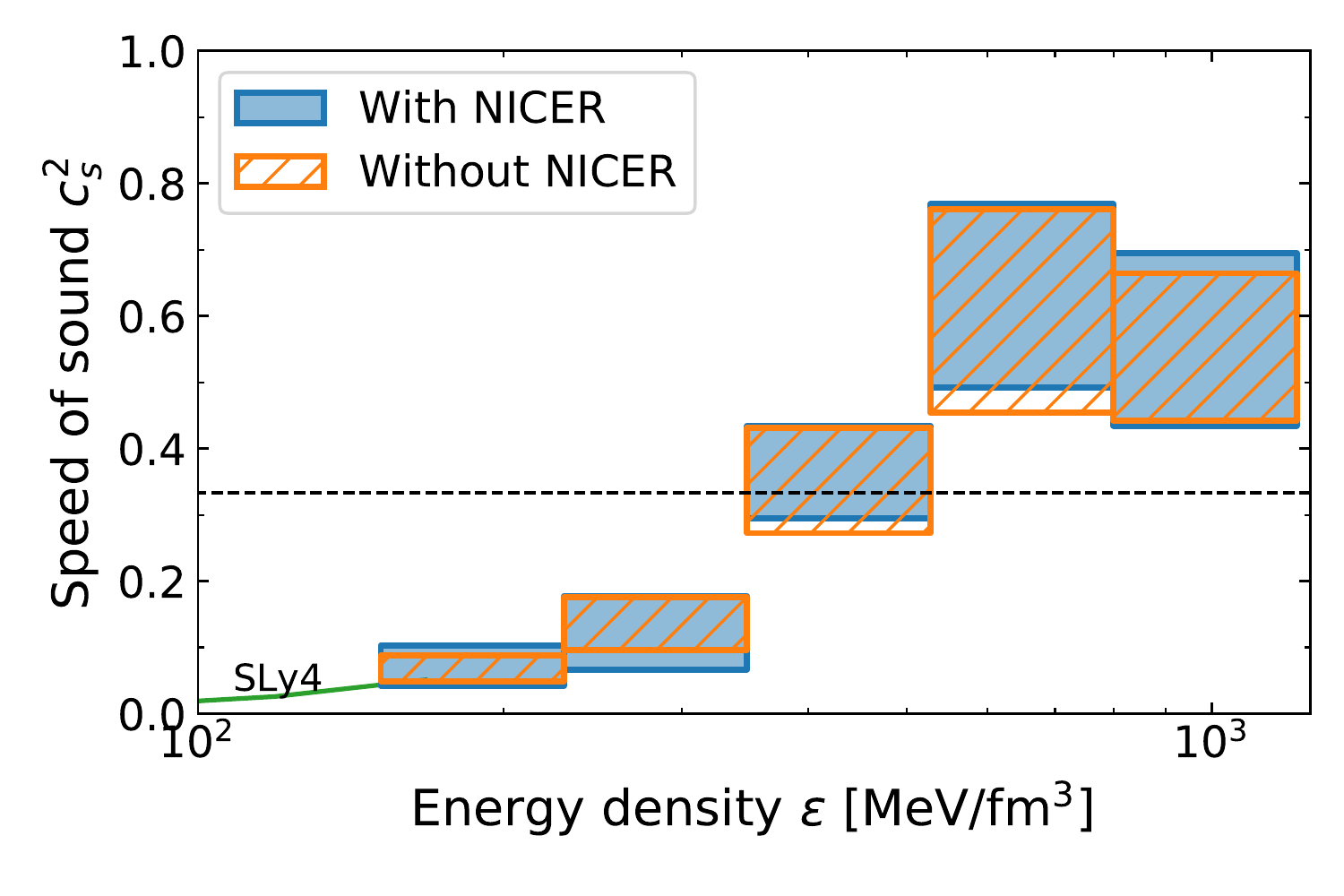} \hspace{0.5em}
  \includegraphics[width=.47\textwidth]{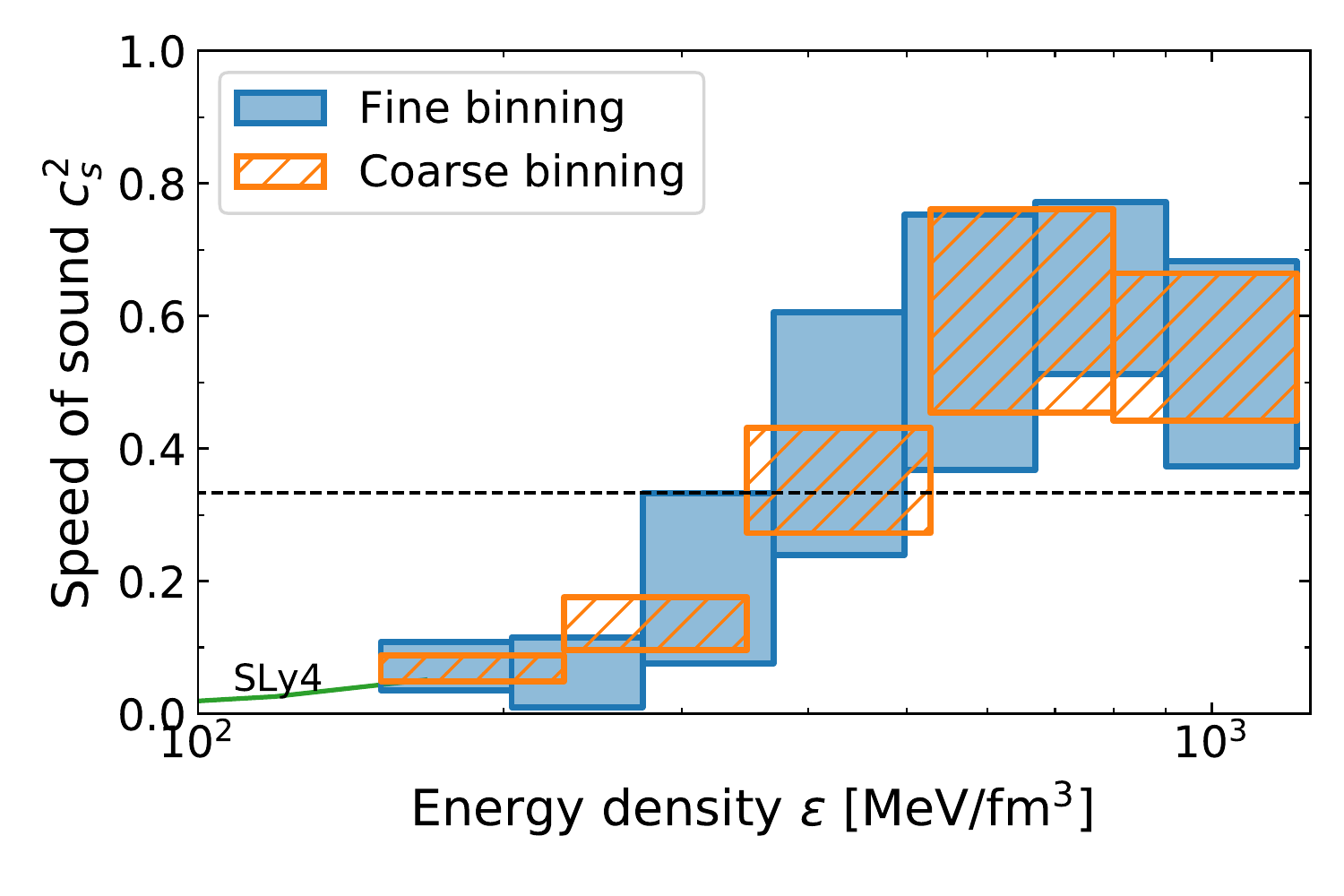}
  \caption{(Left) The speed of sound from our EoSs with and without
    the NICER data by the shaded blue and the hatched orange regions.
    (Right) The speed of sound from fine binning (shaded blue) and
    coarse binning (hatched orange) estimates.}
  \label{fig:cs2}
\end{figure}

Finally we plot the speed of sound in Fig.~\ref{fig:cs2}.  The bare
output from the NN model actually comprises these values of
$c_{s,i}^2$.  It is clear from the plots that the speed of sound
exceeds the conformal limit $c_s^2 = 1/3$; see
Refs.~\cite{Bedaque:2014sqa, Tews:2018kmu} for thorough discussions on
the conformal limit.  The inclusion of the NICER data only slightly widens
uncertainty, and changing the segment number is a more dominant effect
on uncertainty bands as quantified in the left panel of Fig.~\ref{fig:cs2}.

\section{More on the Performance Test:  Taming the Overfitting}
\label{sec:curve}

In Sec.~\ref{sec:learningcurve}, we observed a quantitative difference
between the learning curves for the training data sets with and
without data augmentation by $n_s=100$ as demonstrated in
Fig.~\ref{fig:learningcurve}.  Then, it would be a natural
anticipation to consider that this observational data augmentation may be
helpful to overcome the problems of local minimum trapping and
overfitting that we often meet during the NN training.  This section
is aimed to discuss numerical experiments to understand the behavior
of the learning curve and the role of $n_s$ thereof.  In particular,
we will focus on the overfitting problem here\footnote{We also tested
  the performance to escape from the local minimum trapping, but the
  data augmentation seems not to be useful to resolve the trapping
  problem.}.

\subsection{A toy model}
\label{sec:toy}
\begin{figure}
  \centering
  \includegraphics[width=0.5\textwidth]{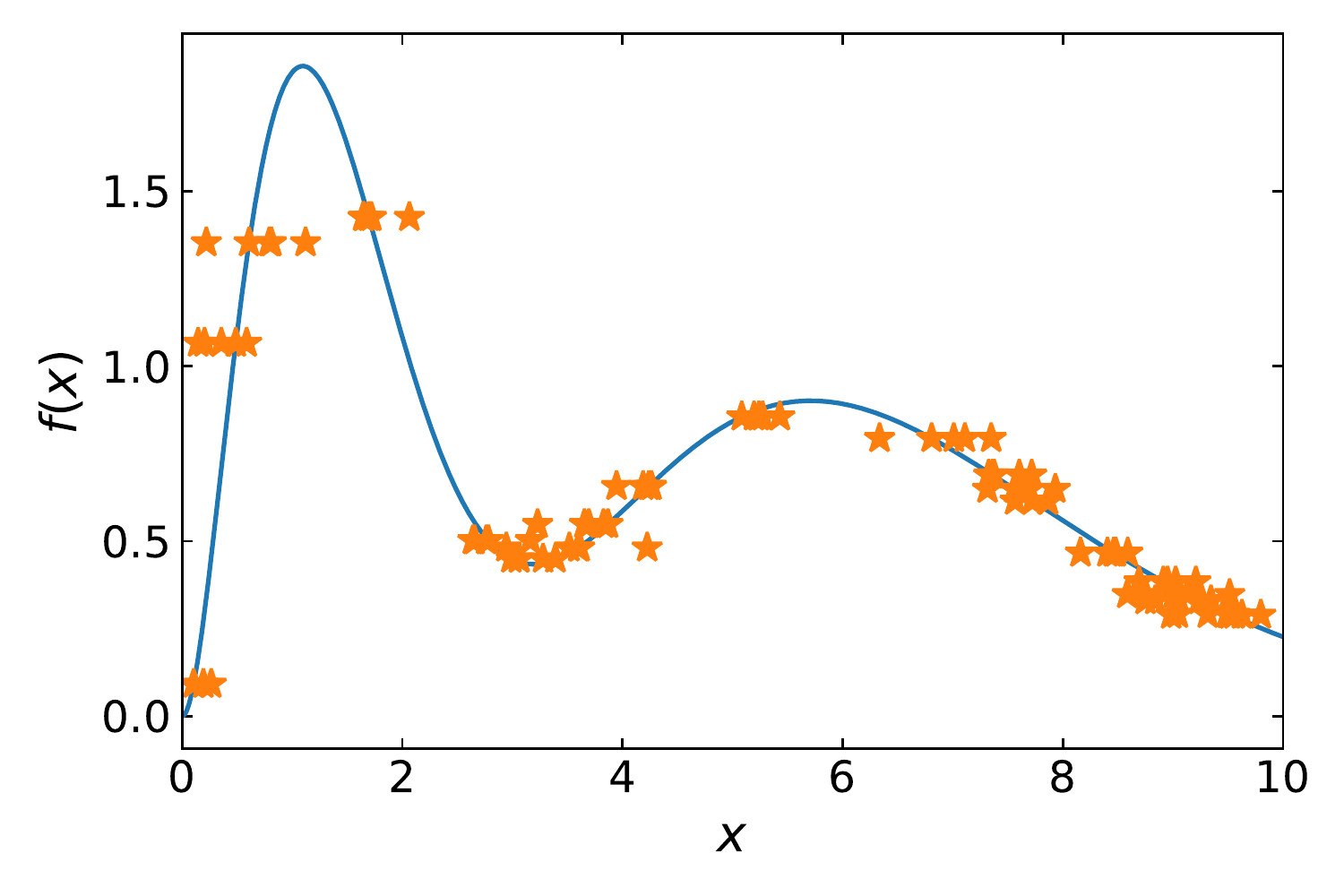}
  \caption{A concrete shape of $f(x)$ of our choice (blue curve) and
    typical training data (orange dots) for $n_{\rm base} = 20$ and
    $n_s = 5$.}
  \label{fig:fit_func}
\end{figure}

Here we consider a specific and simple problem as a toy model.  We
choose the following double-peaked function
$f(x)$\footnote{This choice of $f(x)$ is motivated from spectral
  functions which often appear in the lattice-QCD calculation.}:
\begin{equation}
  f(x) = (10x^2 - 6x^3 + x^4) e^{-x}
  \label{eq:toy}
\end{equation}
in the domain of $x \in [0,10]$.  A concrete shape of above $f(x)$ is
depicted in Fig.~\ref{fig:fit_func}.  In Tab.~\ref{tab:nsfx} we make a
comparison between the neutron star calculation and this toy-model
study of $f(x)$ fitting.

\begin{table}
  \centering
  \begin{tabular}{c|cc} \hline
     & Neutron star ($i=1,\dots,15$) & $f(x)$ fitting \\ \hline
    Input & $(M_i, R_i)$ & $x$ \\
    Output & EoS & $f(x)$ \\
    Training data (input) &  $(M^*_i+\Delta M_i, R^*_i+\Delta R_i)$ & $x^* + \Delta x$ \\
    Number of training data (output) & 1948 & $n_{\rm base} = 20$\\
    Number of training data (input) & 1948 $n_s = $ 1948000 & $n_{\rm base}\cdot n_s = 20\ n_s$ \\ \hline
  \end{tabular}
  \caption{Literal correspondence between the neutron star calculation
    (mock data analysis) and the $f(x)$ fitting.}
  \label{tab:nsfx}
\end{table}

For the training data, we first generate a set of the ``true'' inputs,
denoted by $x^*$, in the interval $[0,10)$ and calculate the
corresponding outputs, $y^* = f(x^*)$.  We denote this base data set
as $T^* = \{ (x^*_p, y^*_p)\}_{p=1}^{n_\text{base}} \subset \mathbb{R}^2$
with $n_\text{base} = |T^*|$ being the size of the base data set.
Then, similarly to the neutron star case, we augment the training data
set by duplicating $n_s$ copies of the base data with the input
fluctuated by ``observational uncertainty'', $\Delta x$.  Here, we set
the distribution of $\Delta x$ as the normal distribution with the
standard deviation fixed as
$\sigma(x^*) = 0.3(x^*+0.5)$ of our choice\footnote{With this choice
  of $\sigma(x^*)$ the uncertainty is smaller in the $x$ region with a
  narrow peak.  For the demonstration purpose the toy model is
  designed in such a way that the problem is not too easy but not
  intractable.}.  Then, the whole training data set can be expressed as
\begin{align}
  T = \bigl\{ (x^*_p + \Delta x_{p,i}, y^*_p)\bigm|
    p \in \{1,\dots,n_\text{base}\},
    i \in \{1,\dots,n_s\},
    \Delta x_{p,i} \sim \calN\bigl(0,\sigma^2(x^*_p)\bigr)\bigr\}\,.
\end{align}
The size of the training data set $T$ is found to be
$|T| = n_s n_\text{base}$.  The trivial case with $n_s=1$ corresponds
to the naive training data set without augmentation, while
$n_s >1$ augments the training data set.  Figure~\ref{fig:fit_func}
exemplifies the training data set with $n_\text{base}=20$ and
$n_s=5$.

Since the task we deal with here is idealized and as simplified as
possible, we can reasonably keep the NN architecture simple and
relatively small (albeit deep).  In the present study we basically use
two types: \texttt{NN1221} and \texttt{NN1991} as tabulated in
Tab.~\ref{tab:NN1221}.  We also choose the simplest loss function
here as \texttt{mae}, i.e., the mean absolute errors:
\begin{align}
  \ell_{\texttt{abs}} (\by, \by')
  &\equiv \left|\by - \by'\right|\,.
    \label{eq:mean-absolute-error}
\end{align}

\begin{table}
  \begin{minipage}{0.47\textwidth}
  \centering
  \begin{tabular}{ccc} \hline
  Layer index & Neurons & Activation \\ \hline
  0 & 1 & N/A \\
  1 & 2 & Leaky ReLU \\
  2 & 2 & Leaky ReLU \\
  3 & 1 & Linear \\ \hline
  \end{tabular}
  (\texttt{NN1221})
  \end{minipage}
  \hfill
  \begin{minipage}{.47\textwidth}
  \centering
  \begin{tabular}{ccc} \hline
  Layer index & Neurons & Activation \\ \hline
  0 & 1 & N/A \\
  1 & 9 & Leaky ReLU \\
  2 & 9 & Leaky ReLU \\
  3 & 1 & Linear \\ \hline
  \end{tabular}
  (\texttt{NN1991})
  \end{minipage}
  \caption{NN architectures used in this work:  \texttt{NN1221} (left)
    and \texttt{NN1991} (right).
    We have chosen the leaky ReLU $f(x) = \alpha x\ (x<0),\ x\ (x\ge 0)$
    to avoid the dying ReLU problem. We fixed the leaky ReLU parameter $\alpha = 0.1$.}
  \label{tab:NN1221}
\end{table}

\subsection{Input noises, parameter fluctuations, and dropout}
\label{sec:paramfluc}

To deepen our understanding of the data augmentation, we argue that
including noises in the input is equivalent to introducing fluctuations to the NN
parameters in the first layer.  For the moment we will not limit
ourselves to the special toy model in Eq.~\eqref{eq:toy} but here we
shall develop a general formulation using the schematic notation
introduced in Sec.~\ref{sec:nnmethod}.

We consider the following case that a noise $\tilde\be$ is added to
input values: $\tilde\bx = \bx + \tilde\be$.   The noise is assumed to
be independent of the input value, i.e.,
$\Pr(\bx,\tilde\be) = \Pr(\bx)\Pr(\tilde\be)$.  In this case the input
noise $\tilde\be$ can be absorbed by redefinition of the first layer
parameters; that is,
\begin{equation}
  \bx^{(1)} = \sigma^{(1)}(W^{(1)}(\bx+\tilde\be) + \bb^{(1)})
  = \sigma^{(1)}(W^{(1)}\bx + \tilde\bb^{(1)})\,,
\end{equation}
where $\tilde\bb^{(1)} \eqdef \bb^{(1)}+W^{(1)}\tilde\be$.
Therefore, we find,
\begin{equation}
  f(\tilde\bx;W^{(1)}, \bb^{(1)}, \dots, W^{(L)}, \bb^{(L)})
  = f(\bx;W^{(1)}, \tilde\bb^{(1)}, \dots, W^{(L)}, \bb^{(L)})\,.
\end{equation}
The loss function with the input noise is given by
\begin{align}
  \mathcal{L}(\{W^{(\ell)}, \bb^{(\ell)}\}_{\ell})
  &= \int d\bx d\tilde\be \Pr(\bx,\tilde\be)
    \ell\bigl(\boldsymbol{y}, f(\tilde\bx;W^{(1)}, \bb^{(1)}, \dots, W^{(L)}, \bb^{(L)})\bigr) \notag\\
  &= \int d\tilde\be\Pr(\tilde\be) \int d\bx\Pr(\bx)
    \ell\bigl(\boldsymbol{y}, f(\bx;W^{(1)}, \tilde\bb^{(1)}, \dots, W^{(L)}, \bb^{(L)})\bigr) \notag\\
  &= \int d\tilde\be\Pr(\tilde\be) \mathcal{L}(W^{(1)}, \bb^{(1)}+W^{(1)}\tilde\be, \dots, W^{(L)}, \bb^{(L)})\,.
\end{align}
This is nothing but the loss function with fluctuations in the
first-layer biases $\bb^{(1)}$;  in summary, the noise $\tilde\be$ in
the input is translated into the fluctuation by $W^{(1)}\tilde\be$ in
the NN parameter $\bb^{(1)}$.  Once we realize this correspondence, we
can generalize this latter prescription of fluctuating NN parameters
not only in the first-layer $\bb^{(1)}$ but also in all the parameters
$\{W^{(\ell)},\bb^{(\ell)}\}_\ell$.  We will test this idea of generalization later.

We can further relate the parameter fluctuations to a standard NN
technique called \textit{dropout}.  Instead of adding random
fluctuations to the parameters (\textit{additive parameter noise}), we
can also think of multiplying random fluctuations to the parameters
(\textit{multiplicative parameter noise}),
i.e., $W_{ij}^{(\ell)} \to \tilde W^{(\ell)}_{ij} = W_{ij}^{(\ell)} \tilde e_{ij}^{(\ell)}$
and $b^{(\ell)}_i \to \tilde b^{(\ell)}_i = b^{(\ell)}_i \tilde e_i^{(\ell)}$
for a fixed $\ell$, where $\tilde e^{(\ell)}_{ij}$ and
$\tilde e^{(\ell)}_i$ are random fluctuations following a certain
distribution.  As a special case,
let us assume that $\tilde e^{(\ell)}_{ij}$ and $\tilde e^{(\ell)}_i$ follow
the simplest discrete probability distribution: the Bernoulli distribution $B(1,p)$,
where 0 and 1 appear with probability $p$ and $1-p$, respectively.
With an additional constraint of
$\tilde e^{(\ell)}_{ij} = \tilde e^{(\ell)}_i$,
putting these multiplicative parameter noise is exactly the procedure
commonly referred to as ``dropout'' with the dropout rate identified
as $p$.

In this way we can understand the parameter noises as generalization
of the data augmentation by noise injection and the dropout.
It is already well known
that the dropout is indeed an established tool to avoid the
overfitting problem.  Thus, we can expect that our observational data augmentation,
which was originally needed to handle the observational uncertainties,
have an effect similar to the dropout and can tame the overfitting
behavior.

\subsection{Numerical tests}
\label{sec:over}

Motivated by the relations between the data augmentation and the
dropout through parameter noise, as argued in the previous subsection,
we shall make a numerical comparison involving all of the data
augmentation, the NN parameter noise, and the dropout using the
toy model defined in Eq.~\eqref{eq:toy}.  For the parameter
fluctuation noise, we consider the additive parameter noise of the
normal distribution $\calN(0, \sigma^2)$ in the first-layer parameters,
$(W^{(1)}, \bb^{(1)})$, or in all the parameters,
$\{W^{(\ell)}, \bb^{(\ell)}\}_\ell$.  To manipulate this special type of
the training presented in this section, we implement the NN and its
optimization from scratch in \texttt{C++}.  As an ideal test case we
want a ``troublesome''  NN suffering from the overfitting problem.
We have first verified that the NN method with \texttt{NN1221} in
Tab.~\ref{tab:NN1221} can reasonably solve the reconstruction problem
of $f(x)$ while \texttt{NN1991} encounters the overfitting problem.
In what follows below, we will present the results
from \texttt{NN1991} to analyze the overfitting in details.
We fix $n_{\rm base} = 20$ and the mini-batch
size is 10.  We repeat carrying out the independent training
procedures with the same setup 100 times and take the average of the
loss function to draw the learning curves with probabilistic
distribution.

\begin{figure}
  \centering
  \begin{tabular}{cc}
    \includegraphics[width=.46\textwidth]{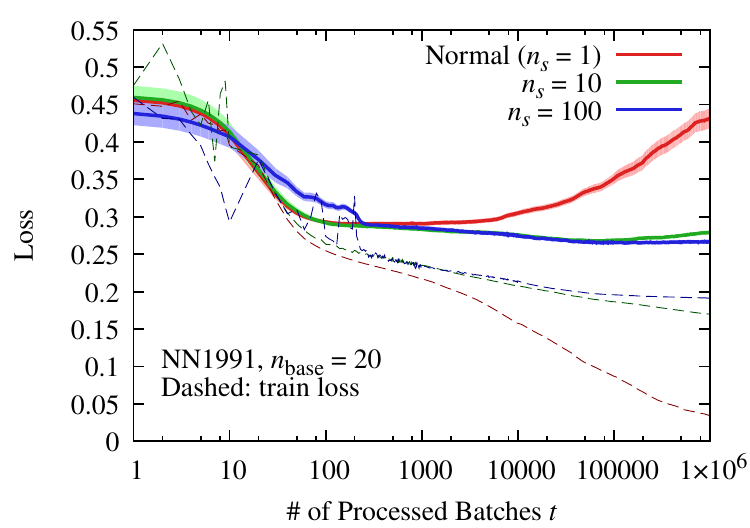} &
    \includegraphics[width=.46\textwidth]{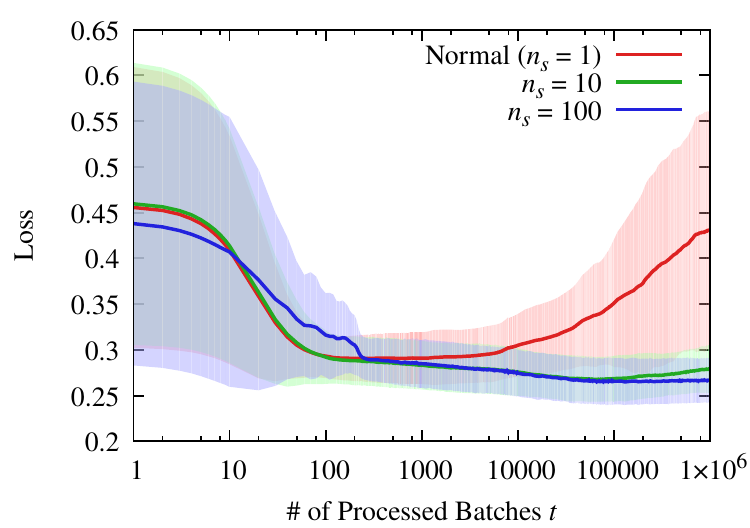} \\
    (a) & (b)
  \end{tabular}
  \caption{(a) Numerical experiment of increasing $n_s$.
    The mean values of the loss functions are represented by the
    dashed and the solid curves for the training and the validation
    losses, respectively.  The bands represents the standard errors.
    (b) The same plot with the bands representing the standard deviations.}
  \label{fig:ns_mean}
\end{figure}

\begin{figure}
  \centering
  \includegraphics[width=0.6\textwidth]{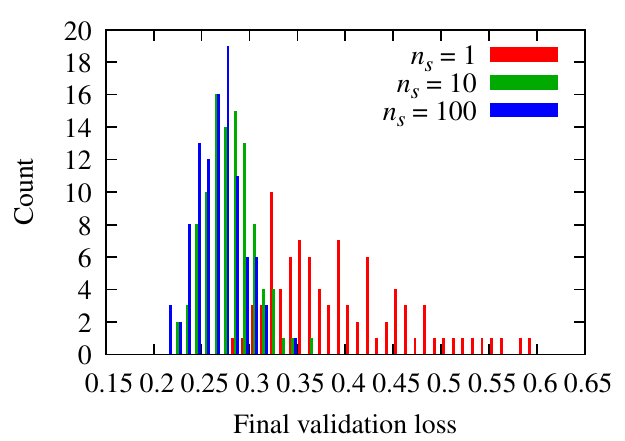}
  \caption{Histogram to show the 100 training distribution of the
    final validation loss at $t=10^6$.}
  \label{fig:ns_histogram}
\end{figure}

In Fig.~\ref{fig:ns_mean} we show the averaged learning curves for
several choices of $n_s$ to check the effect of the data augmentation.
The mean values of the loss function $\langle\mathcal{L}(t)\rangle$
are represented by the dashed (training) and the solid (validation)
curves along with the standard error band,
$\Delta\langle \mathcal{L}(t)\rangle$,
for the validation loss.  Here the (population) mean is estimated by
the sample mean over the 100 training, i.e.,
$\langle \mathcal{L}(t)\rangle
  \simeq \overline{\mathcal{L}(t)}
  = \frac1N \sum_{i=1}^N \mathcal{L}_i(t)$ with $N=100$,
where $t$ is the training time in the unit of the net number of
processed mini-batches, and $i$ is the index of the training.
We should stress that this process with $N=100$ is not the bagging,
but we do so to quantify the performance on average.
We also note that the learning curves of the validation loss
fluctuate, and the standard deviation $\Delta \mathcal{L}(t)$
[as shown in Fig.~\ref{fig:ns_mean}~(b)] is much larger than the
standard error band [as read from Fig.~\ref{fig:ns_mean}~(a)] that
quantifies the typical deviation of the sample mean
$\overline{\mathcal{L}(t)}$ from the ``true'' \textit{population mean}
$\langle \mathcal{L}(t)\rangle$.
Here, the standard deviation and the standard error are estimated by
the unbiased standard variance%
\footnote{The second line can be intuitively understood from
  the fact that the distribution of the mean $\langle \mathcal{L}(t)\rangle$
  narrows with the variance $(\Delta \mathcal{L}(t))^2 / N$
  by the central limit theorem.}:
\begin{align}
  \Delta \mathcal{L}(t)
  &\eqdef \sqrt{[\mathcal{L}(t) - \langle\mathcal{L}(t)\rangle]^2}
  \simeq \sqrt{\frac{N}{N-1}\overline{[\mathcal{L}_i(t)-\langle \mathcal{L}(t)\rangle]^2}}\,, \\
  \Delta\langle \mathcal{L}(t)\rangle
  &\eqdef \sqrt{[\overline{\mathcal{L}(t)} - \langle\mathcal{L}(t)\rangle]^2}
  = \frac1{\sqrt{N}}\Delta \mathcal{L}(t)\,.
\end{align}

As we speculated before, the results in Fig.~\ref{fig:ns_mean}~(a)
evidence that incorporating the data augmentation with $n_s>1$
reduces the overfitting behavior (i.e., the validation loss increasing
while the training loss decreasing).  Obviously the learning curve for
$n_s=1$ indicates the overfitting.  For a sufficiently large $n_s\gtrsim
10$, the validation loss does not increase any more and saturates
towards a certain value.  The improvement with $n_s\gtrsim 10$ can be
understood from the size of the data set,
$n_s n_{\rm base} \gtrsim 10 \cdot 20 = 200$, which is larger
than the number of the NN parameters, i.e., 118 for \texttt{NN1991}.
When $n_s$ is small, the size of the data set is too small as compared
to the NN parameters, so that the NN parameters cannot be well
constrained.  Even for $n_s=100$ there are still discrepancies between
the training and the validation losses as seen in
Fig.~\ref{fig:ns_mean}~(a), which can be explained as follows;  the
size of the independent training base data with $n_{\rm base} = 20$ is
too small.  The data augmentation does not supplement any additional
information on the training data $T$ but the contained information is
the same as the base data set $T^*$.  Therefore, the improvement
should be eventually saturated with increasing $n_s$.

The standard deviation in Fig.~\ref{fig:ns_mean}~(b) at an early time
is large reflecting the variance of initial NN parameters.  The
variance of the initial parameters quickly disappears around
$t \lesssim 100$.  An interesting observation is that the standard
deviation becomes larger again when the overfitting occurs, and its
magnitude is of the same order as the loss increase by the overfitting.
This means that in some trials of training the performance does not
necessarily get worse even in the overfitting time scale.  The actual
distribution of the final validation loss at $t = 10^6$ is shown in
Fig.~\ref{fig:ns_histogram} in a form of histogram.  We can see that
the validation loss spreads to larger values in the overfitting case
($n_s=1$), while the validation loss with the data augmentation
($n_s=10$ and $100$) is well localize around $\mathcal{L} = 0.25$.

\begin{figure}
  \centering
  \begin{tabular}{cc}
    \includegraphics[width=.46\textwidth]{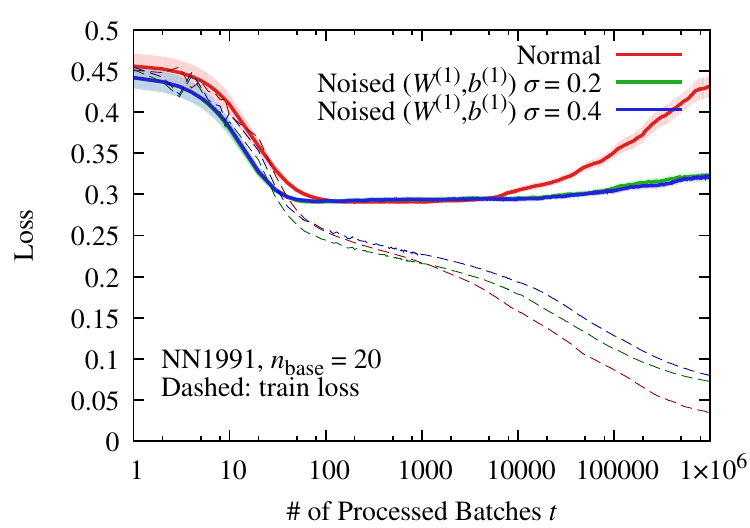} &
    \includegraphics[width=.46\textwidth]{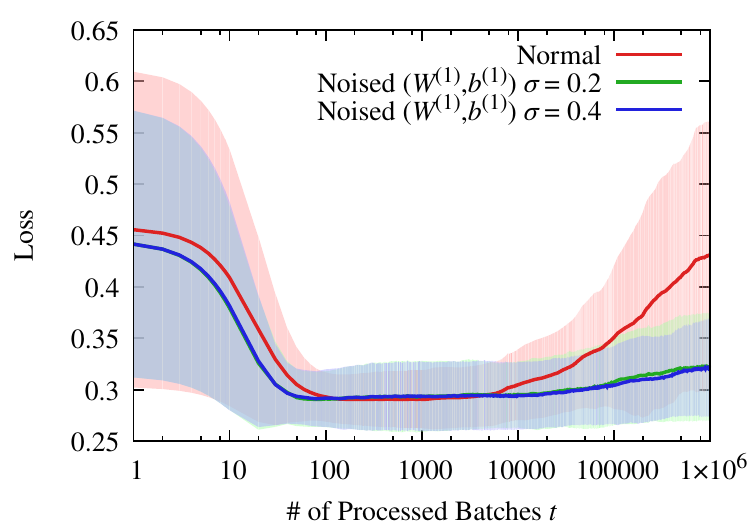} \\
    (a) & (b)
  \end{tabular}
  \caption{(a) Numerical experiment of adding noise to $(W^{(1)}, b^{(1)})$.
    (b) The same plot with the bands representing the standard deviations.}
    \label{fig:fluc_mean}
\end{figure}

In Fig.~\ref{fig:fluc_mean} we next check the effect of noise in the
NN parameters.  We introduce the additive parameter
noise of the normal distribution $\mathcal{N}(0, \sigma^2)$ with
several variances $\sigma = 0$, $0.2$, and $0.4$ to the first-layer
parameters $(W^{(1)}, b^{(1)})$.  From Fig.~\ref{fig:fluc_mean} we can
confirm the same trend as the observational data augmentation, and indeed
the overfitting is evaded similarly.  We have tried different values of
$\sigma = 0.2$ and $0.4$, but they are consistent with each other
within the standard errors.  We can also observe the behavior of the
standard deviation similar to the case of the data augmentation.

\begin{figure}
  \centering
  \begin{tabular}{cc}
    \includegraphics[width=.46\textwidth]{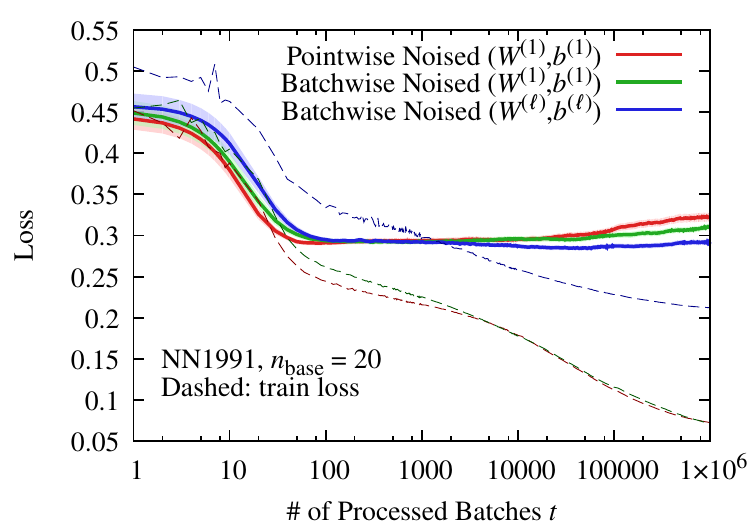} &
    \includegraphics[width=.46\textwidth]{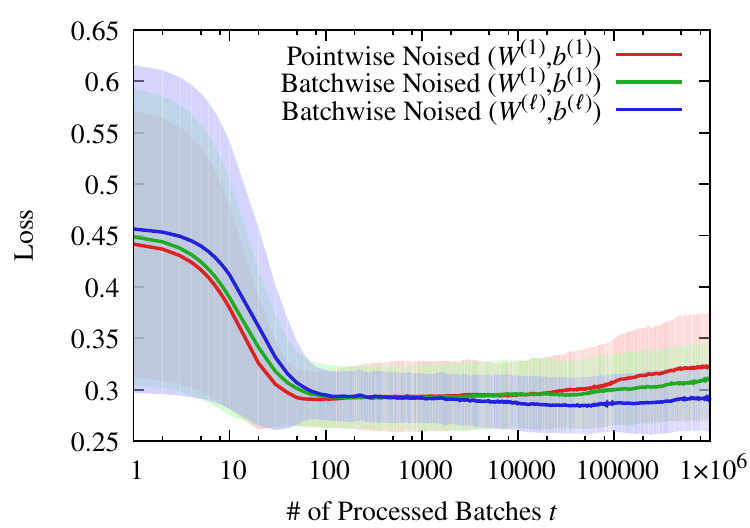} \\
    (a) & (b)
  \end{tabular}
  \caption{(a) Numerical experiment of adding noise in different
    ways;  see the text for more details.
    (b) The same plot with the bands representing the standard deviations.}
  \label{fig:fluc2_mean}
\end{figure}

We also compare different ways to introduce noise in the NN
parameters.   We here consider the following types of the parameter
noise to make a plot in Fig.~\ref{fig:fluc2_mean}:
\begin{enumerate}
\item
  Add random values sampled
  from the normal distribution $\mathcal{N}(0, \sigma^2)$ with
  $\sigma=0.2$ to the first-layer parameters $(W^{(1)}, b^{(1)})$.
  We resample parameter noises every time
  for each sample point in the training data
  [``Pointwise Noised $(W^{(1)}, b^{(1)})$'' in Fig.~\ref{fig:fluc2_mean}].
\item
  Add random values in the same way as (i) but resample
  parameter noises every time a mini-batch starts. We use
  a common set of the parameter noise values throughout a mini-batch.
  [``Batchwise Noised $(W^{(1)}, b^{(1)})$'' in Fig.~\ref{fig:fluc2_mean}].
\item
  Add random values sampled
  from the normal distribution $\mathcal{N}(0, \sigma^2)$ with
  $\sigma=0.2$ to all the NN parameters $(W^{(\ell)}, b^{(\ell)})$.
  We resample parameter noises every time a mini-batch starts as in
  (ii).
  [``Batchwise Noised $(W^{(\ell)}, b^{(\ell)})$'' in Fig.~\ref{fig:fluc2_mean}].
\end{enumerate}
We find that the performance is slightly improved for the batchwise
noise rather than the pointwise noise.  Also, better performance,
i.e., smaller validation losses are reached for all-layer noises as
prescribed in (iii) rather than for the first-layer noises
as in (i) and (ii).

\begin{figure}
  \centering
  \begin{tabular}{ccc}
    \includegraphics[width=.46\textwidth]{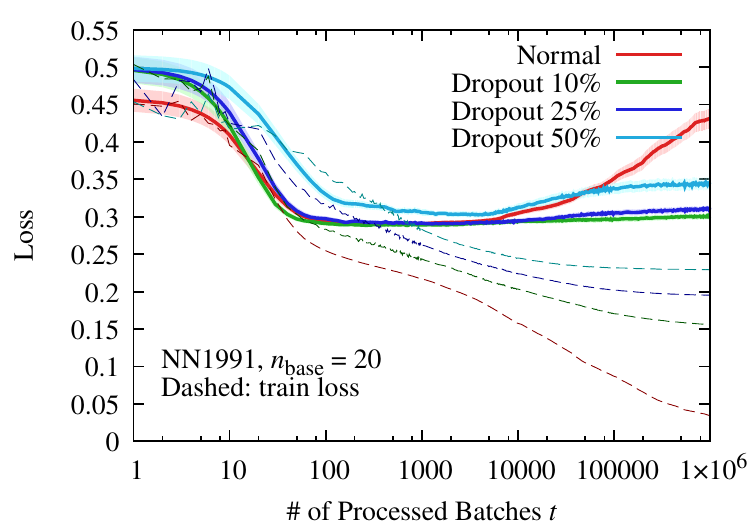} &
    \includegraphics[width=.46\textwidth]{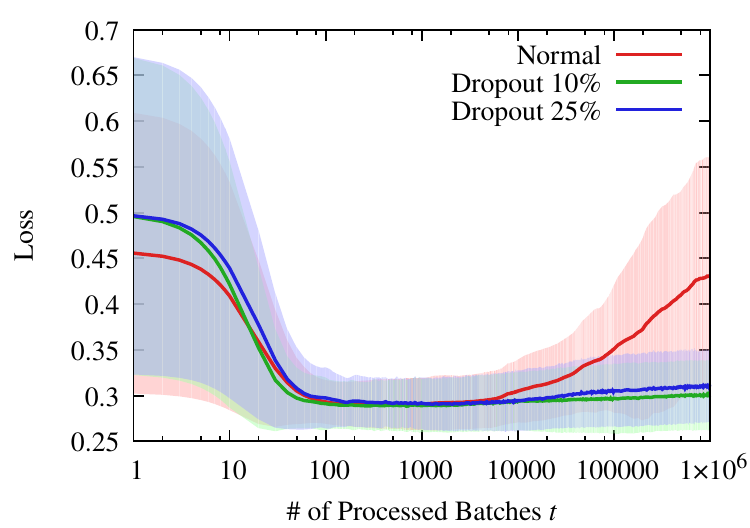} \\
    (a) & (b)
  \end{tabular}
  \caption{(a) Numerical experiment of inserting a dropout.
    (b) The same plot with the bands representing the standard deviations.}
  \label{fig:dropout_mean}
\end{figure}

Finally, we plot the results for the case with a dropout in the NN in
Fig.~\ref{fig:dropout_mean}.  Here again, we can validate the role of
the dropout to overcome the overfitting problem.  We have tried
several different dropout rates; namely, 10\%, 20\%, and 50\%, and
numerically confirmed that the performance is better for small values
of the dropout rate in the current setup.

\begin{figure}
  \centering
  \begin{tabular}{cc}
    \includegraphics[width=.46\textwidth]{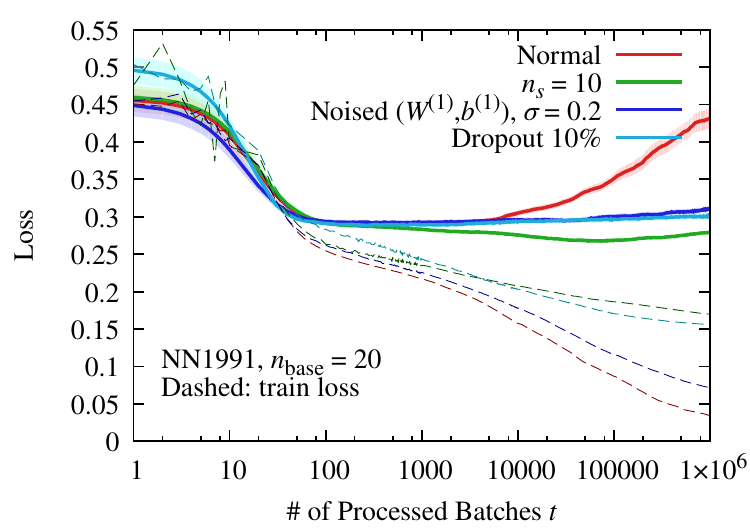} &
    \includegraphics[width=.46\textwidth]{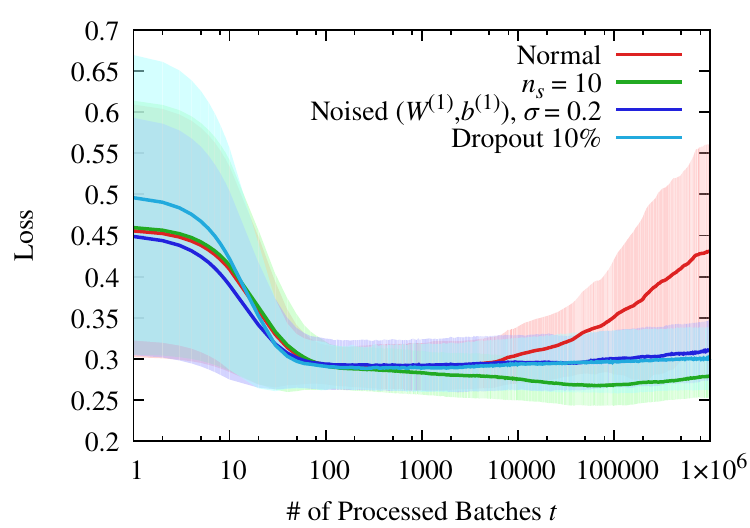} \\
    (a) & (b)
  \end{tabular}
  \caption{(a) Summary of the numerical experiment with the standard
    error bands.
    (b) The same plot with the bands representing the standard deviations.}
  \label{fig:all_mean}
\end{figure}

Figure~\ref{fig:all_mean} is the summary plot of the performance
comparison between three different approaches; namely, the observational data
augmentation, the parameter noise, and the dropout.  The parameter
noise and the dropout results are close to each other and almost
identical within the errors.  Among these three approaches, it has
turned out that the best performance is realized for the observational data
augmentation in the present toy-model setup.  The best strategy can
be different for different model circumstances.  The important point
is not a question of which is the best but the finding that the data
augmentation can tame the overfitting problem at the same level as the
standard procedure of the dropout.  This is a very interesting
observation.  In physics all observational data has uncertainties, and
the data is intrinsically augmented by repeated measurements with limited accuracy.
Then, for physics problems, this NN nature of the
``observational data augmentation'' can naturally evade the
overfitting problem by itself.

\section{Summary and Outlooks}
\label{sec:sum}

In this work, we have performed a comprehensive analysis of
inferring the equation of state (EoS) from the observed masses and
radii of neutron stars using the previously proposed method of deep
machine learning~\cite{Fujimoto:2017cdo, Fujimoto:2019hxv}.
Following the method elaborated in Sec.~\ref{sec:nseos}, we have
subdivided our analyses into three parts:  the examination of the deep
learning methodology using the mock data mimicking the real
observations (Sec.~\ref{sec:mock}), the application of the method to
the actual observational data of neutron stars (Sec.~\ref{sec:obs}),
and the discussion of our \textit{observational data augmentation} by
noise fluctuations in a simple setup idealized for theorization
(Sec.~\ref{sec:curve}).

In the first part (Sec.~\ref{sec:mock}) we have demonstrated how our
method works by presenting the typical two examples of the EoS
prediction.  Then we have quantified the uncertainty or error between
the prediction and the original value as given in
Eq.~\eqref{eq:cs2err} and inspected the prediction reliability for the
whole validation data set.  The upshot is that, by looking at the
histogram and the correlation between the prediction uncertainties in
different energy density regions, we have verified that the EoS
reconstruction fares well for the lower density regions, but the higher
density regions are not sufficiently constrained.
We have also compared the NN method to a more conventional approach,
i.e., the polynomial regression.  We then quantified the results in
terms of $\Delta R_{\rm RMS}$, $\Delta R_{\sigma}$, and
$\Delta R_{2\sigma}$ as given in Eqs.~\eqref{eq:dr} and
\eqref{eq:rsigma}.  We have reached a conclusion that our NN method
leads to more robust outputs, and the polynomial regression may have
outliers that damage the overall performance.

In the second part (Sec.~\ref{sec:obs}) we have applied the EoS
inference for real neutron star data from the various sources; namely,
qLMXBs, thermonuclear bursters, and the pulsar with hot spots on the
surface, at which the NICER mission aims.
We have also introduced natural and intuitive ways for uncertainty
quantification.  One is based on the validation loss and the other is
based on the procedure called the bagging.  We trained multiple NNs
independently and took the average for the bagging.  We have utilized
independent outputs from the multiple NNs to discuss the likelihood of
EoSs with a first-order phase transition.  We made a histogram to
count the rate of EoSs with a first-order transition in each energy
density segment.

In the third part (Sec.~\ref{sec:curve}) we have brought to light a
byproduct from our prescription of incorporating observational
uncertainty.  The training data is augmented with fluctuations
corresponding to observational uncertainty, which we call the
observational data augmentation.  Then, the data augmentation with
$n_s$ fluctuating copies has turned out to tame the overfitting
problem.  We have explicitly confirmed this behavior of the
overfitting reduction by monitoring a time evolution of the validation
loss.

Although we have carried out extensive analyses in this work, there
are still a lot to be done in the future to improve the applicability
of our method to the neutron star EoS inference.
As our NN method reliably predicts the most likely EoS in lower
density region, it would be supplementary to provide information on
the TOV limiting mass (the heaviest observed mass) directly in the NN
input.  At present we have incorporated this information indirectly by just
kicking out the training data that does not reach the TOV limit.
Also, our NN model currently copes with the pairs of mass-radius data
only, but it would be an important next step to extend the NN model so
as to process other forms of data, e.g., the tidal deformability.

Currently we employ the feedforward network.  It is also a natural
future extension to upgrade the NN architecture.  For instance, we can
replace the NN with the convolutional neural network (CNN) as
frequently used in the image processing.  With CNN we can make full
use of the two-dimensional Bayesian posterior distribution for the
input observable data.  On top of the improvement in the NN part, we
can also think of another extension of better EoS parametrization.  It
would be possible to use parametrizations other than the piecewise
polytropes, such as the spectral
representation~\cite{Lindblom:2010bb, Lindblom:2013kra}  and the
quarkyonic parametrization~\cite{Zhao:2020dvu}.  In this work we
jointed our predicted EoS above $n_0$ with the SLy4 below $n_0$.  The
dependence of the joint density and/or choices of the nuclear EoS
should be systematically examined.  Our naive expectation is that none
of these improvements would affect the results drastically.

The uncertainty quantification is another interesting issue.  In this
work we have used the practical but oversimplified techniques; namely,
the bagging and the validation loss.
More rigorous treatments of uncertainties may be possible by adopting
the CNN as mentioned above, or by combining the Bayesian method and NN
approach.  Using such a Bayesian NN, in which the network weights are
sampled from a probability distribution, we can obtain the posterior
probabilities of the outputs and thus we can capture the NN model
uncertainties correctly.  One efficient way to implement this
computation is to set up a dropout as outlined in
Refs.~\cite{pmlr-v48-gal16, NIPS2017_2650d608} (see also
Ref.~\cite{PerreaultLevasseur:2017ltk} for physics applications).  In
analogy to the explanation in Sec.~\ref{sec:paramfluc}, we can regard a
dropout as a Bayesian inference with likelihood functions set to the
Bernoulli distribution $B(1,p)$.  All above-mentioned extensions
deserve further investigations.

\begin{acknowledgments}
  This work was was supported by
  JSPS KAKENHI Grant No.\ 20J10506 (YF)
  and
  Grant Nos.\ 18H01211, 19K21874 (KF).
\end{acknowledgments}

\bibliographystyle{JHEP}
\bibliography{bib_nn}

\end{document}